To the University of Wyoming:

The members of the committee approve the dissertation of Hao Jiang presented on July/22/2014.

Dr. Hertanto Adidharma, Chairperson

Dr. Carrick Eggleston, External Department Member

Dr. Maciej Radosz

Dr. Joseph Holles

Dr. Maohong Fan

APPROVED:

David M. Bagley, Head, Department of Chemical & Petroleum Engineering

Steven F. Barrett, Associate Dean, College of Engineering and Applied Sciences

Hao Jiang, <u>Development of Statistical Associating Fluid Theory for Aqueous Ionic Liquid Solutions by Implementing Monte Carlo Simulations and Ornstein–Zernike Integral Equation: Application to Describing Gas Hydrate Inhibition Performance of Ionic Liquids</u>, Ph.D., Department of Chemical & Petroleum Engineering, August, 2014

A recent version of statistical associating fluid theory (SAFT), namely SAFT2, is coupled with the van der Waals and Platteeuw theory to study the alkane hydrate phase equilibrium conditions. The model is found to provide an accurate representation of the alkane hydrate dissociation conditions with and without inhibitors, such as salts, alcohols, as well as mixed salts and alcohol. Based on SAFT2, a heterosegmented SAFT equation of state is developed to model the thermodynamic properties of aqueous ionic liquid (IL) solutions, which is recently discovered as dual function gas hydrate inhibitors. With transferrable model parameters, the heterosegmented SAFT generally well represents the liquid density, activity coefficient, and osmotic coefficient of aqueous imidazolium IL solutions. The inhibition effects of imidazolium IL on methane hydrate is also studied by the heterosegmented SAFT and the van der Waals and Platteeuw theory. The roles of pressure, anion type, alkyl length of the cation, and IL concentration on the hydrate inhibition performance of the imidazolium ILs are well captured.

The heterosegmented SAFT is then modified to better represent the thermodynamic properties of aqueous IL solutions with the help of Monte Carlo simulation and the solutions of Ornstein–Zernike integral equation. Monte Carlo simulations are conducted on the fluid mixture of charged and neutral hard spheres to obtain its structure and excess energies, the results of which are compared with the thermodynamic properties predicted by solving Ornstein–Zernike equation with the Hypernetted Chain (HNC) and Mean Spherical Approximation (MSA) closures. A simple modification of MSA, referred to as KMSA, is proposed to accurately predict the excess energies of electrolyte system in mixture with neutral component. The KMSA improves the heterosegmented SAFT by taking the effect of neutral alkyl branches on the electrostatic interactions into consideration. Monte Carlo simulations are also conducted on flexible charged hard-sphere chain molecules, and a SAFT model which implements either a dimer or a dimer-monomer approach to account for the charged chain connectivity is proposed. With the SAFT model for charged hard-sphere chain, the cation heads and some anions of ILs are more accurately modeled as charged chains instead of charged spherical segments.

With these improvements to the heterosegmented SAFT, a more accurate representation of activity coefficient and osmotic coefficient is achieved, and the modeling of aqueous IL solutions is extended to ammonium ILs and imidazolium ILs with organic anions. The inhibition effects of ammonium ILs on methane hydrate is investigated using the improved heterosegmented SAFT coupled with the van der Waals and Platteeuw theory, which is demonstrated to be a predictive tool for the screening of effective IL based hydrate inhibitor.

Development of Statistical Associating Fluid Theory for Aqueous Ionic Liquid

Solutions by Implementing Monte Carlo Simulations and Ornstein–Zernike

Integral Equation: Application to Describing Gas Hydrate Inhibition Performance

of Ionic Liquids

by

Hao Jiang

A dissertation submitted to the Department of Chemical and Petroleum

Engineering and the University of Wyoming in partial fulfillment of the other

requirements for the degree of

DOCTOR of PHILOSOPHY

in

CHEMICAL ENGINEERING

Laramie, Wyoming

August, 2014

# COPYRIGHT PAGE

# ACKNOWLEDGMENTS

Dr. Adidharma, I would like to express my most sincere thanks to you for the guidance and encouragement in the past four years. None of my academic achievement is possible without your brilliant advices. You have been an understanding, kind and excellent mentor. I would never forget the support and help I got from you.

I want to express my most sincere thanks to all of my committee members, Dr. Maciej Radosz, Dr. Carrick Eggleston, Dr. Joseph Holles and Dr. Maohong Fan, for the valuable corrections and comments on this dissertation. I also want to thank Dr. Maciej Radosz and Dr. Maohong Fan for their precious recommendations for my postdoctoral application.

I would like to thank all my friends for their endless support and encouragement in the past four years. I also want to thank Heather Warren for her help and kindness.

A special thanks to my fiancée Limin Fu for simply being you. You are and will be the reason that I overcome all the difficulties. Finally, I want to thank my parents for everything they have given me. Their love is the most precious gift I could have in my life.

# TABLE OF CONTENTS

# List of Tables

# List of Figures

# Chapter 1. Introduction

Gas hydrates are clathrate solids that are formed from water and suitable gas molecules at high pressure and low temperature. The gas molecules are trapped in the cavities composed of hydrogen bonded water molecules. A few examples of hydrate-forming gases are methane, ethane, and carbon dioxide. The presence of natural gas hydrate reservoir in marine and frozen area has made gas hydrate a potential energy resource and a promising carbon dioxide sequestration method. Gas hydrate is a problem to the oil and gas industry due to their formation in valve, pipeline, etc., which could lead to catastrophic economic losses and ecological risks. Hence, the prevention of hydrate formation in oil and gas production system is of great importance. This in turn calls for the understanding and prediction of the hydrate phase equilibria in the presence of inhibitors.

Charged systems, also referred to as electrolyte systems or ionic systems, are known to have inhibition effect on gas hydrate. However, the thermodynamics properties of charged systems are difficult to model due to their long range electrostatic interactions. Recently, ionic liquids (IL), as charged fluids, are found to be dual function gas hydrate inhibitors;[1,2] ionic liquids are organic salts with low melting points, which make them liquids at room temperature. ILs are also known to be promising chemicals and solvents for green chemical processes because of their unique physicochemical properties, such as low vapor pressure, low toxicity, and high selectivity.

The gas hydrate phase equilibrium has received lots of attentions and various approaches have been developed to model the hydrate phase equilibrium in the presence of inhibitors, such as salts and alcohols.[3-12] In these models, the solid (gas hydrate) phase is generally described by van der Waals and Platteeuw theory,[13] which is a solid solution theory derived from statistical mechanics, while the fluid phases are usually described by cubic equations of state and activity coefficient models. In general, previous models are able to correlate or predict the hydrate phase equilibrium condition, but most of these models are empirical or semi empirical, which makes them less reliable for the description of gas hydrate inhibitors in wide temperature and concentration ranges. Moreover, these cubic-equation-of-state and activity-coefficient based hydrate modeling methods have difficulties describing the effects of ILs on hydrate phase equilibrium condition due to the IL's complex molecular structure and intermolecular interactions. Our goal in this dissertation is to develop a predictive model for the hydrate phase equilibrium in the presence of charged systems, including ILs. Statistical associating fluid theory (SAFT)[14-19] is selected as the method to describe the fluid phase since it has been successfully applied to many complex fluid systems including brines[20-23] and alcohols,[15,17] while the hydrate phase is modeled by the van der Waals and Platteeuw theory.[13]

Several systematic studies have been conducted in this dissertation before we eventually develop a model for the hydrate equilibrium conditions in IL solutions. In Chapter 2, a recent version of SAFT, referred to as SAFT2,[24] is coupled with the van der Waals and Platteeuw theory[13] to model the hydrate phase equilibria of pure alkanes and alkane mixtures in the absence of inhibitors.[25] In Chapter 3, ion-based SAFT2[21] is used to predict the alkane hydrate dissociation conditions in the presence of single and mixed electrolyte solutions, including NaCl, KCl, and CaCl$_2$.[26] In Chapter 4, the inhibition effects of alcohols including ethanol, ethylene glycol (MEG), and glycerol, and mixtures of MEG with electrolytes are studied using ion-based SAFT2.[27] In these three chapters, we are able to show that the SAFT2 equation of state can be successfully applied to the phase equilibrium modeling of alkane hydrates with or without the presence of electrolytes and alcohols.

In Chapter 5, the heterosegmented version of SAFT2 [28] is then used to study the thermodynamic properties of aqueous imidazolium IL solutions, where heterosegmented SAFT2 is coupled with the van der Waals and Platteeuw theory to predict the inhibition effects of ILs on methane hydrate. [29] We notice that for some aqueous IL solutions, the representation of thermodynamic properties using heterosegmented SAFT2 is not always satisfactory especially in the high concentration range, which indicates that a modification to SAFT2 is needed.

To develop a model that can more accurately represent the thermodynamic properties of ILs in a wider concentration range, the effects of the neutral segments of alkyl branches in the IL cation on the Coulomb interactions and more reasonable molecular depictions of cation and anion of IL may need to be considered, which are discussed in details in Chapters 6 and 7. In Chapter 6, the effects of neutral segments of alkyl branches in the IL cation on the electrostatic interactions are studied by Monte Carlo simulations and integral equation (Ornstein–Zernike equation). [30] An empirical modification to the analytical solution of the mean spherical approximation (MSA) is proposed, which is proved to accurately predict the excess energies of electrolyte systems in the presence of neutral species.

For more reasonable molecular depictions, the cation heads and anions of some ILs can be viewed as charged chain molecules or polyelectrolytes. Therefore, an equation of state, which is based on the thermodynamic perturbation theory of the first order, is developed for charged hard-sphere chain molecules in Chapter 7[31]. The chain connectivity of charged chain molecules is accurately accounted for by the equation of state.

After the effects of neutral species and chain connectivity on the properties of ILs have been studied in Chapters 6 and 7, an improved heterosegmented SAFT model for aqueous ILs solutions is proposed in Chapter 8. In addition, the inhibition effects of several ILs on methane hydrate are simulated and discussed with the improved heterosegmented SAFT model.

# Chapter 2. Hydrate Equilibrium Modeling for Pure Alkanes and Mixtures of Alkanes Using Statistical Associating Fluid Theory

## 2.1 Introduction

Gas hydrates are clathrate solids that are formed from water and gas molecules at high pressure and low temperature. The gas molecules are trapped into cavities composed of hydrogen bonded water molecules. Methane ($CH_4$), ethane ($C_2H_6$), carbon dioxide ($CO_2$), and nitrogen ($N_2$) are examples of hydrate-forming gas molecules. Three hydrate structures have been identified, namely, structures I, II, and H. The structures of hydrate are mainly dependent on the size of the gas molecule. Small molecules, such as $CH_4$ and $C_2H_6$, can form structure I hydrate, while larger molecules, such as propane and isobutane, can form structure II hydrate. Structure H hydrates are more complicated than structure I and II hydrates; examples of structure H formers include isopentane and cycloheptane.

Hydrate present a problem to the oil and gas industry because they usually form in the pipeline and cause plugging, which could result in catastrophic economic losses and ecological risks. Therefore, one of the most important research interests about hydrate is the prevention of its formation in pipelines, flow lines, and valves. This, in turn, calls for the understanding and prediction of the hydrate phase equilibria in the presence of inhibitors.

Many models have been proposed to predict the hydrate phase equilibrium conditions. Most of them are based on van der Waals and Platteeuw model (1959),[1] which describes the hydrate phase based on statistical mechanics. Cubic equations of state and activity coefficient models are the most widely used models for the fluid phase. However, when hydrate inhibitors are present, these fluid models usually need to be modified. Anderson and Prausnitz[2] proposed a model to calculate the inhibition effect of methanol, in which a modified Redlich-Kwong equation of state (EoS) was used for the vapor phase and a modified UNIQUAC model was used for the liquid phase. Mohammadi and Richon[3] predicted the inhibition effect of ethylene glycol on methane hydrate, using a modified Patel-Teja EoS (VPT-EoS[4]). For systems with salt solutions, a different activity coefficient model is often used. Englezos and Bishnoi[5] used the Pitzer activity coefficient model[6] to calculate the activity of water in the presence of electrolyte. In their work, a four-parameter Trebble-Bishnoi[7,8] cubic EoS was used for the vapor phase. Recently, a variant of the cubic EoS, called Cubic Plus Association (CPA),[9] was used to model the hydrate phase equilibria in the presence of alcohol[10] and electrolyte solutions.[11] The association part of this CPA, which was based on Weirtheim's first-order perturbation theory, was used to account for the association due to hydrogen bonding. However, for electrolyte solutions, an activity coefficient model was still needed, coupled with the CPA EoS.[11]

Recently, Li[12,13,14] proposed using a model based on Statistical Associating Fluid Theory (SAFT) to predict the hydrate phase equilibrium condition, both in uninhibited and inhibited systems; the SAFT EoS is an advanced equation of state based on statistical mechanics, and it has been successfully applied to the phase behavior of many industrially important fluids, such as alkanes,[15,16,17,18] alcohols,[19] polymers,[20,21,22,23] electrolyte solutions,[24,25,26] and ionic liquids.[27,28] A complete review of recent advances and applications of SAFT can be found in the literature.[29] The advantage of using SAFT is that the fluid phases, which may contain complex molecules, such as polymers, associating molecules, or charge molecules, can be described using the same theoretical framework. In the hydrate modeling work by Li, the SAFT EoS used was the first version of SAFT namely, SAFT-0 [30] and for alkane mixtures, only the methane/ethane/water and methane/propane/water systems were studied.[13] To the best of our knowledge, there is no hydrate modeling work for ternary gas mixture using SAFT. Furthermore, for systems without inhibitors, SAFT has never been applied to hydrate phase equilibrium modeling at temperatures below the water freezing point.

Since our future goal is to describe the hydrate phase equilibria in the presence of ionic liquids, which have been recently discovered as dual-function inhibitors,[31] we must use a model that has a high probability of success in describing charged molecules and ionic liquids. SAFT2[24] is such a model; it has been successfully applied to aqueous electrolyte solutions and ionic liquids. In this work, therefore, as our test case, we implement SAFT2[24] to describe the hydrate dissociation conditions for three pure alkanes, five binary alkanes, and one ternary alkanes mixtures in the uninhibited systems. For pure and ternary alkanes, the hydrate dissociation pressures are studied at temperatures both above (liquid-vapor-hydrate (LVH) phase equilibria) and below (ice-vapor-hydrate (IVH) phase equilibria) the freezing point of water.

## 2.2 Modeling

In hydrate phase equilibrium calculations, the chemical potential of each species in different phases should be equal. For hydrate-liquid water equilibrium, the chemical potential of water must satisfy the following equation:

$$\Delta \mu_W^H = \mu_W^{MT} - \mu_W^H = \mu_W^{MT} - \mu_W^L = \Delta \mu_W^L \tag{2.1a}$$

where $\mu_W^H$ is the chemical potential of water in the hydrate phase (J/mole), $\mu_W^L$ is the chemical potential of water in the liquid phase, and $\mu_W^{MT}$ is the chemical potential of water in the empty hydrate lattice. For hydrate-ice equilibrium, the following equation must be satisfied:

$$\Delta \mu_W^H = \mu_W^{MT} - \mu_W^H = \mu_W^{MT} - \mu_W^{ice} = \Delta \mu_W^{ice} \qquad (2.1b)$$

where $\mu_W^{ice}$ is the chemical potential of water in ice.

In this work, the chemical potential of water in the hydrate phase is calculated based on the van der Waals-Platteeuw model (1959),[1] while the fugacity of each species in the fluid phase is calculated using SAFT2 EoS.

## 2.2.1 SAFT2 Equation of State

Similar to SAFT-HR,[15,16] the SAFT2[24] EoS is defined in terms of the dimensionless residual Helmholtz energy:

$$\widetilde{a}^{res} = \widetilde{a}^{hs} + \widetilde{a}^{disp} + \widetilde{a}^{chain} + \widetilde{a}^{assoc} \qquad (2.2)$$

where the superscripts on the right side refer to terms accounting for the hard-sphere, dispersion, chain, and association interactions, respectively. We could also add $\widetilde{a}^{ion}$ if we are dealing with charged molecules, but of course this ionic term will not be used in this work. Although, in this work, we address only homosegmented molecules, the SAFT2 EoS is presented in a more general form for heterosegmented molecules, which will enable us to model more-complex molecules in the future. Its application to homosegmented molecules is just a special case, as indicated in the description of each term below.

## 2.2.1.1 Hard Sphere Term

The hard sphere term is the same as the other SAFT versions, which is given by Mansoori[32]

$$\widetilde{a}^{hs} = \frac{6}{\pi N_{Av} \rho_m} \left[ \frac{(\zeta_2)^3 + 3\zeta_1 \zeta_2 \zeta_3 - 3\zeta_1 \zeta_2 (\zeta_3)^2}{\zeta_3 (1-\zeta_3)^2} - \left( \zeta_0 - \frac{(\zeta_2)^3}{(\zeta_3)^2} \right) \ln(1-\zeta_3) \right] \qquad (2.3)$$

where $N_{Av}$ is the Avogadro number, $\rho_m$ is the molar density, and

$$\zeta_k = \frac{\pi}{6} N_{Av} \rho_m \sum_i X_i m_i \sum_\alpha x_\alpha (\sigma_\alpha)^k \qquad (k = 0, 1, 2, 3) \qquad (2.4)$$

where $X_i$ is the mole fraction of component $i$, $m_i$ is the number of segments of component $i$, $\sigma_\alpha$ is the diameter of segment $\alpha$, and $x_\alpha$ is the segment fraction, which is defined as

$$x_\alpha = \frac{\text{number of moles of segments } \alpha}{\text{number of moles of all segments}} \qquad (2.5)$$

For homosegmented molecules, component or molecule $i$ has segments of the same type and, thus, $\sigma_\alpha$ becomes $\sigma_i$ and $x_\alpha$ becomes $x_i \left( = \dfrac{m_i X_i}{\sum_j m_j X_j} \right)$. The summation over $\alpha$ should then be replaced by the summation over $i$.

### 2.2.1.2 Dispersion Term

The dispersion term is given by

$$\tilde{a}^{disp} = \left( \sum_i X_i m_i \right) \left( \frac{1}{k_B T} a_1^{disp} + \frac{1}{(k_B T)^2} a_2^{disp} + \tilde{a}^t \right) \qquad (2.6)$$

where $k_B$ is the Boltzmann constant, and $a_1^{disp}$ and $a_2^{disp}$ are the first- and second-order perturbation terms, respectively. The details of $a_1^{disp}$ and $a_2^{disp}$ can be found elsewhere.[17] The truncation error in Eq. 2.6 is given by the following empirical correlation[24,33]

$$\tilde{a}_t = \sum_{m=2}^{5} \sum_{n=1}^{2} D_{mn} \left( \frac{u}{kT} \right)^m \left( \frac{\zeta_3}{\tau} \right)^n \qquad (2.7)$$

where $D_{mn}$ represents the universal constants,[24] $\tau$ is the closed packing fraction, and $u$ is the temperature-dependent potential depth.

For homosegmented molecules, all of the parameters become those for homosegmented molecules and thus the summations over $\alpha$ and $\beta$ in the calculations of $a_1^{disp}$ and $a_2^{disp}$ should be replaced by the summations over $i$ and $j$.

### 2.2.1.3 Chain Term

The chain term is related to the pair radial distribution function of the reference fluid (square-well fluid) as follows:[17]

$$\widetilde{a}^{chain} = -\sum_i X_i (m_i - 1) \Big[ \ln \overline{g}_i^{SW}(\sigma_{\alpha\beta}) - \ln \overline{g}_{0,i}^{SW}(\sigma_{\alpha\beta}) \Big] \tag{2.8}$$

$$\ln \overline{g}_i^{SW}(\sigma_{\alpha\beta}) = \sum_{\beta \geq \alpha} B_{\alpha\beta,i} \ln g_{\alpha\beta}^{SW}(\sigma_{\alpha\beta}) \tag{2.9}$$

where $g_{\alpha\beta}^{SW}(\sigma_{\alpha\beta})$ is the radial distribution function of square well fluids evaluated at the contact distance, $\overline{g}_0^{SW}$ is $\overline{g}^{SW}$ evaluated at zero density, and $B_{\alpha\beta,i}$ is the bond fraction of type $\alpha\beta$ in molecule of component $i$. The details of the association term can be found elsewhere.[17] For homosegmented molecules, component or molecule $i$ has segments of the same type and thus has only one type of bond. In this case, Eq. 2.9 becomes $\ln \overline{g}_i^{SW}(\sigma_{\alpha\beta}) = \ln g_{ii}^{SW}(\sigma_i)$.

### 2.2.1.4 Association Term

The residual Helmholtz energy that accounts for the association is based on the equation given by Chapman et al:[30]

$$\tilde{a}^{assoc} = \sum_i X_i \sum_\alpha \left[ \sum_{A_{\alpha i}} \left( \ln X^{A_{\alpha i}} - \frac{X^{A_{\alpha i}}}{2} \right) + \frac{n(\Gamma_{\alpha i})}{2} \right] \qquad (2.10)$$

where $n(\Gamma_{\alpha i})$ is the number of association sites on segment $\alpha$ in molecule of component $i$, and $X^{A_{\alpha i}}$ is the mole fraction of molecule of component $i$ not bonded at side $A$ of segment $\alpha$. The details of the association term can be found elsewhere.[19]

For homosegmented molecules, each molecule has segments of the same type and the association is then considered between site $A$ in molecule $i$ and site $B$ in molecule $j$. Therefore, the summation over all segment types in Eq. 2.10 is not needed and $A_{\alpha i}$ and $\Gamma_{\alpha i}$ become $A_i$ and $\Gamma_i$, respectively.

## 2.2.2 Models for Chemical Potential Differences

The chemical potential difference between water in hypothetical empty hydrate lattice and that in hydrate phase is given by the van der Waals and Platteeuw model:[34]

$$\Delta \mu_w^H = \mu_w^{MT} - \mu_w^H = -RT \sum_m \nu_m \ln(1 - \sum_j \theta_{mj}) \qquad (2.11)$$

where $\nu_m$ is the number of cavities of type $m$ per water molecule in the hydrate phase, and $\theta_{mj}$ is the occupancy of molecules $j$ in type $m$ cavities. The occupancy is expressed as[34]

$$\theta_{mj} = \frac{C_{mj} \hat{f}_j^v}{1 + \sum_k C_{mk} \hat{f}_k^v} \qquad (2.12)$$

where $\hat{f}_j^v$ is the fugacity of component $j$ in the hydrate phase, which is equal to that of the vapor and liquid phases, and $C_{mj}$ is the Langmuir constant, which represents the water-gas interaction.

The fugacity of component $j$ is calculated using the SAFT2 EoS by invoking the equilibrium condition between the vapor and liquid phases.

The Langmuir constant can be calculated using the following equation:[35]

$$C_{mj}(T) = \frac{4\pi}{kT} \int_0^{R_m} \exp\left(\frac{-W(r)}{kT}\right) r^2 dr \qquad (2.13)$$

where $R_m$ is the radius of type $m$ cavity and $W(r)$ is the cell potential function developed by McKoy and Sinanoghu,[36] based on the Kihara potential. The details of the cell potential function can be found elsewhere.[34] To simplify the calculation, Parrish and Prausnitz[34] proposed an empirical correlation that could be used to calculate the Langmuir constants in the temperature range of 260-300 K:

$$C_{mj}(T) = \left(A_{mj}/T\right) \exp\left(B_{mj}/T\right) \qquad (2.14)$$

where $A_{mj}$ and $B_{mj}$ are constants for molecule $j$ in cavity of type $m$, which were fitted to match Eq. 2.13. The values of $A_{mj}$ (expressed in units of K/atm) and $B_{mj}$ (given in Kelvin) for some gases are given elsewhere.[34] In this work, we use Eq. 2.14 with the suggested values of $A_{mj}$ and $B_{mj}$ for temperatures down to 240 K. We find that even in the temperature range of 240-260 K, for the gas species that we studied, the maximum deviation of Eq. 2.14 using the suggested constants is still < 0.4%.

The chemical potential difference between water in the empty hydrate lattice and that in the liquid phase is expressed as follows:[37]

$$\frac{\Delta\mu_w^L(T,P)}{RT} = \frac{\Delta\mu_w^0(T_0,0)}{RT_0} - \int_{T_0}^T \left(\frac{\Delta h_w^{MT-L}}{RT^2}\right) dT + \int_0^P \left(\frac{\Delta V_w^{MT-L}}{RT}\right) dP - \ln\gamma_w x_w \qquad (2.15a)$$

where $\Delta\mu_w^0(T_0,0)$ is the reference chemical potential difference between water in an empty hydrate lattice and pure water at 273.15K ($T_0$) and 0 Pa, $\gamma_w$ is the activity coefficient of water, which is assumed to be 1.0, since inhibitors are not present, and $x_w$ is the mole fraction of water. The chemical potential difference between water in the empty hydrate lattice and that in ice is given by

$$\frac{\Delta\mu_w^{ice}(T,P)}{RT} = \frac{\Delta\mu_w^0(T_0,0)}{RT_0} - \int_{T_0}^{T}\left(\frac{\Delta h_w^{MT-ice}}{RT^2}\right)dT + \int_0^P\left(\frac{\Delta V_w^{MT-ice}}{RT}\right)dP \qquad (2.15b)$$

In Eq. 2.15a, $\Delta h_w^{MT-L}$ is the molar enthalpy difference between the empty hydrate lattice and liquid water at zero pressure, which can be calculated by the following equation: [37]

$$\Delta h_w^{MT-L} = \Delta h_w^{MT-ice,0} + \Delta h_w^{ice-L,0} + \int_{T0}^{T}\left(\Delta Cp_w^{MT-L,0} + b^{MT-L}(T-T_0)\right)dT \qquad (2.16a)$$

where $\Delta Cp_w^{MT-L,0}$ and $b^{MT-L}$ are constants given by Holder et al.,[38] $\Delta h_w^{ice-L,0}$ is the molar enthalpy difference between ice and liquid water at $T_0$, which is taken from Sun[37] ($\Delta h_w^{ice-L,0}$ = –6009.5 J/mol), and $\Delta h_w^{MT-ice,0}$ is the molar enthalpy difference between empty hydrate lattice and ice at $T_0$. Similarly, $\Delta h_w^{MT-ice}$ in Eq. 2.15b is the molar enthalpy difference between the empty hydrate lattice and ice at zero pressure, given by

$$\Delta h_w^{MT-ice} = \Delta h_w^{MT-ice,0} + \int_{T0}^{T}\left(\Delta Cp_w^{MT-ice,0} + b^{MT-ice}(T-T_0)\right)dT \qquad (2.16b)$$

where $\Delta Cp_w^{MT-ice,0}$ and $b^{MT-ice}$ are constants also given by Holder et al.[38]

The $\Delta V_w^{MT-L}$ in Eq. 2.15a is the molar volume difference between the empty hydrate lattice and liquid water and the $\Delta V_w^{MT-ice}$ in Eq. 2.15b is the molar volume difference between the empty hydrate lattice and ice. The molar volume of the structure I empty hydrate used in this work is fitted by Sun[39] to the thermal expansion and isothermal compressibility of $CH_4$ hydrate. The molar volumes of liquid water and ice are calculated by using correlations taken from Klauda and Sandler.[40] For the structure II hydrate, the volume difference between the empty hydrate lattice and ice, and the volume difference between the empty hydrate lattice and liquid water used in this work are constants, as given by Yoon.[41]

## 2.3 Model Parameters

### 2.3.1 Parameters for SAFT2

In SAFT2, for non-associating fluid, four parameters are needed: number of segments in a molecule ($m$), the segment volume ($v$), the well depth of the square well potential ($u^0$), and the reduced width of the square well potential ($\lambda$). For associating fluids, two additional parameters i.e., parameter related to the volume available for bonding ($\kappa^{AB}$) and association energy ($\varepsilon^{AB}$), are needed. The parameters for alkane and water are taken from Tan et al.[24] The parameters for SAFT2 are listed in the Table 2.1.

The temperature-dependent binary interaction parameters ($k_{ij}$) between water and alkanes are fitted to the binary vapor-liquid equilibrium data, while the $kij$'s between two different alkanes are set to 0. The binary interaction parameters are listed in Table 2.2.

Table 2.1. Parameters for SAFT2

| | $m$ | $v$ (cc/mol) | $u^0/k$ (K) | $\lambda$ |
|---|---|---|---|---|
| C1 | 1.0 | 15.2581 | 130.5206 | 1.5942 |
| C2 | 1.333 | 16.9740 | 194.6721 | 1.5628 |
| C3 | 1.667 | 18.6313 | 214.1275 | 1.5651 |
| iC4 | 2.096 | 19.758 | 224.5713 | 1.56475 |
| water[‡] | 1.0 | 9.8307 | 311.959 | 1.5369 |

[‡]For water, $\varepsilon^{AB}/k$ =1481.41 K, $\kappa^{AB}$ =0.04682.
[†]$k$ is the Boltzmann constant.

Table 2.2. Binary Interaction Parameters for SAFT2 ($k_{ij} = a + b \cdot T$ [K])

| system | C1/water | C2/water | C3/water | iC4/water |
|---|---|---|---|---|
| $a$ | −0.487 | −0.227 | −0.317 | −0.155 |
| $b$ | 0.0013 | 0.0007 | 0.001 | 0.00049 |

### 2.3.2 Reference Properties

The reference properties, $\Delta\mu_w^0(T_0,0)$ and $\Delta h_w^{MT-ice,0}$, should be obtained before Eq. 2.15 can be used. Since the empty hydrate lattice can never be synthesized in the laboratory, these reference properties are obtained by fitting to experimental data. In this work, for structure I hydrate, the reference properties are fitted to the experimental ice-vapor-methane hydrate phase equilibrium data.

Reference properties can be regressed by using the following equation:[42]

$$\frac{\Delta\mu_w^H}{RT} + \frac{(\Delta Cp_w^{MT-ice,0} - b^{MT-ice}T_0)\ln(T/T_0)}{R} - \frac{2\Delta Cp_w^{MT-ice,0}T_0 - b^{MT-ice}T_0^2}{2R}\left[\frac{1}{T} - \frac{1}{T_0}\right] +$$

$$\frac{b^{MT-ice}}{2R}(T-T_0)-\int_0^P \frac{\Delta v_w^{MT-ice}}{RT}dP=\frac{\Delta\mu_w^0(T_0,0)}{RT_0}+\frac{\Delta h_w^{MT-ice,0}}{R}\left[\frac{1}{T}-\frac{1}{T_0}\right] \qquad (2.17)$$

If the left hand side of the equation is treated as $Y$, and $\left[\dfrac{1}{T}-\dfrac{1}{T_0}\right]$ is treated as $X$, this equation is,

in fact, a linear function, from which the reference chemical potential difference ($\Delta\mu_w^0(T_0,0)$)

and the reference enthalpy difference ($\Delta h_w^{MT-ice,0}$) can be obtained.

Figure 2.1 shows the result of the regression for the structure I hydrate. The reference properties for the structure II hydrate are taken from Yoon.[41] The reference properties for both structures I and II hydrates are given in Table 2.3.

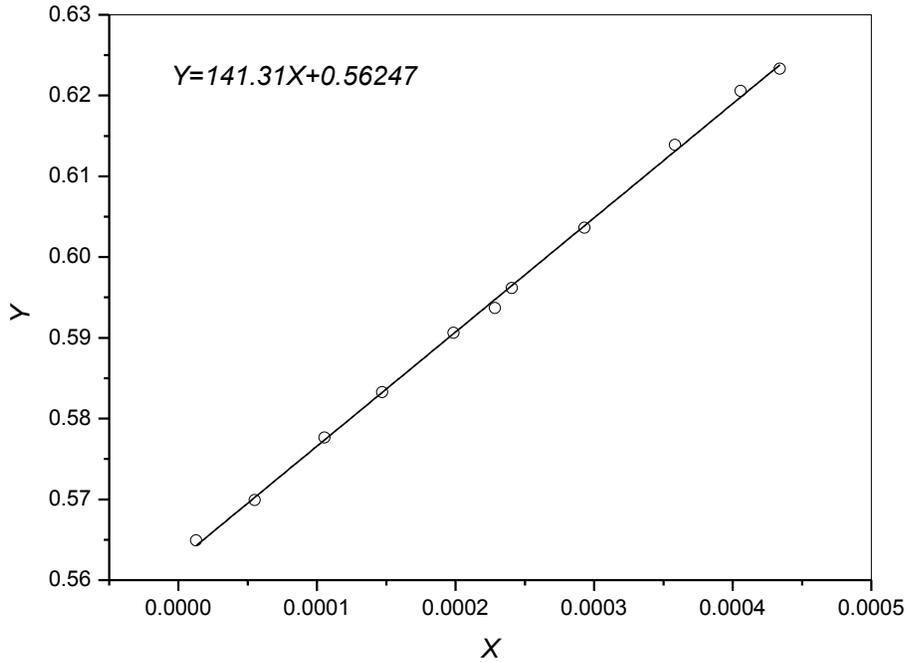

Figure 2.1. Linear regression for the structure I hydrate based on Eq. 2.17 to obtain the reference properties in Eq. 2.15 (data points represent experimental data[43]) and solid line represent the regression line.

Table 2.3. Reference properties for the Structure I and Structure II Hydrates

|  | $\Delta\mu_w^0(T_0,0)$ (J/mol) | $\Delta h_w^{MT-ice,0}$ (J/mol) |
|---|---|---|
| Structure I | 1277 | 1174 |
| Structure II | 883 | 808 |

## 2.4 Results and Discussion

In this work, hydrate dissociation pressures for three pure alkane hydrates, five binary alkane mixture hydrates, and one ternary alkane mixture hydrate are studied. Table 2.4 enlists all of the systems studied, along with the average absolute deviations of the calculated/predicted pressures.

Table 2.4. Systems Studied

| Component 1 | Other component(s) | T-range/K | P-range/bar | Hydrate structure | AAD(P)% | Data source |
|---|---|---|---|---|---|---|
| methane |  | 244.2–272.2 | 9.71–24.71 | I | 0.45 | [43][a] |
|  |  | 275.15–300.15 | 31.7–545.3 | I | 3.14 | [44] |
| ethane |  | 244.9–271.9 | 1.22–4.43 | I | 0.55 | [43] |
|  |  | 273.9–285.4 | 5.08–22.03 | I | 1.61 | [45] |
| propane |  | 245–272.1 | 0.41–1.627 | II | 4.81 | [43] |
|  |  | 274.2–278.4 | 2.07–5.42 | II | 1.38 | [46] |
| ethane | 4.7% methane | 279.4–287.6 | 9.9–29.9 | I | 4.02 | [47] |
|  | 17.7% methane | 281.6–287 | 14.2–30 | I | 2.81 | [47] |
|  | 56.4% methane | 274.8–283.2 | 9.45–24.34 | I | 4.65 | [45] |
|  | 80.9% methane | 288.8–296.4 | 70–234.8 | I | 8.11 | [48] |
|  | 94.6% methane | 284.9–296.6 | 69.3–344.4 | II | 4.93 | [48] |
| propane | 23.75% methane | 274.9–281.4 | 2.63–8.3 | II | 3.84 | [49] |
|  | 36.2% methane | 274.8–280.4 | 2.72–6.87 | II | 3.17 | [45] |
| | 71.2% methane | 274.8–283.2 | 3.65–11.51 | II | 0.91 | [45] |
| | 95.2% methane | 274.8–283.2 | 8.14–22.27 | II | 0.30 | [45] |
| isobutane | 71.4% methane | 273.9–282.7 | 2.08–7.86 | II | 3.06 | [50][b] |
| | 84.8% methane | 274–288.9 | 3.04–20.3 | II | 5.99 | [50] |
| | 97.5% methane | 277.8–289.3 | 10.8–45.6 | II | 6.61 | [50] |
| ethane | 18.6% propane | 273.1–279.6 | 5.4–13 | I | 2.67 | [51] |
| | 34.2% propane | 273.9–277.6 | 4.4–10.6 | II | 5.21 | [51] |
| | 54.1% propane | 275.8–278 | 5–8.5 | II | 2.68 | [51][b] |
| propane | iso-butane | 272.2 | 1.082–1.537 | II | 2.57 | [52] |
| methane | 0.507% ethane 0.0203% propane | 243.7–266.9 | 9.39–20.7 | I | 0.78 | [53] |
| | 7.06% ethane 2.988% propane | 247.5–271 | 2.33–7.11 | II | 3.56 | [53] |
| | 12.55% ethane 2.93% propane | 277.1–296.14 | 11.98–170.34 | II | 2.26 | [54][b] |

[a]Used to obtain reference properties in Eq. 2.15
[b]Used to obtain ω

## 2.4.1 Pure Alkane Hydrates

The model is used to calculate and predict the LVH and IVH phase equilibrium boundaries for methane, ethane, and propane hydrates. Figures 2.2 and 2.3 show the IVH and LVH boundaries of methane hydrate, respectively. Since the reference properties are fitted to the IVH boundary of methane hydrate, Figure 2.2 shows the calculated data based on the regression. For the LVH boundary of methane hydrate, in general, the prediction agrees well with the experimental data,[44] except at high pressure, which might call for pressure-dependent Langmuir constants.[55] We will investigate this issue in our future work. Figures 2.4 and 2.5 show the hydrate equilibrium boundaries of ethane and propane, respectively, and the experimental data[43, 45, 46] are well predicted.

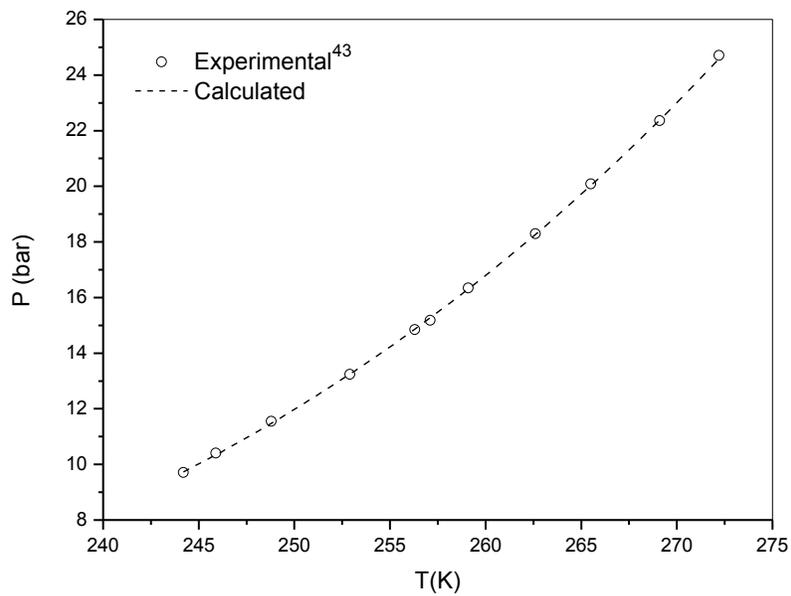

Figure 2.2. Hydrate dissociation pressure for methane (ice-vapor-hydrate (IVH)).

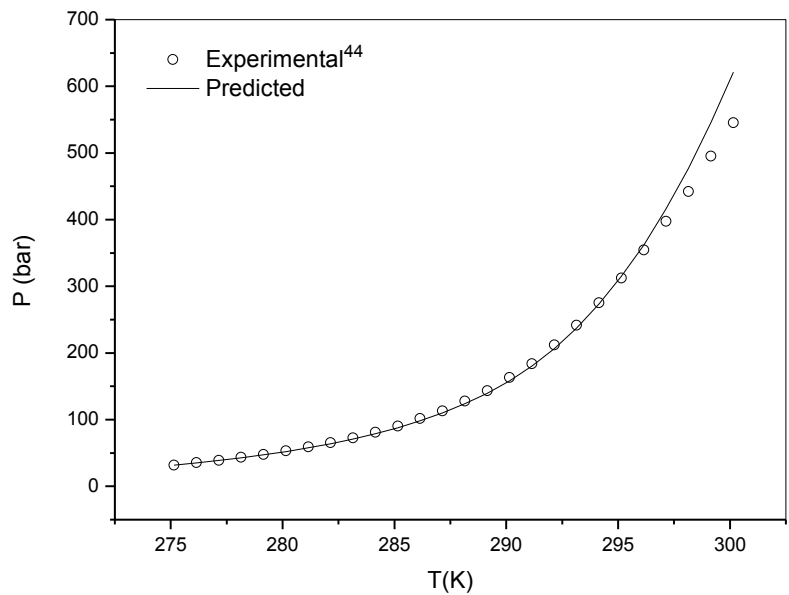

Figure 2.3. Hydrate dissociation pressure for methane (liquid-vapor-hydrate (LVH)).

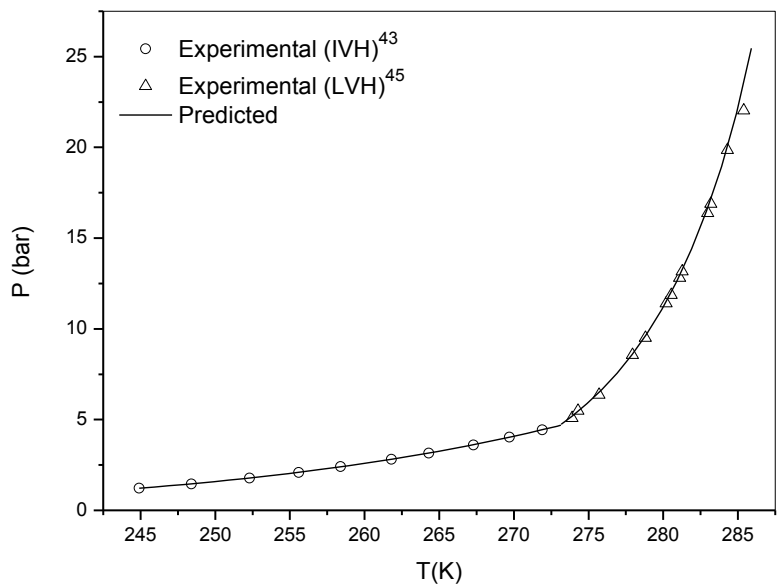

Figure 2.4. Hydrate dissociation pressure for ethane.

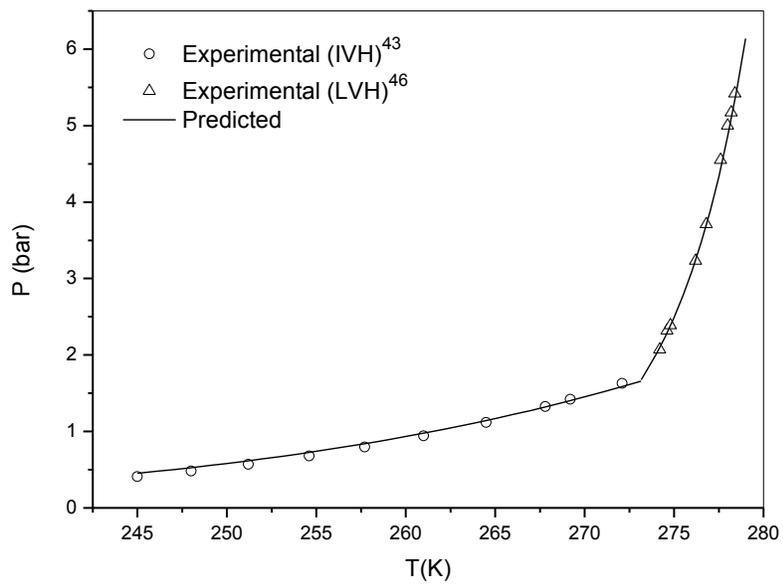

Figure 2.5. Hydrate dissociation pressure for propane.

## 2.4.2 Binary Alkane Mixture Hydrates

For binary alkane mixtures, we consider the LVH boundaries for methane/ethane, methane/propane, methane/isobutane, ethane/propane, and propane/isobutane. These systems involve both structure I and II hydrates.

One could argue that Langmuir constants calculated from the correlation given by Parrish and Prausnitz[34] could result in inaccurate predictions of the phase equilibrium boundaries of binary or ternary gas hydrate systems, because it has been stated that the Kihara potential parameters used by Parrish and Prausnitz[34] did not agree with those obtained from virial coefficient and viscosity data.[56] In our work, it turns out that the hydrate dissociation pressures of alkane mixtures forming structure I hydrate can be predicted by our model that uses the Langmuir constants suggested by Parrish and Prausnitz.[34] For example, Figure 2.6 shows the LVH boundaries for methane/ethane mixtures having different compositions, in which structure I hydrate forms. The model can capture the effect of concentration on the dissociation pressure well.

For alkane mixtures that form the structure II hydrate, however, the model prediction of hydrate dissociation pressure shows some discrepancy. Although Kihara parameters that agree well with those obtained from virial coefficient and viscosity data that are available, such as those obtained by Holder et al.,[57] the average errors of the prediction of hydrate dissociation pressure for certain binary gas mixtures using such parameters are still unsatisfactory. Hence, in this work, an empirical correlation proposed by Ng and Robinson[58] is used for binary alkane mixtures forming structure II hydrates. The correlation for calculating the LVH boundary of binary hydrate mixtures proposed by Ng and Robinson[58] is as follows:

$$\Delta\mu_w^L = RT\left[1 + 3(\omega-1)Y_1^2 - 2(\omega-1)Y_1^3\right] \times \left[\sum_m v_m \ln(1 + \sum_j C_{mj}\hat{f}_j^v) + \ln(\gamma_w x_w)\right] \qquad (2.18)$$

where $Y_1$ is the mole fraction of the more volatile gas component in the vapor phase, and $\omega$ is a constant that should be fitted to experimental data. Thus, the term $\left[1 + 3(\omega-1)Y_1^2 - 2(\omega-1)Y_1^3\right]$ in Eq. 2.18 is the correction factor. Table 2.5 enlists the values of $\omega$ used for binary alkane mixture hydrates. Note that the value of $\omega$ is not dependent on the mixture composition.

The difficulty of the modeling for binary and ternary hydrate systems may also be due to the basic assumption of van der Waals and Platteeuw model,[1] i.e., there is no interaction between guest molecules, but at this stage, we do not attempt to relax this assumption.

Figure 2.7 shows the LVH boundaries for methane/ethane system at higher concentration of methane. At a methane concentration of 80.9%, the structure I hydrate still forms and the dissociation pressure is predicted without using any correction. At a methane concentration of 94.6%, however, the structure II hydrate forms and the correction factor shown in Eq. 2.18 is used. Although the value of $\omega$ is close to 1, this correction is still needed to represent the experimental data well.[48] If $\omega$ were set to 1, the model would overpredict the experimental data.

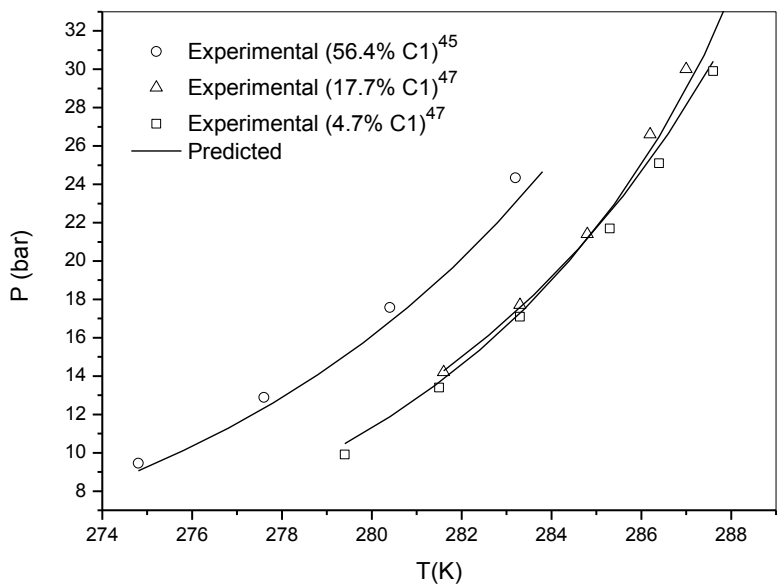

Figure 2.6. Hydrate dissociation pressure for methane/ethane mixtures.

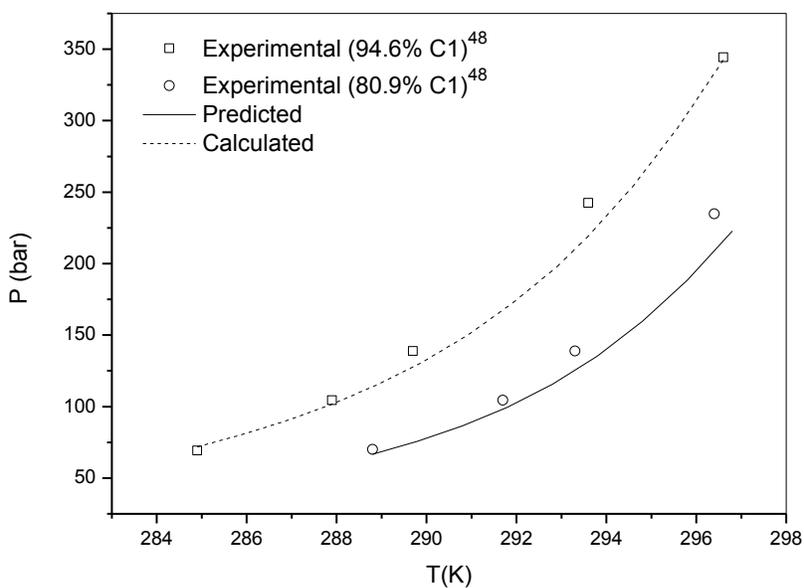

Figure 2.7. Hydrate dissociation pressure for methane/ethane mixtures at higher concentration of methane.

Table 2.5. Constant Used in Eq. 2.18 for Binary Gas Mixture Hydrate

| Binary mixture | C1 + C2 | C1 + C3 | C1 + iC4 | C2 + C3 | C3 + iC4 |
|---|---|---|---|---|---|
| ω | 1.02[†] | 1.0[†] | 1.03[‡] | 1.045[‡] | 1.0[‡] |

[†]Values given by Ng and Robinson[58]

Figure 2.8 shows the hydrate dissociation pressure of methane/propane mixtures having different compositions. Although the structure II hydrate forms in all cases, for this system, no correction is needed ($\omega = 1$) and the dissociation pressures are well-predicted.

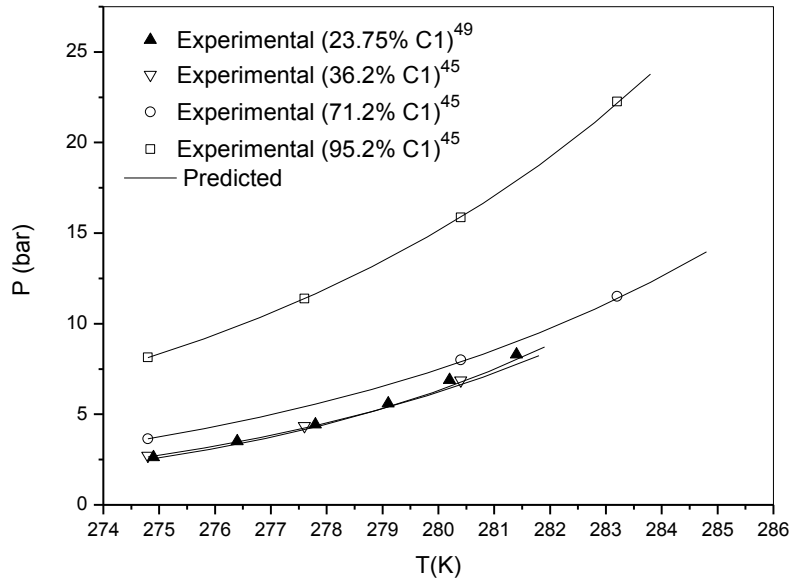

Figure 2.8. Hydrate dissociation pressure for methane/propane mixtures.

Figure 2.9 shows the hydrate dissociation pressure of methane/isobutane mixtures, in which the structure hydrate II forms. The value of $\omega$ is fitted to the experimental data of methane/isobutane mixture having 71.4% methane. This $\omega$ is then used for other mixtures. The effect of concentration on dissociation pressure is well-predicted.

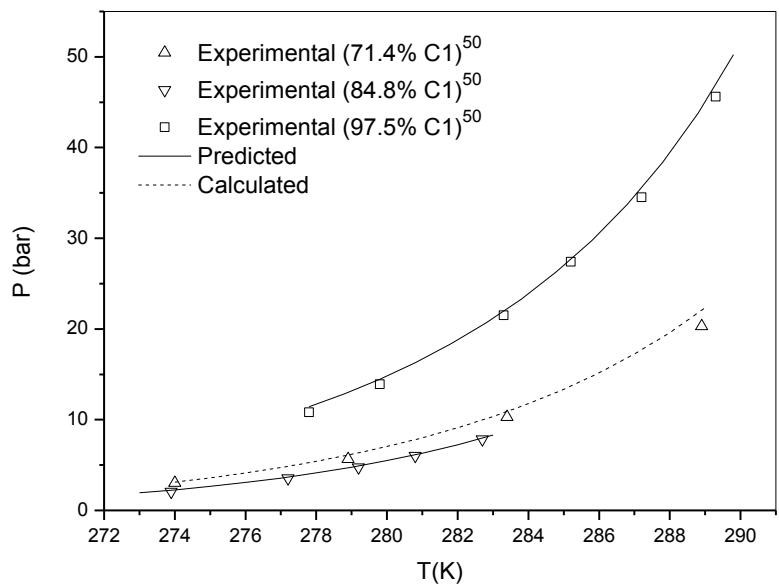

Figure 2.9. Hydrate dissociation pressure for methane/isobutane mixtures.

Figure 2.10 shows the hydrate dissociation pressure for ethane/propane mixtures. When structure I hydrate forms, such as in a mixture with 18.6% propane, no correction factor is needed and the dissociation pressure can be well-predicted by the model. When the structure II hydrate forms, the correction factor is again needed. The value of ω is fitted to the experimental data of ethane/propane mixture having 54.1% propane. By using the value of ω for the other mixture, the model can predict the dissociation pressure well and capture the effect of concentration on the dissociation pressure well.

Figure 2.11 shows the hydrate dissociation pressure for propane/isobutane mixtures at 272.2 K at various compositions. Similar to methane/propane mixtures, no correction is needed (ω = 1) and the effect of concentration on the dissociation pressures are well-predicted.

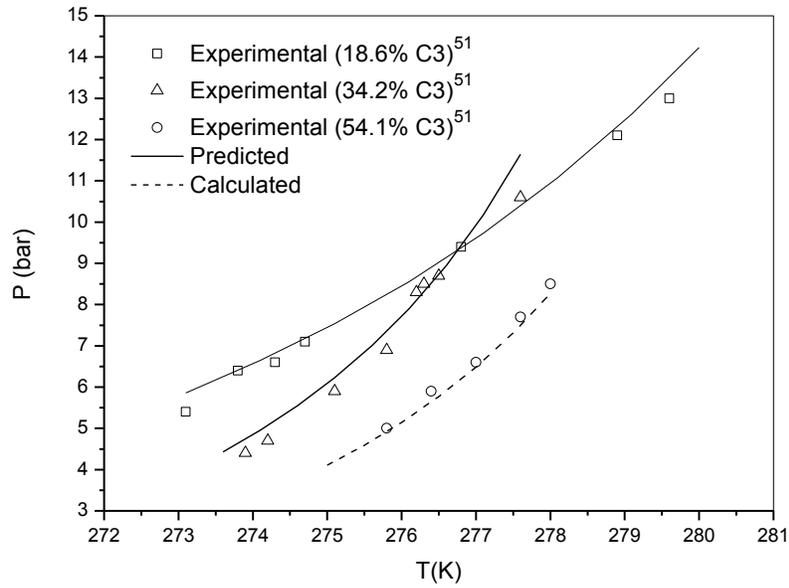

Figure 2.10. Hydrate dissociation pressure for ethane/propane mixtures.

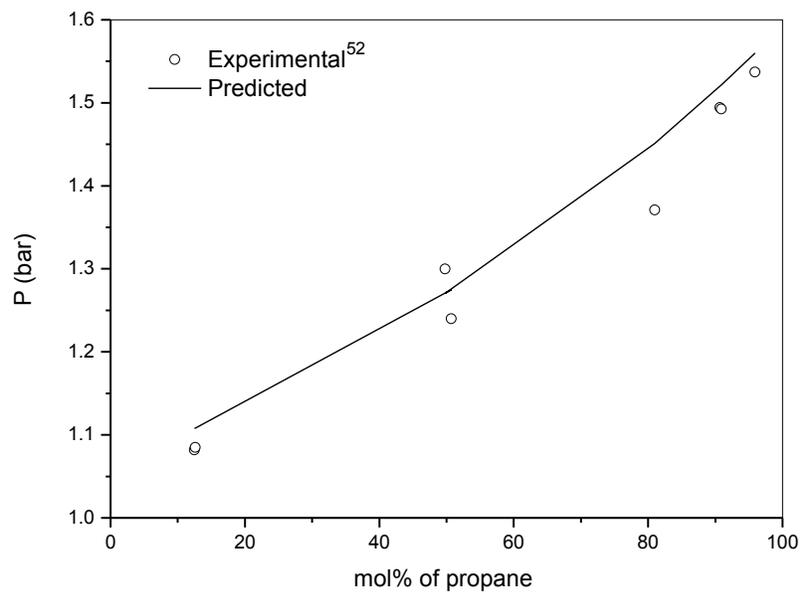

Figure 2.11. Hydrate dissociation pressure for propane/isobutane mixtures at T = 272.2K.

### 2.4.3 Ternary Alkane Mixture Hydrate

For ternary alkane mixtures, the IVH and LVH boundaries for methane/ethane/propane are modeled in this work. When the structure II hydrate forms, similar to the binary alkane hydrate systems, a correlation given by Ng and Robinson[58] must to be introduced. For LVH:

$$\Delta \mu_w^L = RT \prod_{i=1}^{2}\left[1 + 3(\omega - 1)Y_i^2 - 2(\omega - 1)Y_i^3\right] \times \left[\sum_m v_m \ln(1 + \sum_j C_{mj}\widehat{f}_j^v) + \ln(\gamma_w x_w)\right] \quad (2.19a)$$

and for IVH:

$$\Delta \mu_w^{ice} = RT \prod_{i=1}^{2}\left[1 + 3(\omega - 1)Y_i^2 - 2(\omega - 1)Y_i^3\right] \times \sum_m v_m \ln(1 + \sum_j C_{mj}\widehat{f}_j^v) \quad (2.19b)$$

where the product is obtained over the more volatile gas components, which are methane and ethane in this case, and $Y_1$ and $Y_2$ are the mole fractions of methane and ethane, respectively, in the vapor phase. In our work, since the values of $\omega$ for binary alkane mixtures cannot give a satisfactory prediction of $\omega$ for a ternary alkane mixture, the value of $\omega$ for the methane/ethane/propane system is not predicted from those for binary alkane mixtures (methane/propane and ethane/propane), but directly fitted to ternary hydrate experimental data[54], the value of which is found to be 1.041. The experimental and calculated data are depicted in Figure 2.12.

The value of $\omega$ derived from the LVH boundary above is then used to predict the IVH boundary of this system at another composition (89.952% methane/7.06% ethane/2.988% propane). Figure 2.13 shows that the model agrees well with the experimental data. In Figure 2.13, another IVH boundary at different composition is also included, but at that composition, the structure hydrate I forms. As in binary hydrate systems, when structure I hydrate forms, the model can be used directly without introducing any correction factor.

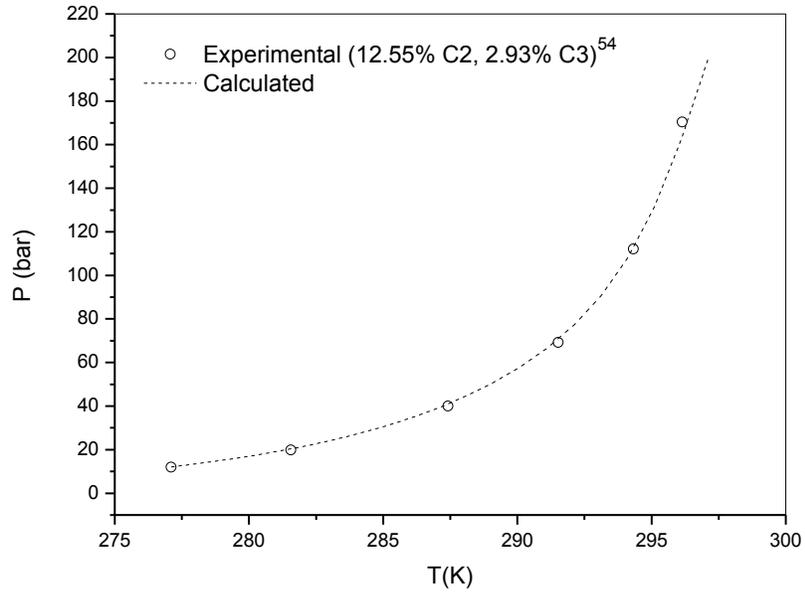

Figure 2.12. Hydrate dissociation pressure for methane/ethane/propane mixture (LVH).

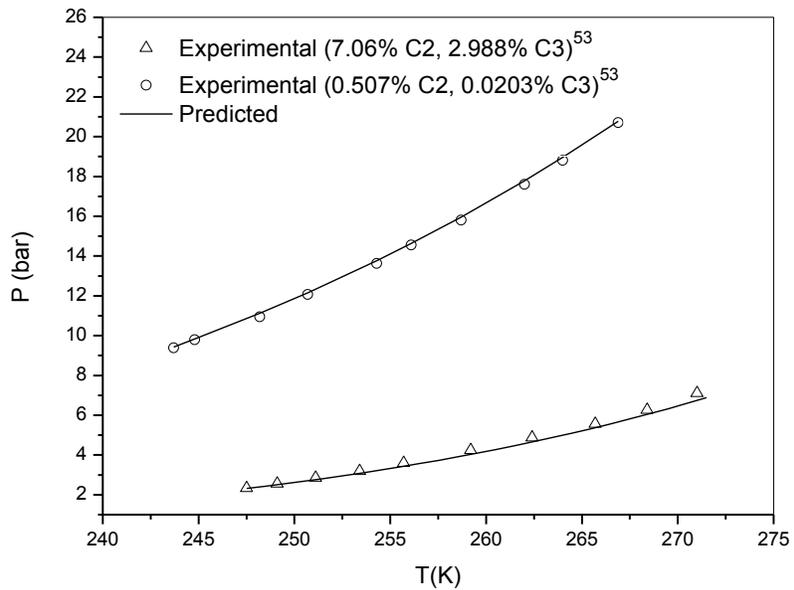

Figure 2.13. Hydrate dissociation pressure for methane/ethane/propane mixtures (IVH).

## 2.5 Conclusions

SAFT2, coupled with the classical van der Waals and Platteeuw model, is used to model the phase equilibrium condition for pure, binary, and ternary alkane hydrates in the absence of inhibitors. The proposed model is capable of representing the LVH and IVH equilibrium data of pure alkane well. For alkane mixtures that form the structure I hydrate, the model predicts the LVH and IVH boundaries of hydrate dissociation well and captures the effect of concentration on the dissociation pressure well, without introducing any correction factor. For alkane mixtures that form the structure II hydrate, a correction factor is needed for some systems and, in that case, the model requires one more additional parameter ($\omega$). This parameter is constant for every alkane mixture and found to be independent of the mixture composition. With this additional parameter, the model can also predict the LVH and IVH boundaries of structure II hydrate well and capture the effect of concentration on the dissociation pressure well.

# Chapter 3. Modeling of Hydrate Dissociation Conditions for Alkanes in the Presence of Single and Mixed Electrolyte Solutions Using Ion-Based Statistical Associating Fluid Theory

## 3.1 Introduction

Gas hydrates are clathrate solids formed by hydrogen-bonded water molecules and guest gas molecules at low temperature and high pressure. The hydrate formation is inhibited when electrolytes are present in the liquid phase, which is the case for example in the hydrate exploration in marine environment or in the flow assurance operation using salt inhibitors. It is important to have an accurate model to predict the hydrate phase equilibrium conditions in the presence of electrolyte solutions.

The thermodynamic properties of electrolyte solutions are more difficult to model than other solutions. The most widely used thermodynamic model for electrolyte solutions is the Pitzer model.[1] Although various activity coefficient models successfully calculated the activity coefficient, by themselves they cannot be used to predict the density of electrolytes. To be able to predict the density of electrolyte solutions, thermodynamic models based on equation of state (EoS) were developed. Statistical associating fluid theory (SAFT) was applied to electrolyte solutions. SAFT-VRE[2] had successfully represented the vapor pressure and density of aqueous electrolytes. Cameretti et al.[3] combined the PC-SAFT EoS[4] with a Debye-Huckle term, namely ePC-SAFT, to calculate the vapor pressures and densities of salt solutions. Held et al.[5] improved ePC-SAFT to deal with mean ionic activity coefficient. Recently, ePC-SAFT was also extended to calculate the liquid densities, osmotic coefficients, and activity coefficients of single-salt alcohol solutions.[6] In our earlier work, SAFT, coupled with restricted primitive model (RPM),[7,8] namely, SAFT1-RPM, was proven to be able to well represent the activity coefficient, osmotic coefficient, vapor pressure, and density of electrolyte solutions. However, SAFT1-RPM does not work well for multivalent salts because of the limited range of the square-well width parameter $\lambda$. Although SAFT2[9,10] was able to deal with mono- and bivalent salts, the parameters in SAFT2 were still salt-based. Hence, it is difficult to transfer the parameters to different salts containing the same ions. Recently, Ion-based SAFT2 was proposed by Ji and Adidharma[11,12,13] to represent the properties of aqueous salt solutions at ambient and elevated temperatures and pressures. Ion-based SAFT2 has less number of parameters compared to the Pitzer model,[1] and since its parameters are ion-specific, the model can handle multiple-salt solutions more easily than the Pitzer model, SAFT1, and SAFT2.

On the basis of the thermodynamic model developed for electrolyte solutions, various approaches have been developed to predict the hydrate phase equilibrium in the presence of electrolyte solutions. In such phase equilibrium calculations, the ability to predict water activity depression is the key to success. Based on Pitzer activity coefficient model,[1] which was used to calculate the water activity in aqueous electrolyte, Englezos and Bishnoi,[14] and Duan and Sun[15] successfully predicted the conditions of gas hydrate dissociation in aqueous salt solutions. In their works, an equation of state was needed to describe the vapor phase while the Pitzer model was used to describe the liquid phase. Hence, different thermodynamic frameworks for the fluid phases were used, which makes the model not self-consistent. Besides the Pitzer model, modified cubic equations of state have also been used to obtain water activity in electrolyte solutions. Zuo and Stenby[16] used a modified Patel-Teja[17] EoS with a modified Debye-Huckle term and obtained satisfactory results. In their work, the mixing rule for energy parameter in the Patel-Teja EoS was concentration dependent and their binary interaction parameters were fitted to osmotic coefficient of salt-water mixture. The Valderrama modification of Patel-Teja EoS (VPT)[18] combined with a non-density-dependent mixing rule[19] was also used to model the hydrate phase equilibria in electrolyte solutions by Tohidi.[20] Recently, Haghighi et al[21] used the cubic plus association (CPA) EoS with the Debye-Huckle term to predict the methane hydrate dissociation conditions in salt solutions. However, in the works of Tohidi and Haghighi, the binary interaction parameters between salt and water were functions of salt concentration and temperature. Although the inhibition effect of electrolyte solutions on hydrate formation can be captured by these models, these models are empirical or semi-empirical. It is generally difficult for these models to represent multiple-salt systems due to the large number of parameters required. Although the Zuo and Stenby's model[16] and the CPA EoS[21] do have the capacity to handle mixed salt systems, the interaction parameters between salts in the Zuo's model were fitted to the water activity data, while the interaction parameters between salts and water in the CPA EoS were obtained by fitting to freezing point depression and vapor pressure of electrolyte solutions. The predictive power of these two models is expected to be limited since the salt-salt or salt-water interaction parameters have to be optimized by fitting to experimental data.

Ion-based SAFT2 has been proven to be reliable in representing the thermodynamic properties of aqueous electrolytes[11, 12] and easy to extend to multiple-salt solutions because of its transferrable parameters.[13] Hence, in this study, Ion-based SAFT2 is used to describe the fluid phases and the van der Waals & Platteeuw model[22] is used to describe the hydrate phase. To the best of our knowledge, the SAFT-based equations of state were only applied to the hydrate modeling in pure water[23,24] or alcohol,[25] but not in electrolyte solutions.

In particular, we apply our model to describe the hydrate dissociation conditions of methane and propane in the presence of single and mixed electrolyte solutions containing NaCl, KCl, and CaCl$_2$. This work also includes ethane hydrate equilibrium in the presence of aqueous NaCl solution.

## 3.2 Modeling

In hydrate phase equilibrium modeling, the chemical potential of each species in different phases should be equal,[23]

$$\Delta\mu_W^H = \mu_W^{MT} - \mu_W^H = \mu_W^{MT} - \mu_W^L = \Delta\mu_W^L \tag{3.1}$$

where $\mu_W^H$ is the chemical potential of water in the hydrate phase, $\mu_W^L$ is the chemical potential of water in the liquid phase, and $\mu_W^{MT}$ is the chemical potential of water in the empty hydrate lattice. The van der Waals and Platteeuw model[22] is applied to calculate the chemical potential of water in the hydrate phase, while Ion-based SAFT2 is applied to describe the fluid phases.

### 3.2.1 Equation of State

The Ion-based SAFT2 EoS is defined in terms of dimensionless residual Helmholtz energy,

$$\tilde{a}^{res} \cong \tilde{a}^{hs} + \tilde{a}^{disp} + \tilde{a}^{chain} + \tilde{a}^{assoc} + \tilde{a}^{ion} \tag{3.2}$$

where the superscripts on the right hand side refer to terms accounting for the hard sphere, dispersion, chain, association, and ion interactions. The details of calculations of hard-sphere, dispersion, chain, and association terms can be found in our earlier work.[23]

The contribution of long-range Coulombic interactions, which is the ionic term in Eq. 3.2, is based on the Restricted Primitive Model (RPM) and expressed as,[7]

$$\tilde{a}^{ion} = -\frac{3x^2 + 6x + 2 - 2(1 + 2x)^{3/2}}{12\pi\rho N_A d^3} \qquad (3.3)$$

where $\rho$ is the molar density, $N_A$ is the Avogadro's number, $d$ is the hydrated diameter, which is an ion property, and $x$ is a dimensionless quantity calculated by,

$$x = \kappa d \qquad (3.4)$$

where $\kappa$ is the Debye inverse screening length. The detailed calculations of Eqs. 3.3 and 3.4 are given elsewhere.[7]

In Ion-based SAFT2, each ion has four parameters: segment volume ($v$), segment energy ($u$), reduced width of square well potential ($\lambda$), and hydrated diameter ($d$). The cations and anions are treated as non-associating spherical segments. Therefore, the association term in Eq. 3.2 is to represent only the water self-association.

The ion parameters are temperature-dependent, and the expressions for calculating these parameters have been obtained, in which their ion-specific coefficients were fitted to the experimental activity coefficients and liquid densities in the temperature range of 308.15 to 473.15 K.[12] In the hydrate equilibrium calculations, however, the system temperature, at which the water activity should be evaluated, could be much lower than 308.15 K.

To make the prediction of water activity in electrolyte solutions more accurate at low temperatures, in this work, we include experimental data at temperatures down to 273.15 K and develop a new set of functions with a new set of ion-specific coefficients, as shown below:

$$v = b_1 \exp\left(b_2 \frac{T - 298.15}{T} + b_3\right) \qquad (3.5)$$

$$u = b_4 \exp\left( b_5 \frac{T - 298.15}{T} + b_6 \right) \qquad (3.6)$$

$$\lambda = b_7 \exp\left( b_8 \frac{T - 298.15}{T} + b_9 \right) \qquad (3.7)$$

$$d = b_{10} \exp\left( b_{11} \frac{T - 298.15}{T} + b_{12} \right) \qquad (3.8)$$

where $b_1$- $b_{12}$ are the ion-specific coefficients that should be fitted to experimental data.

A mixing rule for segment energy describing the short-range interactions between segments $\alpha$ and $\beta$ is applied,[12]

$$u_{\alpha\beta} = u_{\beta\alpha} = \sqrt{u_\alpha u_\beta}\left(1 - k_{\alpha\beta}\right) \qquad (3.9)$$

where $k_{\alpha\beta}$ is the binary interaction parameter. The short-range interactions of cation-cation and anion-anion are neglected, i.e. $k_{\alpha\beta}$=1. For cation-anion interaction, the $k_{\alpha\beta}$ is set to 0.5 in order to reduce the number of parameters in the model.[12] For water-ion and alkane-ion interactions, no binary interaction parameters are used, i.e., the $k_{\alpha\beta}$ is set to 0.

The parameters of water, which are not temperature-dependent, and the binary interaction parameters between alkanes and water used in this study can be found in our earlier work.[23]

**3.2.2 Models for Chemical Potential Differences**

The ions are not incorporated in the hydrate lattice, thus the van der Waals and Platteeuw model used in our earlier work[23] is still valid in this study. The van der Waals and Platteeuw model is defined in terms of water chemical potential. The chemical potential difference between water in the hydrate phase and that in the hypothetical empty hydrate lattice is expressed as,[23]

$$\Delta \mu_w^H = \mu_w^{MT} - \mu_w^H = -RT \sum_m v_m \ln(1 - \sum_j \theta_{mj}) \tag{3.10}$$

where $v_m$ is the number of cavities of type $m$ per water molecule in the hydrate phase and $\theta_{mj}$ is the occupancy of molecules $j$ in type $m$ cavities.

In this work, the lowest temperature studied is 260 K and ice cannot form in the pressure and salt concentration ranges of interest. The chemical potential difference between water in the empty hydrate lattice and that in the liquid water is expressed as,[26]

$$\frac{\Delta \mu_w^L(T,P)}{RT} = \frac{\Delta \mu_w^0(T_0,0)}{RT_0} - \int_{T_0}^T \left( \frac{\Delta h_w^{MT-L}}{RT^2} \right) dT + \int_0^P \left( \frac{\Delta V_w^{MT-L}}{RT} \right) dP - \ln a_w \tag{3.11}$$

where $a_w$ is the water activity calculated from Ion-based SAFT2, $\Delta \mu_w^0(T_0,0)$ is the reference chemical potential difference between water in empty hydrate lattice and pure water at 273.15 K ($T_0$) and 0 Pa, $\Delta h_w^{MT-L}$ is the molar enthalpy difference between empty hydrate lattice and liquid water at zero pressure. The details of Eq. 3.11, including the values of the reference properties and the fitting procedure, can be found elsewhere.[23]

## 3.3 Results and Discussion

### 3.3.1 Parameter Fitting

The ion-specific coefficients in Eqs. 3.5–3.8 for $Na^+$, $K^+$, $Ca^{2+}$, and $Cl^-$ are fitted to the experimental activity coefficients[27,28,29,30] and liquid densities[27,31,32] of aqueous NaCl, KCl, and $CaCl_2$ solutions. The values of ion-specific coefficients obtained are listed in Table 3.1. Table 3.2 lists the average absolute deviations (AADs) of the activity coefficients and densities for NaCl, KCl, and $CaCl_2$ solution, respectively. The concentration of salts, the temperature range, the pressure range, and the source of experimental data used for parameters fitting are also included in the table. With the new expressions for calculating temperature-dependent ion parameters, the model is able to represent the activity coefficients and densities at temperatures down to 273.15 K, as desired.

It is worth mentioning that in the regression, two constraints are imposed to make the ion parameters meet their physical meanings. First, the reduced width ($\lambda$) should be always larger than 1.0 in the whole temperature range and second, the hydrated diameter ($d$) should be larger than the diameters of water and ion segments. The Ion-based SAFT2 parameters calculated from Eqs. 3.5 to 3.8 with the coefficients given in Table 3.1 satisfy these two constraints.

As examples, Figures 3.1 and 3.2 show the goodness of fit of Ion-based SAFT2 for the activity coefficients of aqueous NaCl solutions. Figure 3.3 shows the calculated densities of the same solutions. Similar figures can be obtained for KCl and CaCl$_2$ solutions.

Table 3.1. Coefficients in Eqs. 3.5 – 3.8

| | Na$^+$ | K$^+$ | Ca$^{2+}$ | Cl$^-$ |
|---|---|---|---|---|
| $b_1$ | 0.6045 | 0.2859 | 7.6473 | 7.3058 |
| $b_2$ | −0.6637 | 3.5348 | −0.4869 | −8.8911×10$^{-3}$ |
| $b_3$ | 7.8956×10$^{-2}$ | −0.3057 | 0.1677 | 3.1107×10$^{-3}$ |
| $b_4$ | 1682.96 | 2528.31 | 3477.43 | 1287.50 |
| $b_5$ | 0.1361 | 1.8958 | 0.5346 | −3.2439 |
| $b_6$ | −5.9977×10$^{-2}$ | 0.2571 | 0.1227 | 0.1006 |
| $b_7$ | 1.6588 | 1.9872 | 2.1896 | 1.8973 |
| $b_8$ | −1.4338 | −0.2278 | −0.3104 | 0.6940 |
| $b_9$ | 6.9865×10$^{-2}$ | −0.1861 | 1.9363×10$^{-2}$ | −6.1118×10$^{-2}$ |
| $b_{10}$ | 3.48976 | 3.4132 | 4.9250 | 5.4435 |
| $b_{11}$ | −0.7094 | −0.2243 | −0.8556 | −0.6609 |
| $b_{12}$ | 0.12786 | 0.1442 | 3.3657×10$^{-2}$ | 0.1064 |

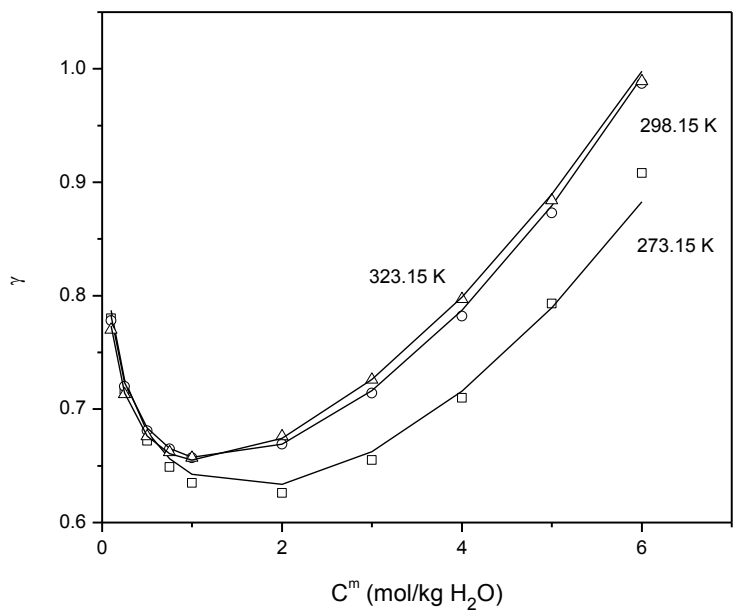

Figure 3.1. Activity coefficients of NaCl aqueous solutions at 273.15, 298.15, and 323.15 K, experimental[27] (points) and calculated (curves).

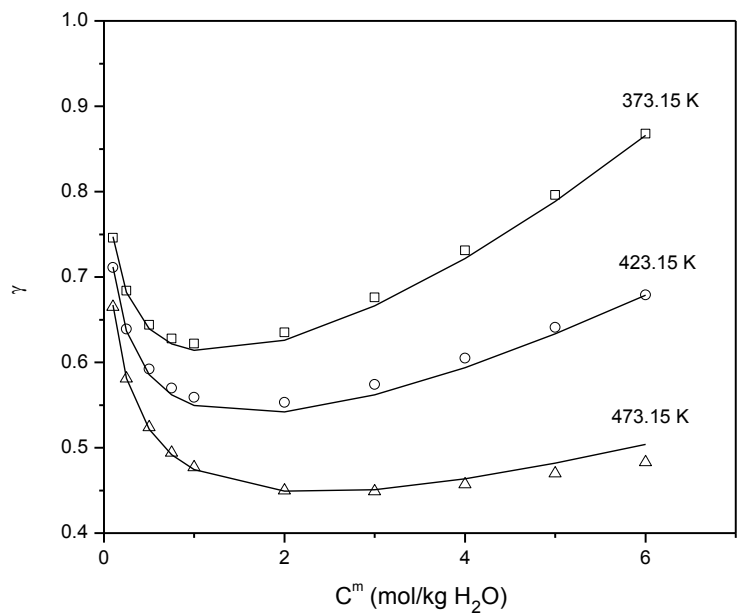

Figure 3.2. Activity coefficients of NaCl aqueous solutions at 373.15, 423.15, and 473.15 K, experimental[27] (points) and calculated (curves).

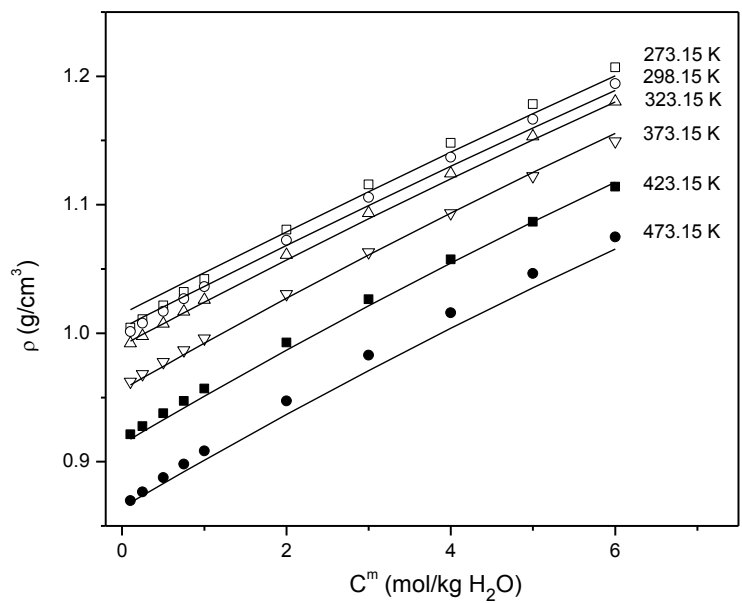

Figure 3.3. Densities of NaCl aqueous solutions, experimental[27] (points) and calculated (curves).

Table 3.2. AADs for the activity coefficients (γ) and densities (ρ) of aqueous NaCl, KCl, and CaCl₂ solutions

|  | Pressure range (bar) | Temperature range (K) | Molality range (mol/kg H₂O) | AAD(γ)[a] % | AAD(ρ)[b] % | Data source |
|---|---|---|---|---|---|---|
| NaCl | 1.0-15.5 | 273.15-473.15 | 0.1-6.0 | 0.84 | 0.47 | 27 |
| KCl | 1.0 | 283.15-473.15 | 0.1-4.0 | 0.70 | 0.61 | 28,29,31 |
| CaCl₂ | 1.0 (for γ) | 273.15-473.15 | 0.1-2.0 (for γ) | 1.45 | 0.52 | 30,32 |
|  | 70.0 (for ρ) |  | 0.24-6.15 (for ρ) |  |  |  |

[a] $AAD(\gamma) = \dfrac{1}{N} \sum_{i=1}^{N} \left| \dfrac{\gamma_i^{cal} - \gamma_i^{exp}}{\gamma_i^{exp}} \right|$

[b] $AAD(\rho) = \dfrac{1}{N} \sum_{i=1}^{N} \left| \dfrac{\rho_i^{cal} - \rho_i^{exp}}{\rho_i^{exp}} \right|$

### 3.3.2 Prediction of Alkane Hydrate Dissociation Pressures

The proposed model is used to predict the methane, ethane, and propane hydrate phase equilibrium in the presence of single and mixed salt solutions containing NaCl, KCl, and $CaCl_2$. Table 3.3 summarizes all of the systems studied along with the average absolute deviations (AAD) of the predicted pressures.

Table 3.3. Systems studied

| Gas | Concentration of salt in the aqueous phase (wt %) | T-range/K | P-range/bar | AAD(P)[a] % | Data source |
|---|---|---|---|---|---|
| methane | 11.7% NaCl | 268.3-278.05 | 26.9-75.5 | 1.25 | 33 |
| | 17.1% NaCl | 263.35-274.95 | 23.9-85.7 | 3.83 | |
| | 21.5% NaCl | 261.85-272.85 | 29.4-110 | 6.18 | |
| | 24.1% NaCl | 263.05-268.65 | 47.8-95.5 | 12.6 | |
| | 5% KCl | 271.6-283.2 | 27.1-86.9 | 3.44 | 34 |
| | 10% KCl | 270.1-281.5 | 27.8-88.2 | 4.96 | |
| | 5% $CaCl_2$ | 279.9-284.4 | 60.5-101.2 | 3.48 | 35 |
| | 10% $CaCl_2$ | 278.4-282.3 | 68.8-102.2 | 1.47 | |
| | 15% $CaCl_2$ | 273.5-278.8 | 49.2-97 | 6.78 | |
| | 20% $CaCl_2$ | 270.1-273.5 | 63.3-102.9 | 3.37 | |
| | 3% NaCl + 3% KCl | 271.4-279.2 | 27.04-58.57 | 1.01 | 36 |
| | 5% NaCl + 5% KCl | 270.3-281.5 | 28.29-93.79 | 1.23 | |
| | 5% NaCl + 10% KCl | 267.5-279 | 25.69-90.46 | 2.19 | |
| | 5% NaCl + 15% KCl | 266.3-276.2 | 29.14-86.89 | 1.05 | |
| | 10% NaCl + 12% KCl | 264.6-274.4 | 29.89-88.19 | 2.66 | |
| | 15% NaCl + 8% KCl | 264.4-272.1 | 36.14-88.39 | 3.34 | |
| | 3% NaCl + 3% $CaCl_2$ | 270.4-281.8 | 25.04-81.59 | 1.91 | |
| | 6% NaCl + 3% $CaCl_2$ | 271.3-280.1 | 31.34-78.39 | 1.63 | |
| | 10% NaCl + 3% $CaCl_2$ | 269.4-277.3 | 32.14-74.44 | 1.09 | |
| | 3% NaCl + 10% $CaCl_2$ | 268.8-279.7 | 30.19-96.64 | 1.67 | |
| | 6% NaCl + 10% CaCl₂ | 268.6-277.1 | 36.89-95.14 | 3.85 | |
| | 10% NaCl + 6% CaCl₂ | 266-274.3 | 28.19-68.99 | 3.38 | |
| ethane | 10% NaCl | 273.7-280.4 | 8.83-21.65 | 1.32 | 37 |
| | 15% NaCl | 272.7-277.1 | 10.82-21.51 | 1.99 | |
| | 20% NaCl | 266.2-271.4 | 6.89-15.24 | 7.19 | |
| propane | 3% NaCl | 272.2-276.2 | 1.79-4.55 | 2.02 | 37 |
| | 7.5% CaCl₂ | 271.6-274.2 | 2.34-4.27 | 2.12 | |
| | 10% KCl | 271-273.4 | 2.28-4.21 | 2.48 | |
| | 4.9% NaCl + 3.8% CaCl₂ | 270.9-273.5 | 2.34-4.41 | 3.55 | 38 |
| | 5% NaCl + 5% KCl | 270.9-273.5 | 2.34-4.41 | 3.36 | |
| | 7.5% KCl + 7.5 CaCl2 | 266.3-270.1 | 1.81-4.32 | 6.19 | 39 |
| | 7.5% NaCl + 7.5% KCl | 265.2-269 | 1.57-3.72 | 7.30 | |

$$^{\text{a}}\ AAD(P) = \frac{1}{N} \sum_{i=1}^{N} \left| \frac{P_i^{pre} - P_i^{\exp}}{P_i^{\exp}} \right|$$

### 3.3.2.1 Systems with Single-Salt Solutions

Figure 3.4 shows the prediction of methane hydrate dissociation pressure in aqueous NaCl solutions, which is found to be reasonable; experimental data from de Roo et al[33] are used to verify our prediction. It can be noticed that as the temperature decreases, the discrepancy between the model predictions and the experimental data increases. This could be due to the use of the coefficients in Table 3.1 far beyond the temperature range in which they are fitted.

Figure 3.5 presents the inhibition effect of KCl solutions on methane hydrate dissociation conditions. Our model can give a satisfactory prediction compared to the experimental data.[34] The methane hydrate phase equilibrium in the presence of CaCl₂ solutions is shown in Figure 3.6. Similar to the prediction with other electrolyte solutions, our model captures the effect of salt concentration on the hydrate dissociation conditions. The data of methane hydrate dissociation pressure in CaCl$_2$ solution are taken from Kharrat et al.[35]

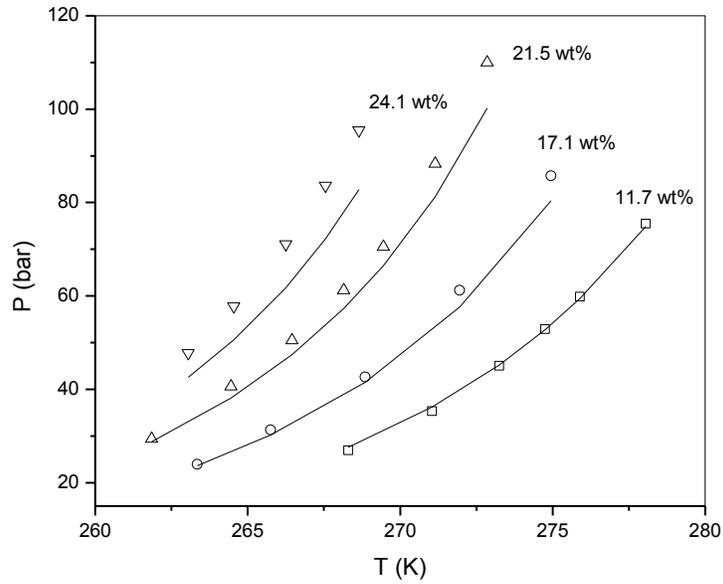

Figure 3.4. Methane hydrate dissociation pressures in the presence of NaCl solutions of different concentrations, experimental[33] (points) and predicted (curves).

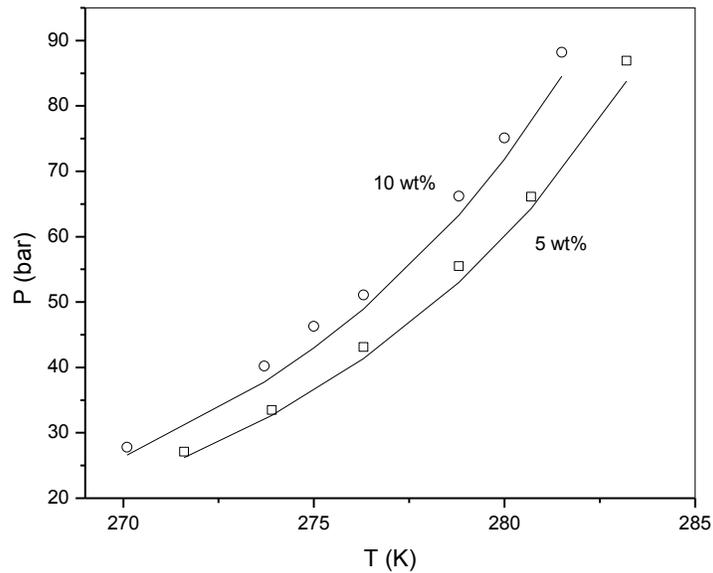

Figure 3.5. Methane hydrate dissociation pressure in the presence of KCl solutions of different concentrations, experimental[34] (points) and predicted (curves).

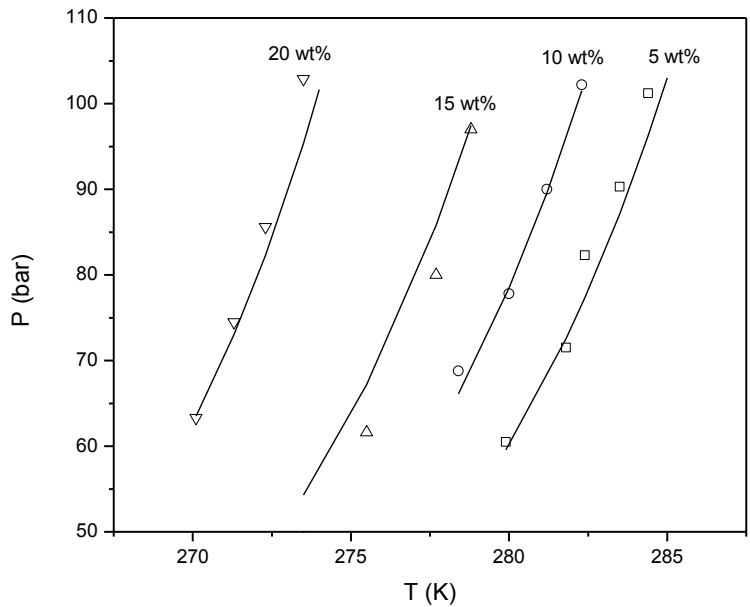

Figure 3.6. Methane hydrate dissociation pressure in the presence of CaCl$_2$ solutions of different concentrations, experimental[35] (points) and predicted (curves).

Figure 3.7 shows the experimental ethane hydrate dissociation pressures in aqueous NaCl solutions[37] along with the model prediction. The model can well capture the inhibition effect of aqueous NaCl solutions. Propane hydrate dissociation conditions in aqueous NaCl, KCl, and CaCl$_2$ solutions are shown in Figure 3.8. Propane forms structure II hydrate and its dissociation pressure is much lower than those of methane and ethane hydrates. The experimental data from Tohidi et al[37] are used to verify our prediction. Again, reasonable prediction is obtained.

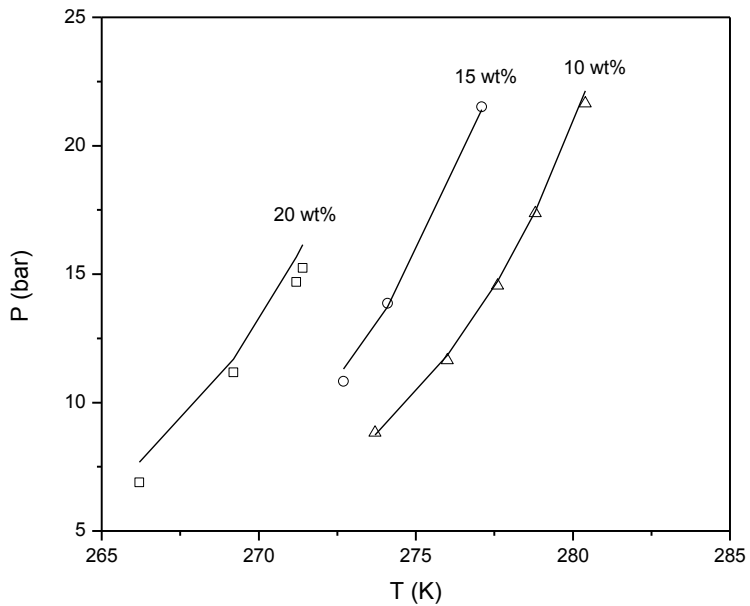

Figure 3.7. Ethane hydrate dissociation pressures in the presence of NaCl solutions of different concentrations, experimental[37] (points) and predicted (curves).

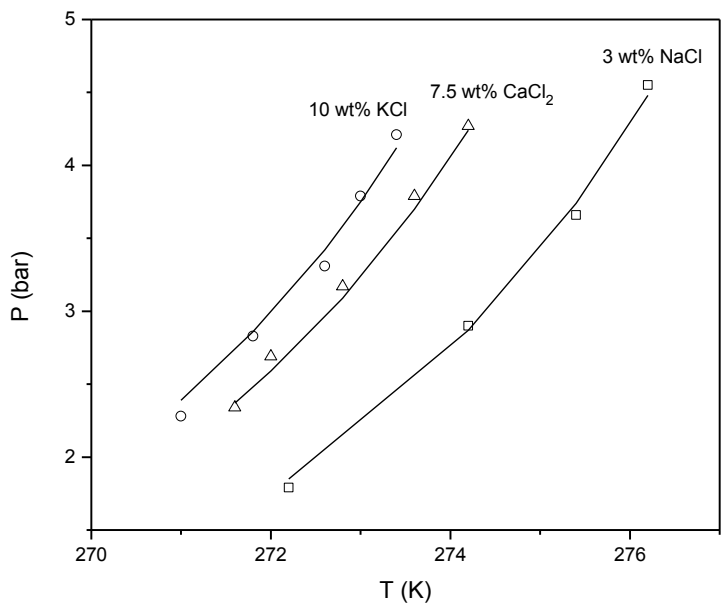

Figure 3.8. Propane hydrate dissociation pressures in NaCl, KCl, and CaCl$_2$ solutions, experimental[37] (points) and predicted (curves).

**3.3.2.2 Systems with Mixed Salt Solutions**

The model can be conveniently applied to mixed electrolyte solutions with no additional parameters to fit. In this work, the model prediction of the inhibition effect of mixed aqueous salt solutions on methane and propane hydrates is explored.

Figures 3.9 and 3.10 show the hydrate dissociation pressures of methane in the presence of NaCl + KCl solutions. The experimental data are taken from Dholabhai et al.[36] The model prediction is satisfactory and the effect of different salt concentrations can be accurately captured.

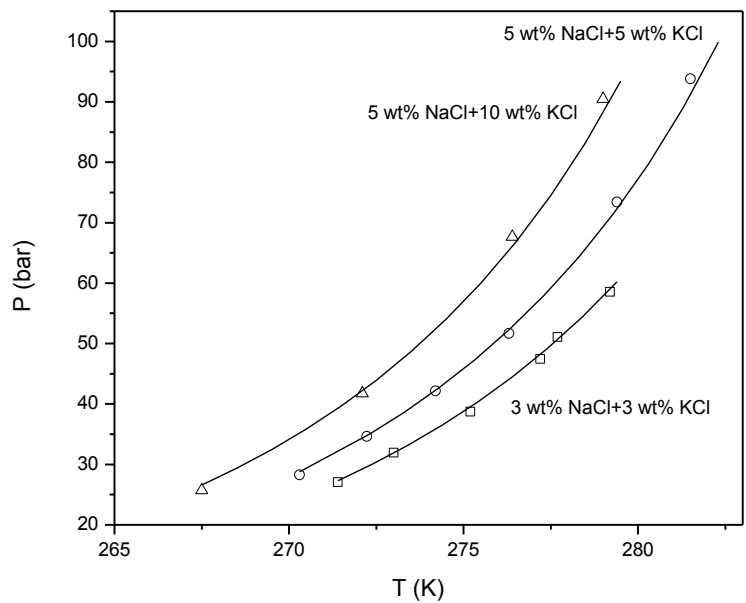

Figure 3.9. Methane hydrate dissociation pressures in the presence of NaCl+KCl solutions, experimental[36] (points) and predicted (curves).

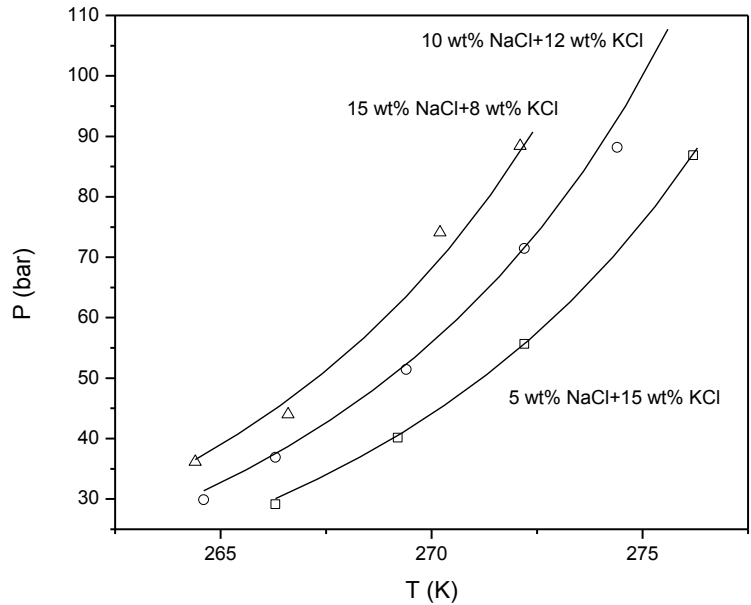

Figure 3.10. Methane hydrate dissociation pressures in the presence of NaCl+KCl solutions, experimental[36] (points) and predicted (curves).

Figures 3.11 and 3.12 show the hydrate dissociation pressures of methane in the presence of NaCl+CaCl$_2$ solutions. The effect of different NaCl and CaCl$_2$ concentrations on the hydrate dissociation pressure is well predicted. As shown in Figure 3.12, in agreement with the experimental data,[36] the predicted inhibition effects of 10 wt% NaCl+6 wt% CaCl$_2$ and 6 wt% NaCl+10 wt% CaCl$_2$ solutions are almost the same.

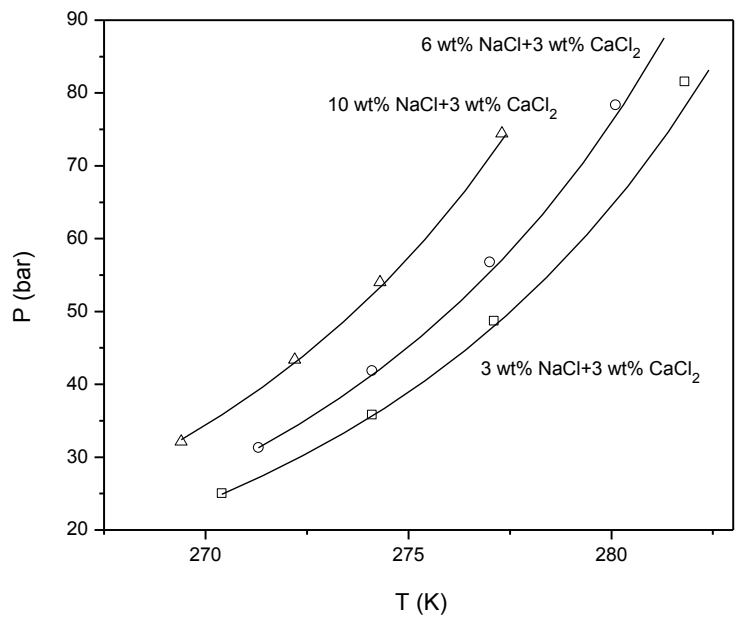

Figure 3.11. Methane hydrate dissociation pressures in the presence of NaCl+CaCl$_2$ solutions, experimental[36] (points) and predicted (curves).

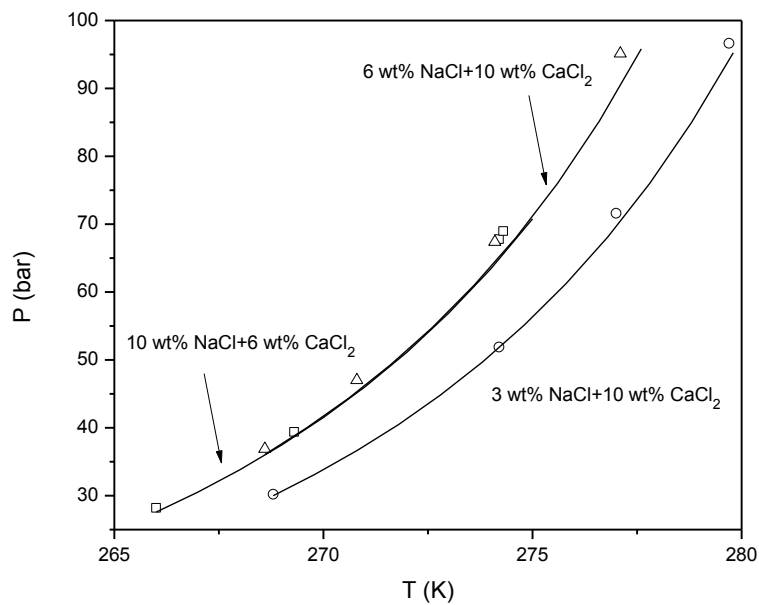

Figure 3.12. Methane hydrate dissociation pressures in the presence of NaCl+CaCl$_2$ solutions, experimental[36] (points) and predicted (curves).

Figure 3.13 shows the hydrate dissociation pressures of propane in the presence of mixed salt solutions, including a mixture of KCl and CaCl$_2$. Again, a good agreement between experimental data and our prediction is demonstrated. For the record, of course, the model predictions can be improved by fitting the binary interaction parameters between water and ions or between two ions to experimental hydrate dissociation pressure. However, since we are to explore the predictive power of our model, which turns out to be reasonable, we do not attempt to adjust these parameters.

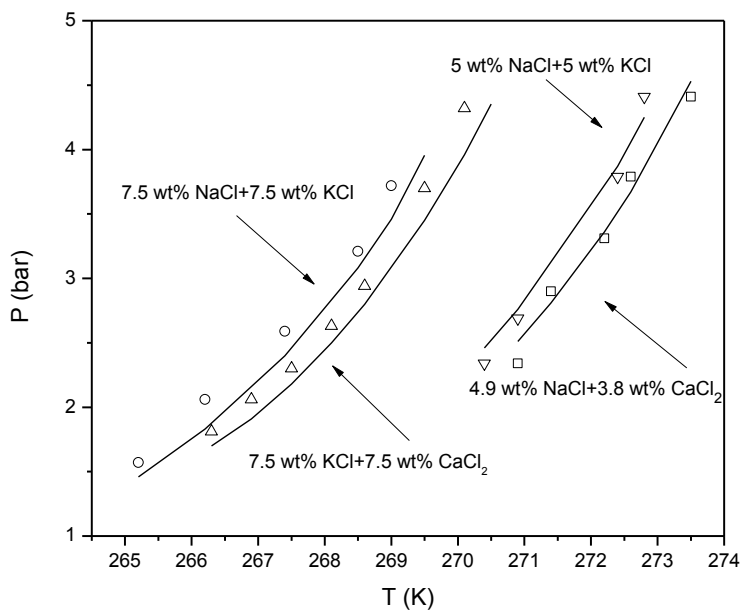

Figure 3.13. Propane hydrate dissociation pressures in the presence of mixed salt solutions, experimental[39] (points) and predicted (curves).

## 3.4 Conclusions

Ion-Based SAFT2, coupled with the van der Waals and Platteeuw model, is applied to describe the alkane hydrate dissociation conditions in the presence of single and mixed electrolyte solutions containing NaCl, KCl, and CaCl₂. In this model, each ion has 4 parameters: segment volume ($v$), segment energy ($u$), reduced width of square well potential ($\lambda$), and hydrated diameter ($d$), which are temperature-dependent. A new set of expressions with a new set of ion-specific coefficients for calculating these ion parameters has been developed. The new expressions enable us not only to fit well the activity coefficients and densities of single-salt solutions in a wider temperature range than the expressions obtained in our previous work, but also to predict water activity in electrolyte solutions better at low temperatures, which is critical in the modeling of hydrate phase equilibria.

The model can be conveniently applied to mixed electrolyte solutions with no additional parameters to fit. We find that the model accurately predicts the alkane hydrate dissociation conditions in the presence of electrolyte solutions studied and well captures the effect of different salt concentrations.

Experimental Measurements and Predictions of Dissociation Conditions. *J. Chem. Thermodyn.* **2008**, 40, 1693-1697.

# Chapter 4. Prediction of Hydrate Dissociation Conditions for Alkanes in the Presence of Alcohol and Electrolyte Solutions Using Ion-Based Statistical Associating Fluid Theory

## 4.1 Introduction

Gas hydrates are clathrate solids formed by water molecules and suitable gas molecules at low temperature and high pressure. Hydrate formation in the pipeline can cause blockage and safety problem in the natural gas production process. To avoid any adverse circumstances associated with this hydrate formation, alcohol and electrolyte solutions are widely used to inhibit the formation of gas hydrate. Despite their wide applications, however, the inhibition effects of alcohols and electrolytes, especially of their mixtures, on gas hydrate are still difficult to model due to the existence of hydrogen bonding and electrostatic interactions.

Anderson and Prausnitz[1] proposed a model to calculate the inhibition effect of methanol on gas hydrate. In their model, the Redlich-Kwong equation of state was used to calculate the fugacities of components in the vapor phase, and the UNIQUAC model was used to describe the liquid phase. Englezos predicted the methanol inhibition effect by the Trebble-Bishnoi equation of state.[2] A modified Patel-Teja equation of state, namely, VPT equation of state,[3] was also implemented by Mohammadi and Richon to predict the inhibition effect of ethylene glycol.[4] The modeling of the inhibition effects of solutions containing both alcohol and electrolyte on gas hydrate was also studied. For example, Nasrifar et al. modeled the hydrate phase equilibrium in the presence of methanol, ethylene glycol (MEG), and electrolytes, including NaCl, KCl, and $CaCl_2$.[5] In their work, based on the assumption of no interaction between alcohol and electrolyte, the activity of water is calculated by summing its separate contributions from electrolyte and alcohol. However, Jager et al. suggested that the interaction between electrolyte and alcohol should be considered to give an accurate hydrate inhibition prediction.[6] Guo et al.[7] used a modified Patel-Teja equation of state[8] to predict the inhibition effects of methanol and salt mixture, in which to get a reasonable prediction, a concentration-dependent binary interaction parameter between water and methanol must be used. Tohidi et al.[9] combined the VPT equation of state[3] with non-density dependent mixing rules and a modification of Debye-Huckle electrostatic term to predict the hydrate formation conditions in the presence of alcohols and salt mixtures. In their work, the binary interaction parameters between salt and alcohol are obtained by fitting water vapor pressure data, and the binary interaction parameters between salt and gas are dependent on the salt concentration. The cubic plus association equation of state (CPA) was also employed by Tohidi et al.[10,11,12] to predict the hydrate dissociation conditions in the presence of methanol, MEG, or salts. In the CPA equation of state, the temperature-dependent binary interaction parameters between non-associating and associating molecules are fitted to vapor-liquid equilibrium data, while concentration-dependent interaction parameters are needed when the mixture contains salts.

Due to its theoretical basis, statistical associating fluid theory (SAFT) could be a potential model to describe these complex systems. Its predictability has been successfully demonstrated in the thermodynamic property calculations of alcohols and electrolytes. For example, Li and Englezos successfully employed the SAFT equation of state to predict the vapor-liquid equilibrium containing alcohol.[13] For the modeling of aqueous electrolyte solutions, ion-based SAFT2[14,15,16] has been proven to be a successful tool for both single and mixed salt systems. Recently, Sadowski et al.[17] implemented ePC-SAFT[18] to model the thermodynamic behavior of salts in methanol and ethanol. However, to the best of our knowledge, SAFT has been applied only to hydrate phase equilibrium modeling in pure water[19,20], alcohols[21,22], or electrolytes[23,22], but not in aqueous solutions containing both alcohol and electrolyte.

It is the purpose of this work to apply ion-based SAFT2, coupled with van der Waals & Platteeuw theory,[24] to describe the methane, ethane, and propane hydrate dissociation conditions in the presence of aqueous ethanol, MEG, and glycerol solutions, as well as mixed MEG and electrolyte solutions, including NaCl, KCl, and CaCl$_2$.

## 4.2 Modeling

In hydrate phase equilibrium modeling, the chemical potential of water in different phases should be equal,[19]

$$\Delta \mu_w^H = \mu_w^{MT} - \mu_w^H = \mu_w^{MT} - \mu_w^L = \Delta \mu_w^L \qquad (4.1)$$

where $\mu_w^H$ is the chemical potential of water in the hydrate phase, $\mu_w^L$ is the chemical potential of water in the liquid phase, and $\mu_w^{MT}$ is the chemical potential of water in the empty hydrate lattice. The chemical potential of water in the hydrate phase can be obtained from the van der Waals and Platteeuw theory,[24] while the fluid phases are described by ion-based SAFT2.

### 4.2.1 Equation of State

The ion-based SAFT2 is expressed in terms of dimensionless residual Helmholtz energy,

$$\tilde{a}^{res} = \tilde{a}^{hs} + \tilde{a}^{disp} + \tilde{a}^{chain} + \tilde{a}^{assoc} + \tilde{a}^{ion} \qquad (4.2)$$

where the terms on the right side of the equation are the dimensionless residual Helmholtz energy accounting for hard sphere, dispersion, chain, association, and ion interactions. The details of calculations of hard sphere, dispersion, chain, and ionic terms can be found elsewhere.[19,23]

The association term in Eq. 4.2 can be expressed as,[25]

$$\widetilde{a}^{\,assoc} = \sum_i X_i \sum_\alpha \left[ \sum_{A_{\alpha i}} \left( \ln X^{A_{\alpha i}} - \frac{X^{A_{\alpha i}}}{2} \right) + \frac{n(\Gamma_{\alpha i})}{2} \right] \tag{4.3}$$

where $n(\Gamma_{\alpha i})$ is the number of association sites on segment $\alpha$ in molecule of component $i$, and $X^{A_{\alpha i}}$ is the mole fraction of molecule of component $i$ not bonded at site $A$ of segment $\alpha$. The details of the association term can be found elsewhere.[26] Similar to water, in this work, the alcohols are modeled as associating chain molecules with a certain number of associating sites corresponding to the number of hydroxyl groups. We assign two types of association sites: one type represents the lone pairs of electrons of oxygen (O), and the other type represents the hydrogen (H). Only hydrogen-lone pair interactions are allowed.

Each associating molecules are described by 6 parameters: segment number ($m$), segment volume ($v$), well depth of square well potential ($u$), the reduced width of square well potential ($\lambda$), association energy ($\varepsilon^{AB}$), and parameter related to association volume ($\kappa^{AB}$).

For electrolytes, the cation and anion are treated as nonassociating spherical segments. Each ion segment has 4 parameters: segment volume ($v$), segment energy ($u$), reduced width of square well potential ($\lambda$), and hydrated diameter ($d$). The ion parameters are temperature-dependent, which were obtained in our previous work by fitting activity coefficient and liquid density data of aqueous salt solutions.[23] It is worthy to notice that, although the model is now applied to aqueous solutions containing MEG and salt, new ion parameters are not needed.

To deal with mixtures, a few combining rules are implemented. For two different segments

$\alpha$ and $\beta$, as in our previous work,[19] the combining rule for the short-range dispersive interaction energy is given by

$$u_{\alpha\beta} = u_{\beta\alpha} = \sqrt{u_\alpha u_\beta}\left(1 - k_{\alpha\beta}\right) \tag{4.4}$$

where $k_{\alpha\beta}$ is the binary interaction parameter. For two different associating molecules, the following combining rules for the cross-association parameters are used.[27]

$$\varepsilon^{A_{\alpha i}B_{\beta j}} = \sqrt{\varepsilon^{A_{\alpha i}B_{\alpha i}}\varepsilon^{A_{\beta j}B_{\beta j}}} \tag{4.5}$$

$$\kappa^{A_{\alpha i}B_{\beta j}} = \frac{\sqrt{\kappa^{A_{\alpha i}B_{\alpha i}}\kappa^{A_{\beta j}B_{\beta j}}}\sqrt{v_{\alpha i}v_{\beta j}}}{\left(\frac{1}{2}\left[v_{\alpha i}^{1/3} + v_{\beta j}^{1/3}\right]\right)^3} \tag{4.6}$$

### 4.2.2 Models for Chemical Potential Differences

The van der Waals and Platteeuw theory[24] is used to calculate the chemical potential of water in the hydrate phase. The chemical potential difference between water in a hypothetical empty hydrate lattice and that in the hydrate phase is given by

$$\Delta\mu_w^H = \mu_w^{MT} - \mu_w^H = -RT\sum_m v_m \ln(1 - \sum_j \theta_{mj}) \tag{4.7}$$

where $v_m$ is the number of cavities of type $m$ per water molecule in the hydrate phase and $\theta_{mj}$ is the occupancy of molecules $j$ in type $m$ cavities. The detailed description of the hydrate phase based on this van der Waals and Platteeuw theory can be found elsewhere.[19] For the record, the fugacity of molecule $j$ needed in Eq. 4.7 is calculated using the ion-based SAFT2.

## 4.3 Results and Discussion

### 4.3.1 SAFT2 Parameters for Alcohols and Water

For ethanol, MEG, and glycerol, the SAFT2 parameters are obtained by fitting the saturated vapor pressure and saturated liquid density data of pure fluid.[28] The segment parameters, along with the association schemes, of alcohols and water are given in Table 4.1. The water parameters are taken from our previous work.[19]

### 4.3.2 Dielectric Constant of Aqueous MEG Solution

For a mixture of salt and aqueous MEG solution, the dielectric constant ($\varepsilon$) of aqueous MEG solution is needed in the ionic term of Eq. 4.2. This dielectric constant for an MEG concentration up to 40 wt% is developed in this work by fitting the experimental dielectric constant data of MEG and water mixture:[29]

$$\varepsilon = -72.56 \cdot x + 81.41 \qquad (4.8)$$

where $x$ is the mole fraction of MEG in the aqueous solution on a salt-free basis.

### 4.3.3 Binary Interaction Parameters

The binary interaction parameters of MEG-water and glycerol-water are obtained by fitting binary vapor-liquid equilibrium data, which turn out to be temperature-dependent. The goodness of fitting is shown in Figures 4.1 and 4.2. The binary interaction parameters between alkanes and water, which were also obtained by fitting vapor-liquid equilibrium data, can be found in our previous work.[19] The equations of these temperature-dependent binary interaction parameters are given in Table 4.2. The $k_{\alpha\beta}$'s between water and ethanol, and between alkanes and alcohols are set to 0.

Table 4.1. Ion-based SAFT2 parameters for associating fluids

| | Ethanol | MEG | Glycerol | Water |
|---|---|---|---|---|
| Association Scheme | 2 O 1 H | 2 O 2 H | 3 O 3 H | 2 O 2 H |
| $m$ | 1.6730 | 1.5902 | 2.2772 | 1.000 |
| $v$ (cc/mol) | 15.459 | 19.213 | 18.613 | 9.8307 |
| $u/k$ (K) | 322.309 | 316.650 | 255.011 | 311.959 |
| $\lambda$ | 1.347 | 1.610 | 1.709 | 1.537 |
| $\varepsilon^{AB}/k$ (K) | 2195.418 | 2066.432 | 1997.107 | 1481.41 |
| $\kappa^{AB}$ | 0.02108 | 0.01983 | 0.02663 | 0.04682 |
| AAD(P)% [a] | 0.90 | 0.20 | 3.5 | 0.26 |
| AAD($\rho$)% [b] | 0.79 | 0.13 | 0.55 | 0.33 |

$$^a\ AAD(P) = \frac{1}{N}\sum_{i=1}^{N}\left|\frac{P_{exp}^i - P_{cal}^i}{P_{exp}^i}\right| \qquad ^b\ AAD(\rho) = \frac{1}{N}\sum_{i=1}^{N}\left|\frac{\rho_{exp}^i - \rho_{cal}^i}{\rho_{exp}^i}\right|$$

The $k_{\alpha\beta}$'s between ions and MEG are first set to 0, but could be adjusted if needed. As described in our previous work,[23] the short-range interactions of cation-cation and anion-anion are neglected, i.e. $k_{\alpha\beta} = 1$. For cation-anion interaction, the $k_{\alpha\beta}$ is set to 0.5, while for water-ion and alkane-ion interactions, no binary interaction parameters are used, i.e., the $k_{\alpha\beta}$ is set to 0.

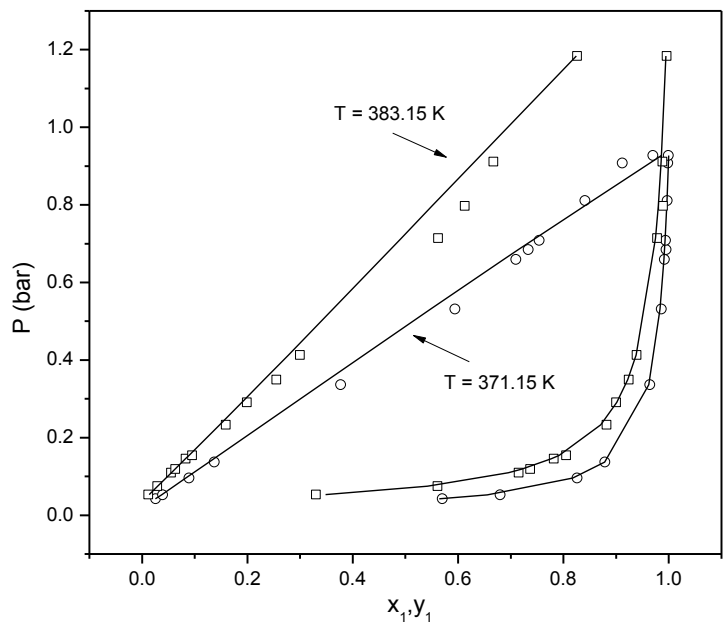

Figure 4.1. Vapor-liquid equilibria for water (1)/ MEG (2) system at 371.15 K and 383.15 K, experimental (points)[30] and calculated (curves).

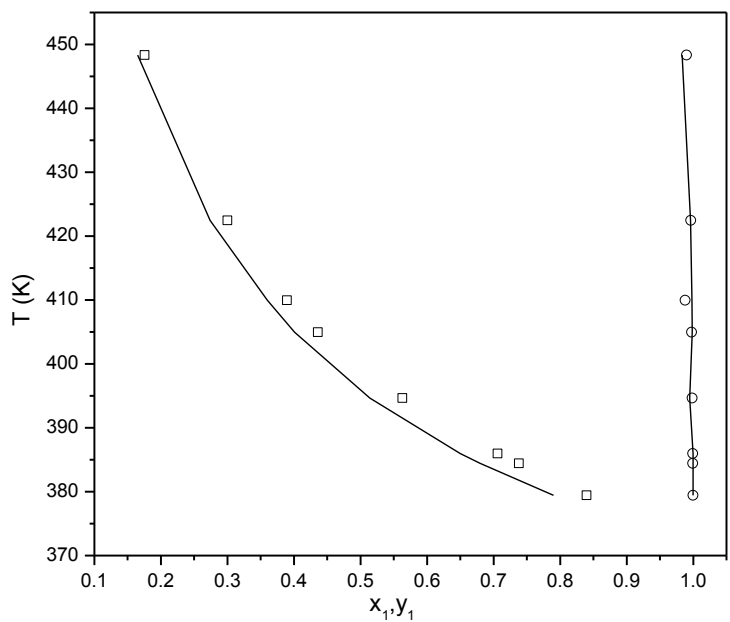

Figure 4.2. Vapor-liquid equilibria for water (1)/glycerol (2) system at P = 1.01325 bar, experimental (points)[31] and calculated (curves).

Table 4.2. Temperature-dependent binary interaction parameters $(k_{\alpha\beta})^*$ for dispersion energy

|  | $a$ | $b$ |
|---|---|---|
| Water/Ethylene Glycol | $-2.6667\times10^{-4}$ | $8.097\times10^{-2}$ |
| Water/Glycerol | $-4.3315\times10^{-4}$ | 0.15469 |
| Water/ Methane | $1.3\times10^{-3}$ | $-0.487$ |
| Water/Ethane | $7\times10^{-4}$ | $-0.227$ |
| Water/Propane | $1\times10^{-2}$ | $-0.317$ |

$^*$ $k_{\alpha\beta} = aT + b$   ($T$ is in Kelvin)

**4.3.4 Prediction of Alkane Hydrate Dissociation Pressure**

Table 4.3 summarizes all of the systems studied along with the average absolute deviations (AAD) of the predicted pressures.

Table 4.3. Systems studied

| Gas | Inhibitor | AAD(P)%$^*$ | Data Source |
|---|---|---|---|
| Methane | 5 wt% Ethanol | 4.96 | [32] |
|  | 10 wt% Ethanol | 7.67 |  |
|  | 10 wt% MEG | 2.63 | [33] |
|  | 30 wt% MEG | 3.27 |  |
|  | 50 wt% MEG | 8.76 |  |
|  | 25 wt% Glycerol | 1.23 | [34] |
|  | 50 wt% Glycerol | 4.49 |  |
|  | 20 wt% MEG + 15 wt% NaCl | 3.70 | [35],[36] |
|  | 12.07 wt% MEG + 15.01 wt% NaCl | 2.12 |  |
| | 3.77 wt% MEG + 15.67 wt% NaCl | 2.74 | |
| | 15.36 wt% MEG + 3.77 wt% NaCl | 3.45 | |
| | 23.88 wt% MEG + 3.77 wt% NaCl | 3.41 | |
| | 5.77 wt% MEG + 3.77 wt% NaCl | 3.81 | |
| | 35 wt% MEG + 8 wt% KCl | 9.90 | [32], [37] |
| | 23 wt% MEG + 10 wt% KCl | 3.97 | |
| | 20 wt% MEG + 10 wt% KCl | 9.63 | |
| | 26 wt% MEG + 14 wt% $CaCl_2$ | 3.03 | [38] |
| | 14 wt% MEG + 18 wt% $CaCl_2$ | 6.40 | |
| | 13.4 wt% MEG + 15.3 wt% $CaCl_2$ | 1.29 | |
| Ethane | 5 wt% Ethanol | 1.57 | [32] |
| | 10 wt% Ethanol | 2.82 | |
| Propane | 5 wt% Ethanol | 4.12 | |
| | 10 wt% Ethanol | 8.74 | |
| | 10 wt% MEG | 2.05 | [39] |
| | 20 wt% MEG | 1.02 | |
| | 10 wt% Glycerol | 1.67 | |
| | 20 wt% Glycerol | 2.84 | |
| Methane + Ethane + Propane | 10 wt% MEG + 10 wt% NaCl | 6.28 | [40] |
| | 20 wt% MEG + 10 wt% NaCl | 14.9 | |
| | 10 wt% MEG + 10 wt% $CaCl_2$ | 1.00 | |
| | 20 wt% MEG + 5 wt% NaCl + 5 wt% $CaCl_2$ | 4.69 | |
| | 15 wt% MEG + 5 wt% NaCl + 5 wt% $CaCl_2$ | 12.76 | |

$$^* AAD(P) = \frac{1}{N} \sum_{i=1}^{N} \left| \frac{P_{\exp}^i - P_{pre}^i}{P_{\exp}^i} \right|$$

### 4.3.4.1 Hydrate in the Presence of Alcohol

Figure 4.3 shows the prediction of methane hydrate dissociation pressures in the presence of aqueous ethanol solution, which is found to be reasonable. The experimental data are taken from Mohammadi and Richon.[32] Figure 4.4 shows the inhibition effects of ethanol on ethane and propane hydrates. Note that propane forms structure II hydrate. The binary interaction parameter between ethanol and water, which is set to 0, is not further adjusted since satisfactory predictions have been obtained.

The inhibition effects of MEG on methane and propane hydrates are also studied. Figure 4.5 shows the prediction of methane hydrate dissociation pressures in the presence of MEG at different concentrations. Compared to the experimental data,[33] the model with the binary interaction parameters given in Table 4.2 gives accurate predictions, in which the temperature and concentration effects are well captured. Propane hydrate dissociation condition in the presence of MEG is presented in Figure 4.6. Satisfactory prediction is also found for this structure II hydrate former.

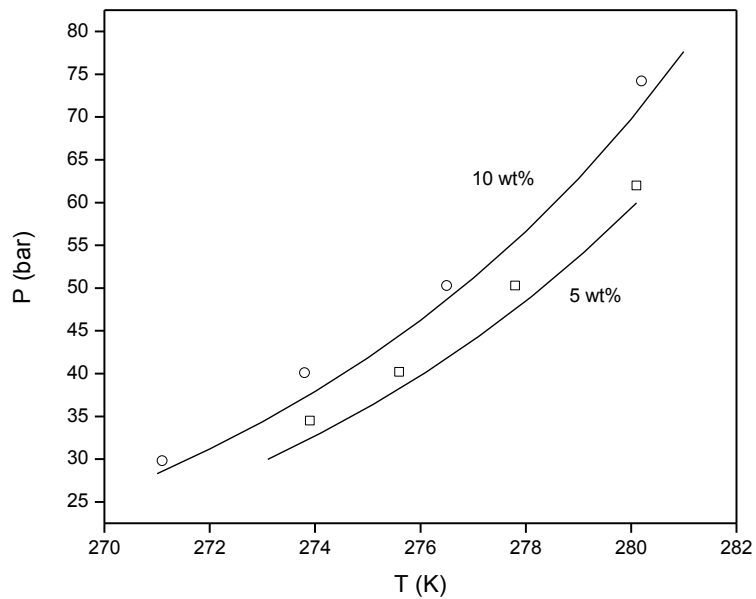

Figure 4.3. Methane hydrate dissociation pressure in the presence of ethanol of different concentrations, experimental (points)[32] and predicted (curves).

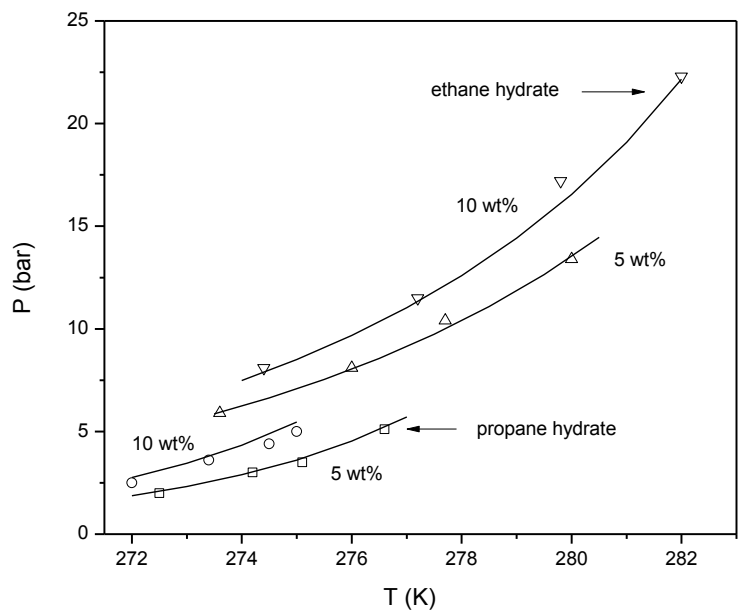

Figure 4.4. Ethane and propane hydrate dissociation pressures in the presence of ethanol of different concentrations, experimental (points)[32] and predicted (curves).

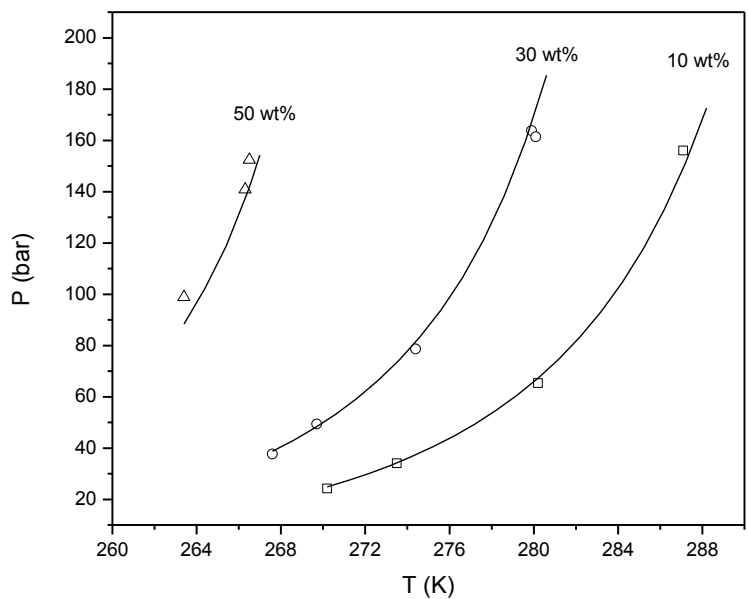

Figure 4.5. Methane hydrate dissociation pressures in the presence of MEG of different concentrations, experimental (points)[33] and predicted (curves).

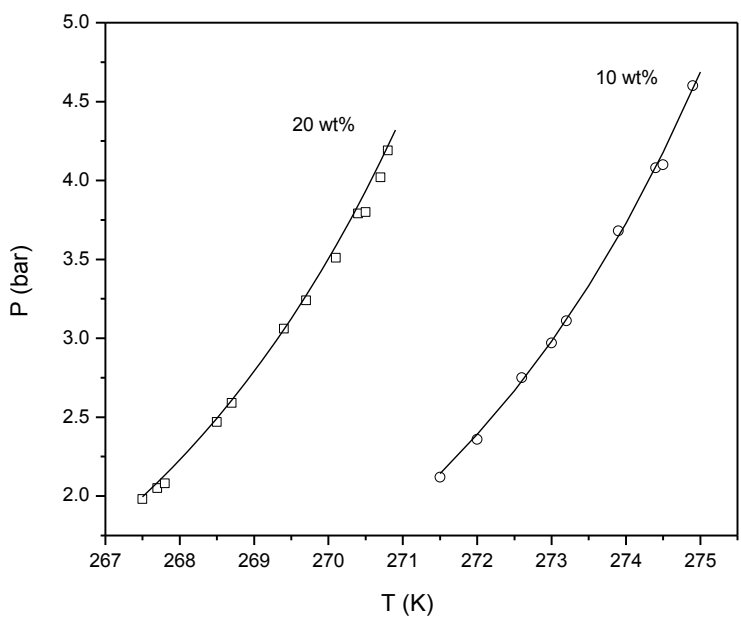

Figure 4.6. Propane hydrate dissociation pressures in the presence of MEG of different concentrations, experimental (points)[39] and predicted (curves).

Methane hydrate dissociation pressures in aqueous glycerol solutions are predicted, as shown in Figure 4.7. Again, good agreement between our prediction and experimental data[34,39] is demonstrated. The inhibition effect of glycerol on propane hydrate is also well represented by the model, as shown in Figure 4.8.

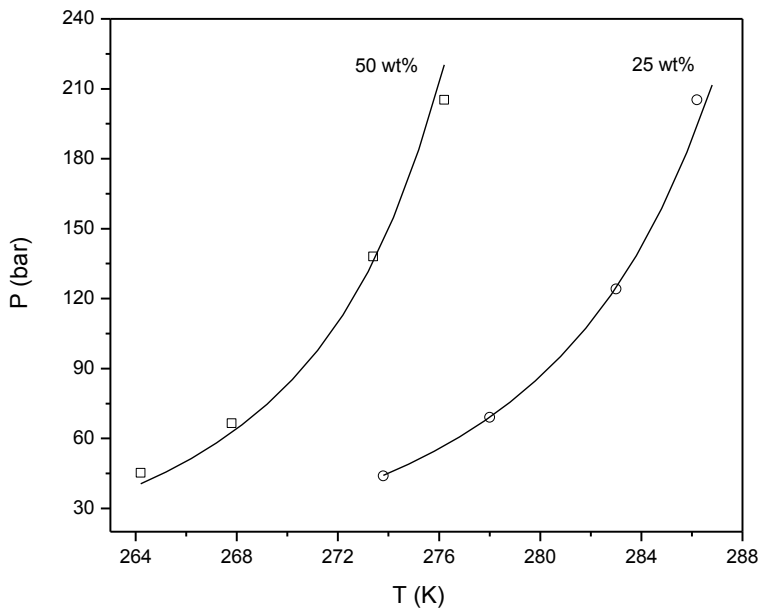

Figure 4.7. Methane hydrate dissociation pressures in the presence of glycerol of different concentrations, experimental (points)[34] and predicted (curves).

### 4.3.4.2 Hydrate in the Presence of MEG and Electrolytes.

In this work, methane hydrate dissociation pressures in the presence of MEG and electrolytes, including NaCl, KCl, CaCl$_2$, are predicted. The inhibition effects of MEG and electrolytes on a ternary methane, ethane, and propane mixture hydrate are then considered.

Figures 4.9 and 4.10 present the methane hydrate dissociation pressures in the presence of MEG and NaCl mixture. Although the parameters of Na$^+$ and Cl$^-$ used were obtained in our previous work[23] by fitting activity coefficients and liquid density of aqueous NaCl solution in the absence of alcohol, the model gives good predictions for all conditions considered. This result shows that the ion parameters of the model are not solvent-specific and could be used with confidence in other solvent as long as the dielectric constant of that solvent is known. Note also that the binary interaction parameters between Na$^+$ and MEG, and between Cl$^-$ and MEG are 0. As shown in Figures 4.9 and 4.10, the concentration and temperature effects of NaCl and MEG are well captured.

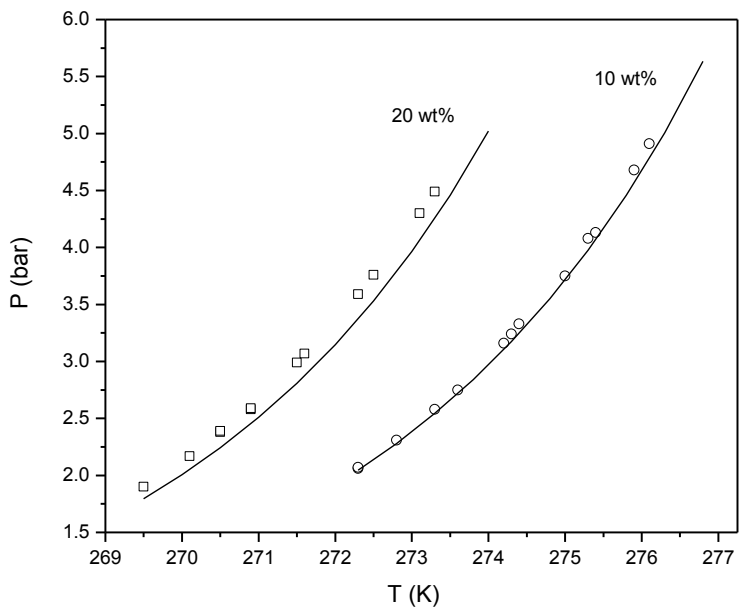

Figure 4.8. Propane hydrate dissociation pressure in the presence of glycerol of different concentrations, experimental (points)[39] and predicted (curves).

Figure 4.11 shows the prediction of methane hydrate dissociation pressures in the presence of MEG and KCl mixture. With the K$^+$ parameters obtained from our previous work,[23] good agreement can also be found between our prediction and experimental data[35,37]. The binary interaction parameter between K$^+$ and MEG is 0.

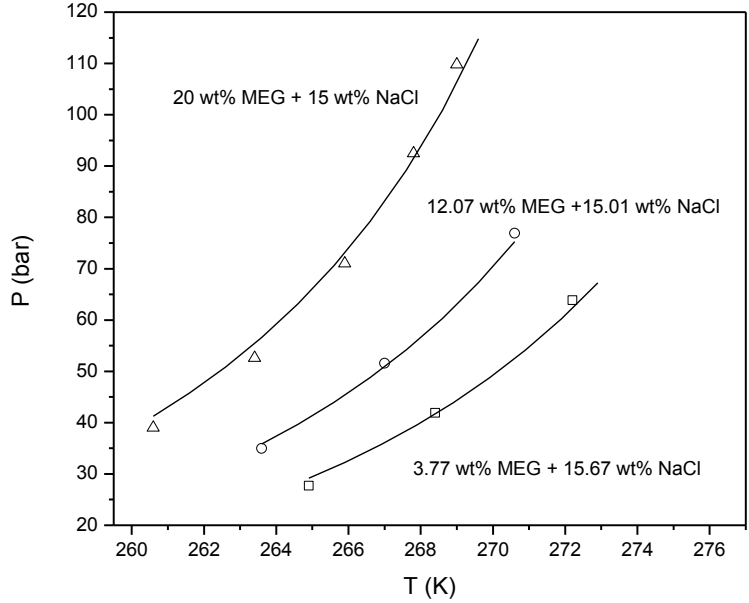

Figure 4.9. Methane hydrate dissociation pressures in the presence of MEG + NaCl of different concentrations, experimental (points)[35,36] and predicted (curves).

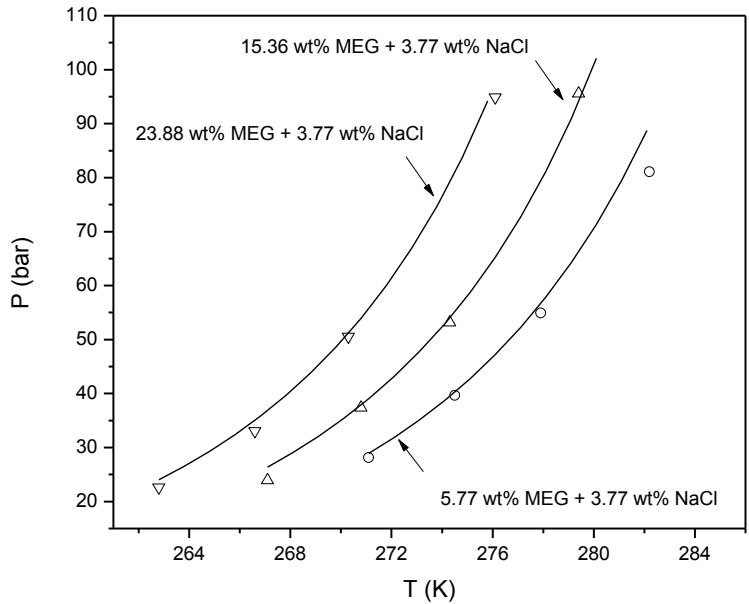

Figure 4.10. Methane hydrate dissociation pressures in the presence of MEG + NaCl of different concentrations, experimental (points)[36] and predicted (curves).

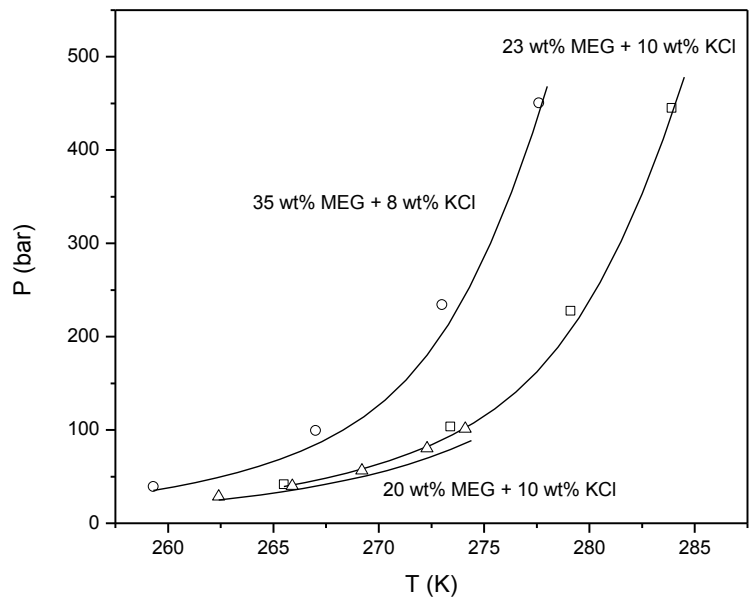

Figure 4.11. Methane hydrate dissociation pressures in the presence of MEG + KCl of different concentrations, experimental (points)[35,37] and predicted (curves).

The inhibition effect of MEG and $CaCl_2$ mixture on methane hydrate is also investigated. As in other systems, the binary interaction parameter between $Cl^-$ and MEG for this mixture is also 0. However, since the dispersion energy of $Ca^{2+}$ obtained from our previous work[23] is much higher than that of MEG, a non-zero $k_{\alpha\beta}$ is needed to adjust the dispersive interaction energy between $Ca^{2+}$ and MEG. Since we do not have activity coefficient data or other data for $Ca^{2+}$ in MEG, the $k_{\alpha\beta}$ is obtained by fitting the methane hydrate dissociation pressures at a mixture concentration of 13.4 wt% MEG+15.3 wt% $CaCl_2$. Figure 4.12 shows the fitting curve (dashed curve), where the value of $k_{\alpha\beta}$ is found to be −0.235. With the same $k_{\alpha\beta}$, the methane hydrate dissociation pressures at other MEG and $CaCl_2$ concentrations are well predicted, as depicted in Figure 4.12.

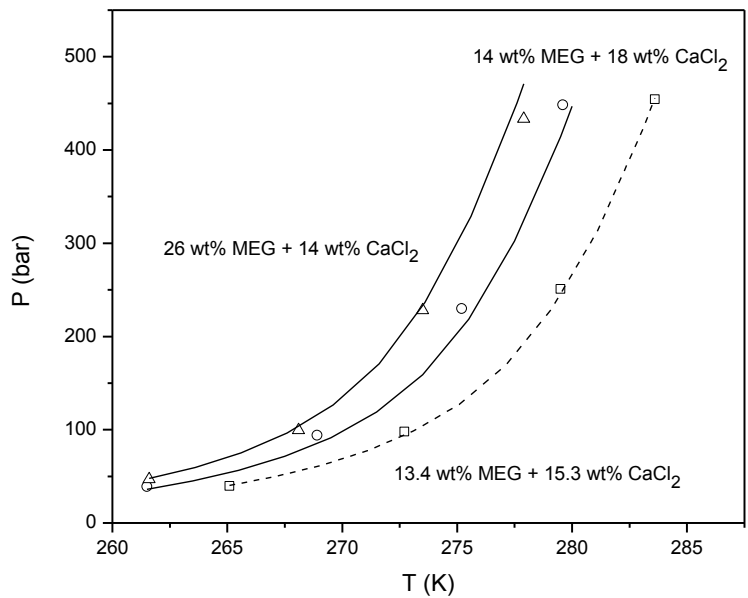

Figure 4.12. Methane hydrate dissociation pressures in the presence of MEG + CaCl₂ of different concentrations, experimental (points),[38] predicted (solid curves), and calculated (dashed curve).

The dissociation condition of a ternary methane, ethane, and propane mixture hydrate in the presence of MEG and salt is considered. This ternary gas mixture contains 91.96 mol% methane, 5.13 mol% ethane, and 2.91 mol% propane, which forms structure II hydrate. As before, the binary interaction parameters between $Na^+$ and MEG, and between $Cl^-$ and MEG are 0, while the binary interaction parameter between $Ca^{2+}$ and MEG is −0.235. Figure 4.13 presents the inhibition effects of MEG and NaCl, and MEG and CaCl₂ on methane, ethane, and propane mixture hydrate. Compared to the experimental data,[40] the model provides accurate predictions, except for 20 wt% MEG, where the model tends to give higher but still reasonable predictions.

To analyze further, the inhibition effect of MEG without salt on this ternary gas mixture hydrate is also predicted. Similar to the results shown in Figure 4.13, the model also gives reasonable predictions. However, as shown in Figures 4.5 and 4.6, the predictions of the dissociation pressures of the individual methane or propane hydrate in the presence of MEG, even at high MEG concentration, are accurate, and hence, the inaccuracy of Figure 4.13 is arguably resulted from the van der Waals and Platteeuw theory.[24] Actually, it is widely recognized that the van der Waals and Platteeuw theory, when it is applied to the modeling of gas mixture hydrate, cannot give very accurate results.[41] Since the prediction is reasonable for this ternary gas mixture hydrate, in this work we do not attempt to make correction to the van der Waals and Platteeuw theory.

Figure 4.14 shows the inhibition effect of MEG + NaCl + CaCl$_2$ solution on this ternary gas mixture hydrate. While not perfect, the prediction is again reasonable. At higher MEG concentration, the model tends to give a higher prediction.

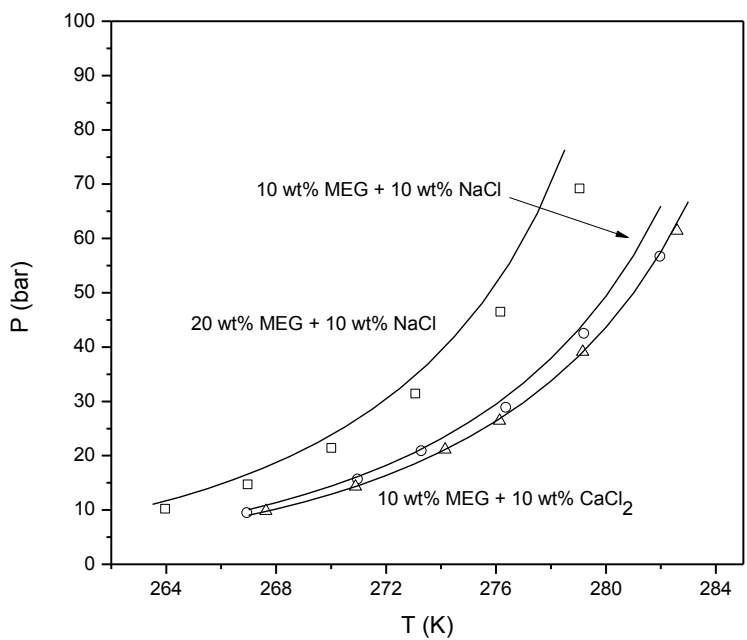

Figure 4.13. Methane + ethane + propane hydrate dissociation pressures in the presence of MEG + NaCl and MEG + CaCl$_2$, experimental (points)[40] and predicted (curves).

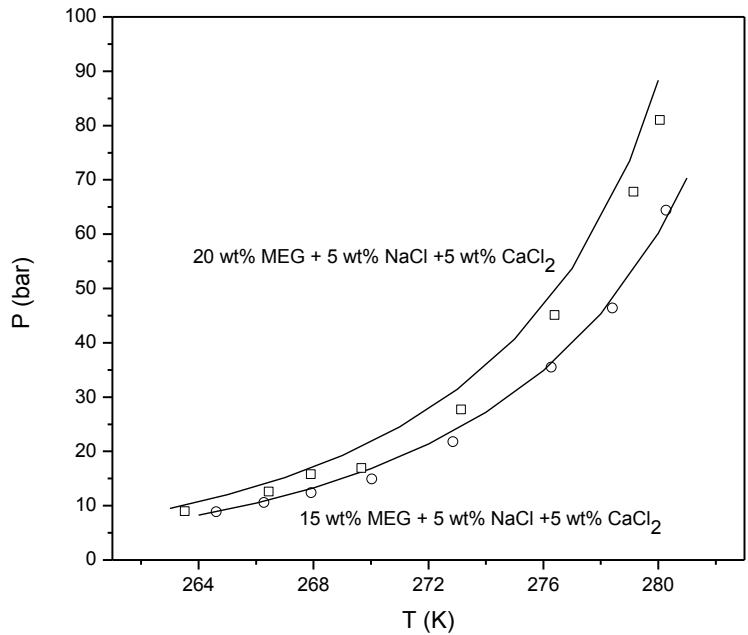

Figure 4.14. Methane + ethane + propane hydrate dissociation pressures in the presence of MEG + NaCl + CaCl$_2$, experimental (points)[40] and predicted (curves).

## 4.4 Conclusions

Ion-based SAFT2, coupled with van der Waals and Platteeuw theory, is used to predict the methane, ethane, and propane hydrate dissociation conditions in the presence of aqueous ethanol, MEG, and glycerol solutions, as well as mixed MEG and electrolyte solutions, including NaCl, KCl, and CaCl$_2$. The SAFT2 parameters for ethanol, MEG, and glycerol are fitted to the saturated vapor pressure and saturated liquid density data of pure fluid. The model is found to reliably predict the inhibition effects of these alcohols on alkane hydrate. It is even more convincing to find that with the same ion parameters developed for aqueous solutions in our previous work, the model can predict the alkane hydrate dissociation pressures in the presence of both MEG and salts, including NaCl, KCl, and CaCl$_2$. This result also demonstrates that the ion parameters of the model are not solvent-specific and could be used with confidence in other solvent as long as the dielectric constant of that solvent is known. In all cases studied, the effects of temperature and concentration on hydrate dissociation conditions are well captured.

# Chapter 5. Thermodynamic Modeling of Aqueous Ionic Liquid Solutions and Prediction of Methane Hydrate Dissociation Conditions in the Presence of Ionic Liquid

## 5.1 Introduction

Ionic liquids (ILs), also known as liquid salts or ionic fluids, are organic salts with low melting point, which makes them liquids at ambient or relatively low temperature. ILs are promising chemicals or solvents for green chemical processes because of their unique physicochemical properties, such as low vapor pressure, low viscosity, high electric conductivity, excellent solvation behavior, non–flammability, and high selectivity.

Recently, ILs were also found to be a new class of dual function hydrate inhibitors, which not only shift the hydrate dissociation curve to higher pressures and lower temperatures, but also slow down the rate of hydrate formation[1,2]; hydrate is clathrate solid formed by water and suitable gas molecules at high pressure and low temperature. Thus, IL could be a new effective hydrate inhibitor that is promising for the offshore oil and gas production process. The use of IL as inhibitor has been experimentally studied for both methane and carbon dioxide hydrates.[3,4,5] The huge combination of cations and anions that can be tailored to form ILs would enable us to find more effective gas hydrate inhibitors that are inexpensive and biodegradable. However, at the same time this also poses a real challenge. Due to the huge number of ILs potentially used as hydrate inhibitors, experimental study is not only inadequate but also expensive. Hence, it is extremely important to have thermodynamic models to describe their properties and behavior. It is the purpose of this work to introduce a thermodynamic model that can be used to describe the properties and the thermodynamic hydrate inhibition performance of this promising class of inhibitors.

Several thermodynamic models have been proposed to describe the properties of pure IL and fluid mixture containing IL. An early work on the thermodynamic modeling of IL by Shariati and Peters[6] was based on Peng-Robison (PR) equation of state.[7] Carvalho et al.[8] studied the solubility of $CO_2$ in IL by combining the PR EOS and the UNIQUAC model, and Chen et al.[9] modeled the liquid-liquid equilibrium of mixtures containing IL and organic solvent, such as alcohol and benzene, using NRTL model.[10] However, most of these models are empirical. Although the experimental data can be correlated, the predictive power of these models is limited.

Recently, models with theoretical basis were developed to describe the thermodynamic properties of ILs. Liu and coworkers[11] developed a square-well chain fluid (SWCF) EOS and successfully described the density of pure IL and gas solubility in IL. In their model, IL was treated as a homosegmented chain molecule with a square-well potential. Liu and coworkers[12] also proposed a heterosegmented SWCF EOS and modeled the binary vapor-liquid equilibria of systems containing IL. In this hetero-SWCF EOS, IL was modeled as a di-block molecule, with the alkyl chain as one block and the cation head-anion pair as the second block. Statistical associating fluid theory (SAFT) was also applied to the thermodynamic modeling of ILs. Kroon et al.[13] developed a truncated perturbed chain polar statistical associating fluid theory (tPC-PSAFT) to model the $CO_2$ solubility in IL, where the dipolar interactions between IL molecules and the Lewis acid-base type of association between the IL and the $CO_2$ molecules were accounted for. Ji et al.[14] used ePC-SAFT to correlate the density of pure IL and predict gas solubility in IL. They proposed and compared six different strategies for the IL modeling, and the one with the Debye-Huckle term to account for the Coulomb interaction was found to be the most predictive. In this strategy, the cation and anion of an IL were modeled as homosegmented chains without association. Heterosegmented SAFT models were also developed to describe the thermodynamic behavior of IL. Ji and Adidharma[15] proposed a heterosegmented model to correlate and predict the density of pure IL. This heterosegmented SAFT EOS was then applied to correlate and predict the solubility and partial molar volume of $CO_2$ in IL.[16,17] In this model, the cation was treated as a chain molecule while the anion was represented as a spherical segment. The chain molecule of cation was composed of one effective segment representing the cation head and different groups of segments representing the alkyl chains. The cation head and anion each was assumed to have one association site, which could only cross associate to each other.

Although several theoretical thermodynamic models have been developed for IL, most of their applications focused on the description of pure IL properties and gas solubility in IL. The modeling work of aqueous IL solutions is quite rare although many ILs, as electrolytes, can totally or partially dissolve in water. This could be due to the fact that the modeling of these systems is even more challenging. Soft-SAFT was implemented to correlate the density of pure IL and predict the binary phase equilibria of IL and alcohol (or water) mixtures.[18] In their model, IL molecules were modeled as homosegmented chains with three associating sites to account for the interaction between cation and anion. Liu and coworkers[19] used the SWCF EOS to model the thermodynamic properties of aqueous IL solutions. PC-SAFT was also implemented to describe the properties of aqueous solutions of IL.[20] In these two models, cross association between cation and anion was also considered. The electrostatic interaction between cation and anion was described by using an electrostatic term based on the primitive model (PM) of the mean spherical approximation (MSA).[21] Their model parameters were obtained by fitting to the density and osmotic coefficients of aqueous IL solutions. Since these two works are based on a homosegmented model, the model parameters are not transferrable, which means that new parameters need to be fitted when new IL species having the same constituents are involved in the modeling.

Due to the complex molecular structure and thermodynamic behavior of IL, the modeling of IL as hydrate inhibitor is still very limited. Kaniki et al.[22] studied the inhibition effect of tributhylmethylphosphonium methylsulfate on methane and carbon dioxide hydrates. In this model, the Peng-Robinson EOS[7] and the NRTL model[10] were used to describe the fluid phase while the van der Waals and Platteeuw theory[23] was applied to model the hydrate phase. Marziyeh et al.[24] used the electrolyte cubic square well (ECSW) EOS[25] and the van der Waals and Platteeuw theory[23] to predict the methane hydrate phase equilibria in the presence of IL. In their model, the parameters of the ECSW EOS were adjusted to the osmotic coefficients of aqueous IL solutions. A recent work by Partoon et al.[26] implemented a Pitzer-type model to represent the methane hydrate dissociation conditions in the presence of ionic liquid. In our previous work, the heterosegmented SAFT EOS has been successfully applied to predict the n-alkane hydrate dissociation conditions in pure water and in solutions containing conventional inhibitors, including salt solutions.[27,28,29] In this work, we extend this heterosegmented SAFT to describing the thermodynamic properties of aqueous imidazolium IL solutions containing [$C_x$MIM][Br] ($2 \leq x \leq 6$), [$C_4$MIM][Cl], or [$C_4$MIM][BF$_4$], and apply it, coupled with the van der Waals and Platteeuw theory,[23] to predict the dissociation conditions of methane hydrate in the presence of imidazolium ILs.

## 5.2 Thermodynamic Model

An IL consists of an organic cation and an organic or inorganic anion. The cation of the imidazolium IL is composed of an aromatic ring (imidazolium head/cation head) and several alkyls. In this heterosegmented SAFT model, the cation is treated as a chain with a spherical segment representing the cation head and different groups of segments representing the alkyls, as depicted in Figure 5.1. The positive charge is assumed to be located on the head segment of the cation chain. The anion of the IL is modeled as a spherical segment with a negative charge.

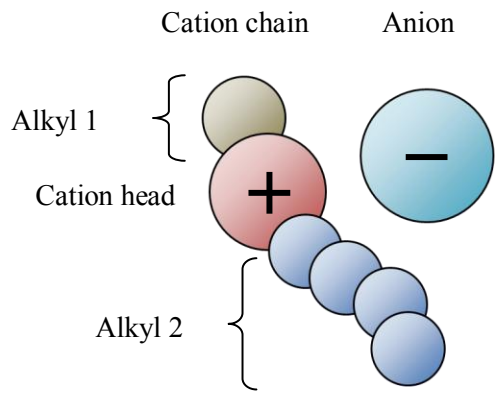

Figure 5.1. Heterosegmented SAFT model for an ionic liquid.

In recent years, molecular simulations have been done to study and understand the bulk and structural properties of imidazolium IL and water mixtures.[30,31] In these molecular dynamic simulations, the radial distribution functions of cation-anion, cation-water, anion-water, anion-anion, and cation-cation pairs, the average number of hydrogen bonds, and the average size of the biggest water clusters at different water concentrations can be obtained and analyzed. Niazi et al.[31] drew a general conclusion that is very important for our modeling purposes. At low to intermediate water concentrations, the imidazolium IL structural property is found to be similar to that of pure ILs, where the IL molecules take the form of ionic pairs due to the strong interaction between the cation and the anion. As the water concentration increases, however, a greater solvation around the IL cations and anions exists and finally, at a water concentration greater than about 70 mol%, the system experiences a transition into an aqueous solution with the IL as solute.

Based on those molecular simulation studies, since the systems investigated in this work contain a large amount of water, our model should be developed to account for the long-range Coulomb interaction as well as the other interactions. Note that in our previous works dealing with pure ILs and gas solubility in ILs,[15,16,17] the Coulomb interaction was not included because the IL molecules could be considered to take the form of ionic pairs. In that case, the cation-anion interaction was described by association only.

## 5.2.1 Equation of State

The heterosegmented SAFT is expressed in terms of dimensionless residual Helmholtz energy,

$$\tilde{a}^{res} \cong \tilde{a}^{hs} + \tilde{a}^{disp} + \tilde{a}^{chain} + \tilde{a}^{assoc} + \tilde{a}^{ion} \tag{5.1}$$

where the superscripts on the right hand side refer to terms accounting for the hard sphere, dispersion, chain, association, and ion interactions. The details of calculations of hard-sphere, dispersion, and chain terms can be found in our earlier work.[27]

The common approach to account for the Coulomb interaction is to use a primitive model of MSA, in which the solvent is represented by a continuous medium of uniform dielectric constant. This approach has been proven to work even if the solvent is not a single component.[29]

Since in our model the cation of the imidazolium IL comprises of a positively charged spherical segment representing the cation head and neutral segments representing the alkyls, the presence of these neutral segments could affect the Coulomb interaction between the cation head and anion. The neutral segments are parts of the cation, not the solvent; if they were parts of the solvent, they would be treated as parts of the continuous medium (background), which only alter the dielectric constant of the medium. To investigate this issue, we calculate the internal energies and osmotic coefficients of equal- and unequal-sized mixtures of charged and neutral segments using the restricted primitive model (RPM) in the MSA formalism, and compare the results with the available Monte Carlo (MC) simulation data.[32] Although RPM is derived in the absence of neutral segments, it is found that in the electrolyte concentration range of the MC data, the agreement between the calculated and MC data is surprisingly good. However, due to the unavailable MC data, we cannot affirm that RPM would also be applicable at high electrolyte concentrations.

Based on this finding, to a first approximation, RPM is assumed to be able to describe the electrostatic interaction between cation and anion in the presence of neutral segments. The ionic term is then expressed as,[21]

$$\widetilde{a}^{ion} = -\frac{3x^2 + 6x + 2 - 2(1+2x)^{3/2}}{12\pi\rho N_A d^3} \tag{5.2}$$

where $\rho$ is the molar density, $N_A$ is the Avogadro's number, $d$ is the hydrated diameter, $x$ is a dimensionless quantity calculated by,

$$x = \kappa d \tag{5.3}$$

where $\kappa$ is the Debye inverse screening length.

The association term in Eq. 5.1 should in principle accounts for the hydrogen-bond interactions between water-water, water-anion, water-cation, and cation-anion pairs. Following our previous works[15,27], in this work the water-anion and water-cation hydrogen-bond interactions are not explicitly accounted for. If a non-primitive model were instead used, these water-ion hydrogen-bond interactions would be needed to describe the ion solvation.

Therefore, the association in Eq. 5.1 considers only water-water and cation-anion interactions. One association site each is assigned to the cation head and anion, and thus only cross association is considered. As used in our previous works,[27] water molecule is modeled as a spherical segment with two types of association sites. The details of the association term can be found elsewhere.[29]

As mentioned earlier, in this heterosegmented SAFT model, IL molecule is divided into several groups of segments: a spherical segment representing the cation head, a spherical segment representing the anion, and groups of segments of different types representing the alkyls. Each of these groups of segments has five segment parameters: segment number ($m$), segment volume ($v^{oo}$), segment energy ($u/k$), reduced width of the square well potential ($\lambda$), and bond number ($n_B$), which is used to calculate the bond fraction in the cation chain. The details about the segment number ($m$) and bond fraction of alkyl groups can be found in our previous work.[15] Note that the bond numbers of the cation head and anion are 0 because they are spherical segments ($m = 1.0$). Since the positive charge of the cation is assumed to be located on the head segment of the cation chain and the negative charge is located on the anion segment, the cation head and anion each has one additional ion parameter: the hydrated diameter ($d$). Besides these segment/ion parameters, there are also two additional cross association parameters, i.e., association energy ($\varepsilon/k$) and association bonding volume parameter ($\kappa$), which are the properties of the cation head-anion pair.

To capture the temperature effect on the properties of aqueous IL solution, the segment energy ($u/k$) of the cation head is allowed to be temperature-dependent, given by

$$\frac{u}{k} = c_1 + c_2 \cdot T + c_3 \cdot T^2 \tag{5.4}$$

where $T$ is the temperature in Kelvin, and $c_1$, $c_2$, and $c_3$ are the coefficients that should be fitted to experimental data. The rest of the model parameters are constant, not temperature-dependent. The combining rules for the model parameters are the same as in our earlier work.[28,29] In this work, the binary interaction parameter ($k_{ij}$) is only used to correct the short-range cation head–water and alkyl–water interactions.

### 5.2.2 Models for Chemical Potential Differences

The proposed heterosegmented SAFT EOS, coupled with van der Waals and Platteeuw theory,[23] is applied to study the methane hydrate dissociation conditions in the presence of imidazolium IL inhibitor. In hydrate phase equilibrium modeling, the chemical potential of water in different phases should be equal,[27]

$$\Delta \mu_w^H = \mu_w^{MT} - \mu_w^H = \mu_w^{MT} - \mu_w^L = \Delta \mu_w^L \tag{5.5}$$

where $\mu_w^H$ is the chemical potential of water in the hydrate phase, $\mu_w^L$ is the chemical potential of water in the liquid phase, and $\mu_w^{MT}$ is the chemical potential of water in the empty hydrate lattice. The chemical potential of water in the hydrate phase can be obtained from the van der Waals and Platteeuw theory,[23] while the fluid phases are described by the heterosegmented SAFT EOS. The detailed description of the hydrate phase based on this van der Waals and Platteeuw theory can be found elsewhere.[27]

## 5.3 Results and Discussions

### 5.3.1 Parameter Estimation

In a heterosegmented model, the parameters are assigned to each group of segments. The model parameters are transferrable and can be applied to other imidazolium ILs having ions that have been included in this work.

Some of the model parameters had been obtained in our previous works. The parameters for the alkyl groups, except $v^{oo}$, $u/k$, and $\lambda$ of the ethyl group (C2−), are taken from the work of Ji and Adidharma,[15] in which the alkyl parameters were derived from those of the corresponding n-alkanes. The parameters of Cl− are taken from our previous work derived from the density and activity coefficient of NaCl aqueous solution.[28] The water and methane parameters are also taken from our previous work without modification.[27] In addition, the $k_{ij}$'s of methyl-water, ethyl-water, and hexyl-water are set to zero because the results are found to be insensitive to these parameters.

With these available parameters, the following parameters still need to be fitted to experimental data: (1) $v^{oo}$, $u/k$, and $\lambda$ of the ethyl group (C2−), (2) the segment/ion parameters of the cation head (IMI+) and anions (BF4−, Br−), (3) the cross association parameters between cation head and anions (Cl−, Br−, BF4−), (4) the coefficients $c_1$, $c_2$, and $c_3$ of Equation 5.4, and (5) the binary interaction parameters ($k_{ij}$'s) of cation head-water, propyl-water, butyl-water, and pentyl-water.

The density ($\rho$)[33,34] (Bhajan et al., 2012; Mohammed et al., 2010) and mean activity coefficient ($\gamma_{\pm}$)[35,36,37] data of aqueous solutions of [C2MIM][Br], [C4MIM][Cl], [C4MIM][BF4], and [C4MIM][Br] are used to fit the model parameters described above for the groups involved.

Whenever possible, mean activity coefficient data is used instead of osmotic coefficient data ($\phi$) in the parameter estimation since the EOS parameters are more sensitive to activity coefficient data.[38] The experimental activity coefficient data of [C$_2$MIM][Br] used in the fitting are taken from literature.[35] Due to the unavailability of experimental data, however, the activity coefficient data for aqueous solutions of [C$_4$MIM][Cl], [C$_4$MIM][BF$_4$], and [C$_4$MIM][Br] are obtained by using the correlations proposed by Gonzalez et al.[36] and Shekaari et al.[37] The binary interaction parameters of propyl-water and pentyl-water are obtained by fitting to the osmotic coefficient data of [C$_3$MIM][Br] and [C$_5$MIM][Br] solutions at 298.15 K.[39] All of the parameters obtained are listed in Tables 5.1, 5.2, 5.3, and 5.4. The average absolute deviations (AAD) of the model from the fitted data are given in Table 5.5. The AAD in this work is defined as

$$AAD(\theta) = \frac{1}{N} \sum_{i=1}^{N} \left| \frac{\theta_i^{\text{calc}} - \theta_i^{\text{exp}}}{\theta_i^{\text{exp}}} \right| \tag{5.6}$$

where $\theta$ denotes the density or mean activity coefficient or osmotic coefficient, and $N$ is the number of data points.

Table 5.1. Parameters of cation head and anions (Br$^-$, BF$_4^-$)

| | $\nu^{oo}$ (cc/mol) | $u/k$ (K) | $\lambda$ | $d$ (Å) |
|---|---|---|---|---|
| IMI$^+$ | 22.28 | 987.7- 6.02×10$^{-4}$·T+ 5.23×10$^{-3}$·T$^2$ | 1.680 | 12.21 |
| Br$^-$ | 11.95 | 1094.45 | 1.542 | 7.74 |
| BF$_4^-$ | 16.26 | 221.96 | 1.668 | 15.59 |

Table 5.2. Cross association parameters between cation head and anions ($Cl^-$, $Br^-$, $BF_4^-$)

| | $IMI^+$ - $Cl^-$ | $IMI^+$ - $BF_4^-$ | $IMI^+$ - $Br^-$ |
|---|---|---|---|
| $\varepsilon/k$ (K) | 2104.02 | 887.79 | 1734.09 |
| $\kappa$ (cc/mol) | 0.0033 | 0.01475 | 0.03824 |

Table 5.3. Parameters for ethyl

| $n$ | $MW$ (g/mol) | $v^{oo}$ (cc/mol) | $u/k$ (K) | $\lambda$ |
|---|---|---|---|---|
| 2 | 29.062 | 22.73 | 325.69 | 1.961 |

Table 5.4. Binary interaction parameters

| | $IMI^+$-$H_2O$ | methyl-$H_2O$ | ethyl-$H_2O$ | propyl-$H_2O$ | butyl-$H_2O$ | pentyl-$H_2O$ | hexyl-$H_2O$ |
|---|---|---|---|---|---|---|---|
| $k_{ij}$ | 0.296 | 0 | 0 | 0.0778 | −0.0084 | −0.027 | 0 |

As shown in Tables 5.1 to 5.4, 7 parameters are assigned to the cation head of imidazolium ionic liquid; each anion has 4 parameters, and 2 parameters are used to describe the association between cation head and anion. For each alkyl group, 4 parameters are needed including 1 binary interaction parameter between alkyl group and water.

It is worthy to mention that although the number of parameters seems to be excessive, these model parameters are transferable. The number of parameters to describe a group of ILs is in fact smaller than that of homosegmented models. The same for the binary interaction parameters in this work, they are transferable to any IL having alkyl, cation head, and anion that have been considered.

### 5.3.2 Liquid Density and Activity Coefficient

The liquid density of aqueous [C$_2$MIM][Br] solution fitted by the heterosegmented SAFT EOS is shown in Figure 5.2. The concentration and temperature effects are well captured with an AAD of 0.397%. Similar calculation results of density are also found for [C$_4$MIM][Cl], [C$_4$MIM][BF4], and [C$_4$MIM][Br] solutions. It is worthy to mention that the water parameters are taken from our previous work[40] without any modification, and with these water parameters the model slightly overestimates the pure water density when the temperature is lower than 323.15 K. Therefore, the inaccuracy of the density calculation of the aqueous solution is primarily attributed to that of pure water.

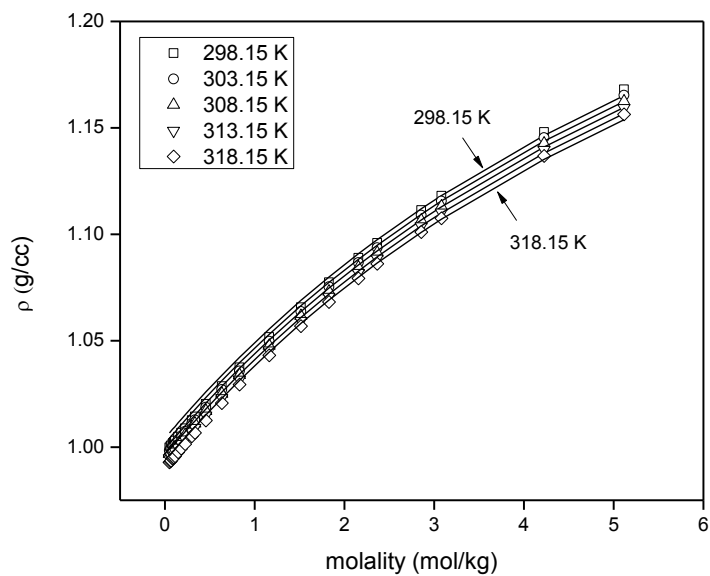

Figure 5.2. Density of aqueous [C$_2$MIM][Br] solution, experimental (points)[33] and calculated (curves).

Table 5.5. Systems studied and average absolute deviations

|  | T (K) | Max molality (mol/kg) | AAD($\rho$)% | AAD($\gamma_\pm$)% | AAD($\phi$)% |
|---|---|---|---|---|---|
| [C$_4$MIM][Cl] | 298, 333 | 2.0 | 0.725 | 0.948 | 0.63[a] |
| [C$_4$MIM][BF$_4$] | 318 | 1.02 | 0.886 | 2.321 | 11.6[a] |
| [C$_2$MIM][Br] | 298.15 | 0.44 | 0.397 | 13.3 | 6.41[a] |
| [C$_3$MIM][Br] | 298-318 | 2.70 | - | - | 1.36 |
| [C$_4$MIM][Br] | 298-318 | 2.02 | 0.655 | 0.817 | 0.93[a] |
| [C$_5$MIM][Br] | 298-318 | 2.00 | - | - | 1.45 |
| [C$_6$MIM][Br] | 298-318 | 2.39 | - | - | 6.65[a] |

[a] Systems that are predicted

The activity coefficients of [C$_4$MIM][Cl] and [C$_4$MIM][BF$_4$] solutions fitted by our model are shown in Figure 5.3. As mentioned earlier, the activity coefficient data used here are obtained by using correlations provided in the literature, which were constructed by fitting to the experimental osmotic coefficient data.[36,37] As shown in Figure 5.3, the activity coefficients are well represented except that of [C$_4$MIM][BF$_4$] at low concentrations. The discrepancy between our calculation and activity coefficient data is arguably due to the inaccuracy of the correlation.

Similar good model representation of activity coefficient can also be found for aqueous [C$_4$MIM][Br] solution, as shown in Figure 5.4. However, Figure 5.4 also shows that for aqueous [C$_2$MIM][Br] solution, despite the correct trend, the activity coefficient calculation is not satisfactory. The unsatisfactory representation of activity coefficient data usually indicates that the ionic term describing the Coulomb interactions might need further consideration. In the current model, the cation head is assumed to be a spherical segment and the effect of neutral segments on the cation-anion interaction is neglected in the whole concentration range of charged segments. Thus, one could attempt to relax these assumptions and obtain a more accurate approximation. This will be addressed in our future works.

In a different perspective, in fact one could obtain a good representation of [C$_2$MIM][Br] activity coefficient with an AAD of 0.91% by allowing the association strength between the cation head of $[C_2MIM]^+$ and $Br^-$ to be a non-transferable parameter that is exclusive for $[C_2MIM][Br]$, as indicated by the dashed line in Figure 5.4. Nevertheless, we will not follow this route because having a set of transferable parameters is more desirable. Moreover, although the activity coefficient calculation of this IL is not satisfactory, as shown later, our current model gives a reasonable prediction for the thermodynamic inhibition effect of $[C_2MIM][Br]$ on methane hydrate.

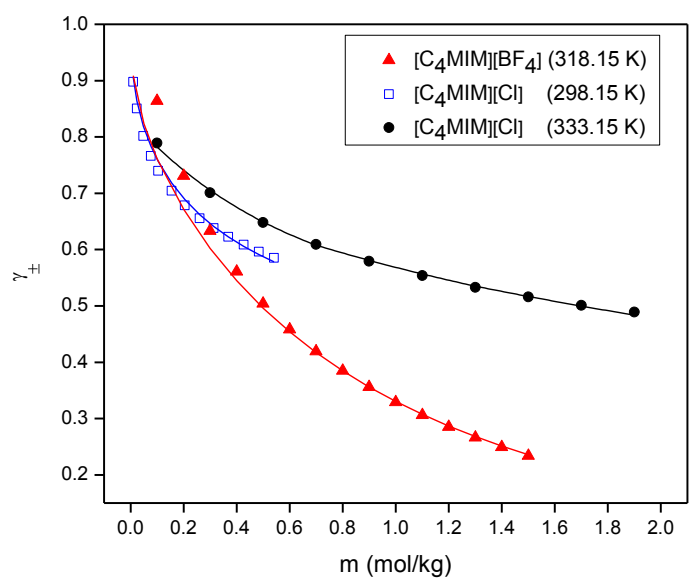

Figure 5.3. Mean activity coefficient of $[C_4MIM][Cl]$ solution at 298.15 K and 333.15 K, and $[C_4MIM][BF_4]$ solution at 318.15 K, experimental (points),[36,37] calculated (curves).

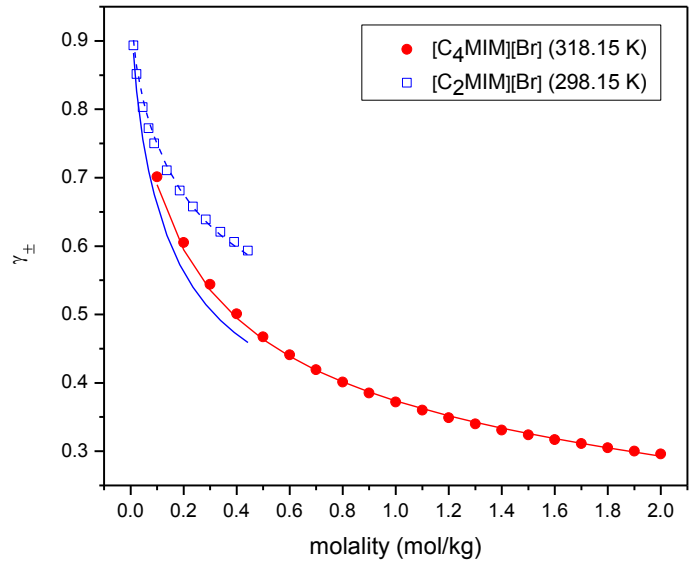

Figure 5.4. Activity coefficients of [C$_2$MIM][Br] solution at 298.15 K, and [C$_4$MIM][Br] solution at 318.15 K, experimental (points),[35,37] calculated (curves).

### 5.3.3 Osmotic Coefficient

Figure 5.5 shows the osmotic coefficients of aqueous [C$_4$MIM][Cl] and [C$_4$MIM][BF$_4$] solutions predicted by the proposed model. The prediction of the osmotic coefficients of [C$_4$MIM][Cl] solutions at two different temperatures is satisfactory. The discrepancy between the model prediction and experimental data for [C$_4$MIM][BF$_4$] is believed to be due to the inaccurate representation of its activity coefficient at low concentration, as shown in Figure 5.3.

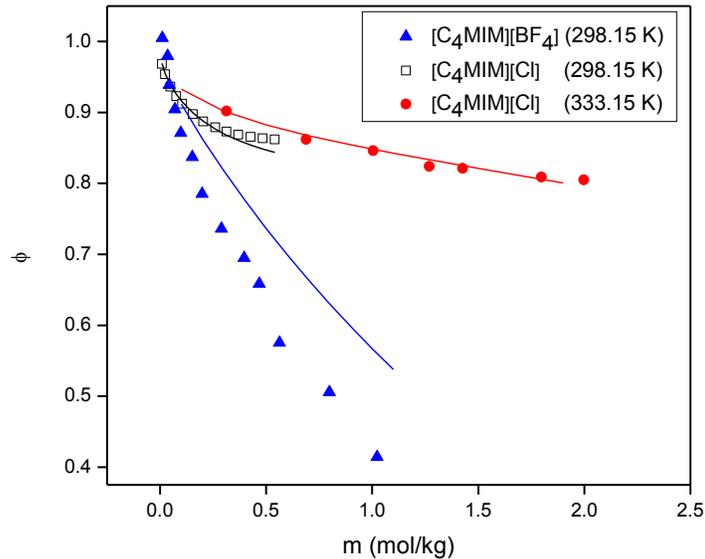

Figure 5.5. Osmotic coefficients of [$C_4$MIM][Cl] solution at 298.15 K and 333.15 K , and [$C_4$MIM][BF$_4$] solution at 318.15 K, experimental (points),[36,37] predicted (curves).

Figure 5.6 shows the osmotic coefficients of [$C_x$MIM][Br] ($3 \leq x \leq 6$) solutions predicted or calculated by our model. The osmotic coefficients of [$C_4$MIM][Br] at 318.15 K are well predicted using parameters fitted to the corresponding activity coefficient data. For [$C_6$MIM][Br], reasonable model prediction of osmotic coefficient can be found without using $k_{ij}$ between hexyl and water. Figure 5.7 shows the osmotic coefficients of [$C_5$MIM][Br] at different temperatures. By using the $k_{ij}$ between pentyl and water obtained at 298.15 K, the model prediction well captures the temperature effect on the osmotic coefficient of this IL. By using the $k_{ij}$ between propyl and water obtained at 298.15 K, similar results can also be found for the osmotic coefficients of [$C_3$MIM][Br] at different temperatures.

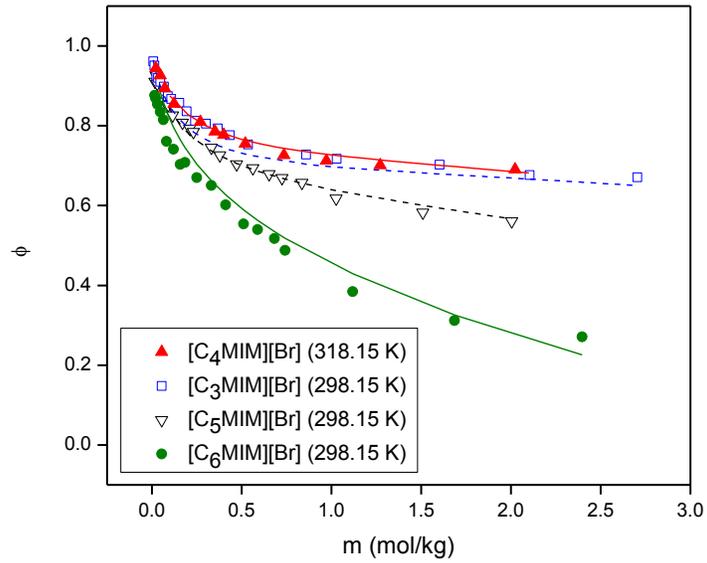

Figure 5.6. Osmotic coefficients of [C$_x$MIM][Br] ($3 \leq x \leq 6$) solutions at 298.15 K and 318.15 K, experimental (points),[37,39] predicted (solid lines) and calculated (dashed lines).

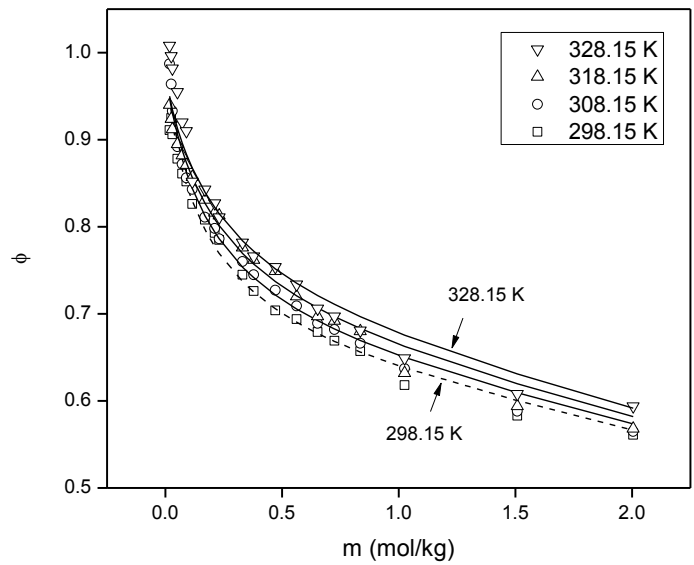

Figure 5.7. Osmotic coefficients of [C$_5$MIM][Br] solution at 298.15 K, 308.15 K, 318.15 K, and 328.15 K, experimental (points),[39] predicted (solid lines) and calculated (dashed line).

### 5.3.4 Methane Hydrate Dissociation Conditions in the Presence of Ionic Liquid Inhibitors

Methane hydrate dissociation conditions in the presence of IL inhibitors are predicted by the heterosegmented SAFT EOS coupled with the van der Waals and Platteeuw theory.[23] At first, we predict the methane hydrate dissociation conditions in the presence of ILs that are used in the parameter estimation, i.e., [C$_4$MIM][BF$_4$], [C$_2$MIM][Br], and [C$_4$MIM][Cl], the data of which are available; as described in the parameter estimation section, the parameters are fitted only to the density and activity coefficient data. Then, we proceed to predict the methane hydrate dissociation conditions in the presence of other ILs that are not used in the parameter estimation, i.e., [MIM][Cl] and [C$_2$MIM][Cl], the data of which are also available. The parameters of the constituents of these ILs are just transferred from other systems to check their transferability.

Figures 5.8 and 5.9 show the methane hydrate dissociation conditions in the presence of 10 wt% [C$_4$MIM][BF$_4$] and 10 wt% [C$_2$MIM][Br], respectively. As seen in these figures, the model slightly overestimates the hydrate dissociation conditions for [C$_4$MIM][BF$_4$], but well predicts the hydrate dissociation conditions for [C$_2$MIM][Br]. In Figure 5.9, the dissociation conditions in the presence of 10 wt% [C$_2$MIM][Cl] is also included to observe the effect of the anion type. In agreement with the experimental data, the model predicts that an imidazolium IL with chloride anion is more effective than that with bromide anion.

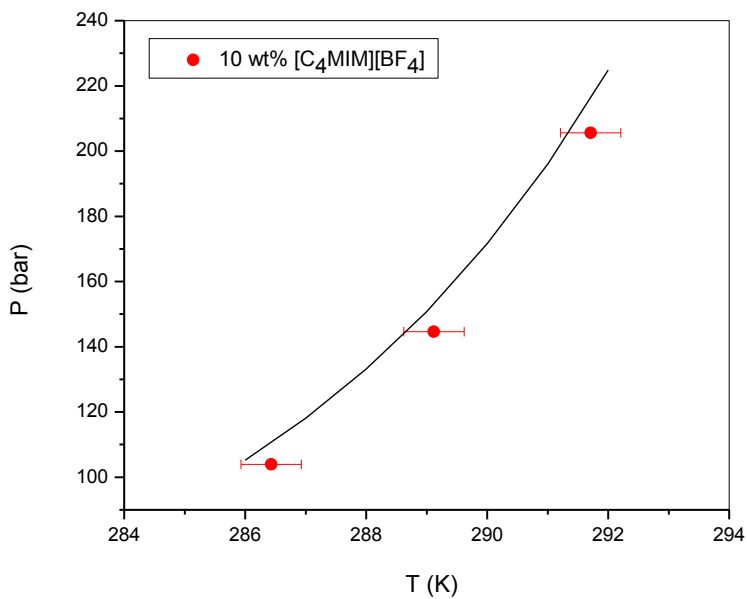

Figure 5.8. Methane hydrate dissociation conditions in the presence of 10 wt% [C$_4$MIM][BF$_4$], experimental (points)[2] and predicted (curves).

Figure 5.10 shows the effect of the alkyl length in the cation on the effectiveness of imidazolium IL inhibitors. Although the cation head of [MIM][Cl] is in fact different from that of the other ILs because it has only one branch, in this work its parameters are considered the same as those of cation head with two branches. As demonstrated in Figure 5.10, although the model representation is not perfect, it captures the effect of the alkyl length on the methane hydrate dissociation conditions. As the alkyl length of the cation increases, the effectiveness of the IL in inhibiting hydrate formation decreases. Note that at this concentration, the effectiveness of [MIM][Cl] is very similar to that of ethylene glycol, which is the conventional inhibitor commonly used.

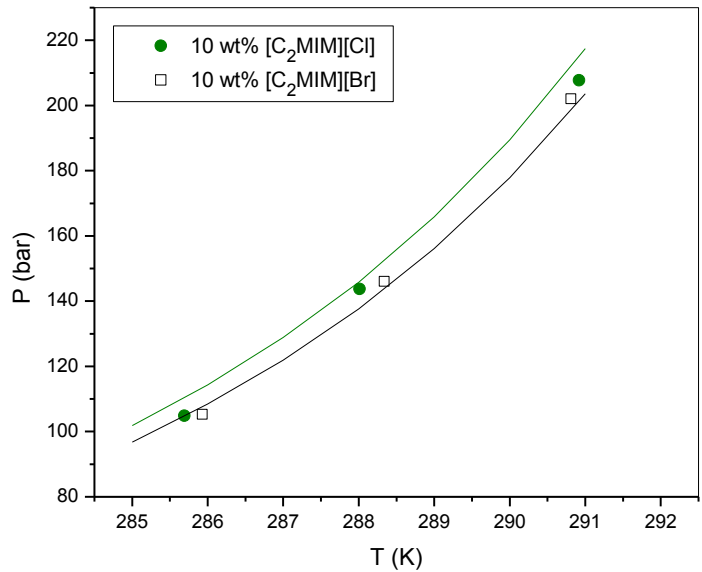

Figure 5.9. The effect of anion type on the effectiveness of imidazolium IL inhibitors, experimental (points)[2] and predicted (curves).

Finally, Figure 5.11 shows the methane hydrate dissociation conditions in the presence of [$C_2$MIM][Cl] at different concentrations. The concentration effect of [$C_2$MIM][Cl] on methane hydrate dissociation conditions is well captured. For the record, the methane hydrate dissociation pressure at 40 wt% (4.55 mol/kg water) of [$C_2$MIM][Cl] is underestimated by the proposed model (not shown). This could be due to the fact that the structural properties of imidazolium IL/water mixture will change as the concentration of water becomes smaller than some critical water concentration, as discussed by Niazi et al.[31] The number of water clusters will start dropping at concentrations below this critical water concentration, and thus a complete ion solvation cannot be assumed anymore. Since the number of water clusters drops, which makes methane hydrate more difficult to form, it is understandable that the effectiveness of the IL in inhibiting methane hydrate is greater than that predicted by the model.

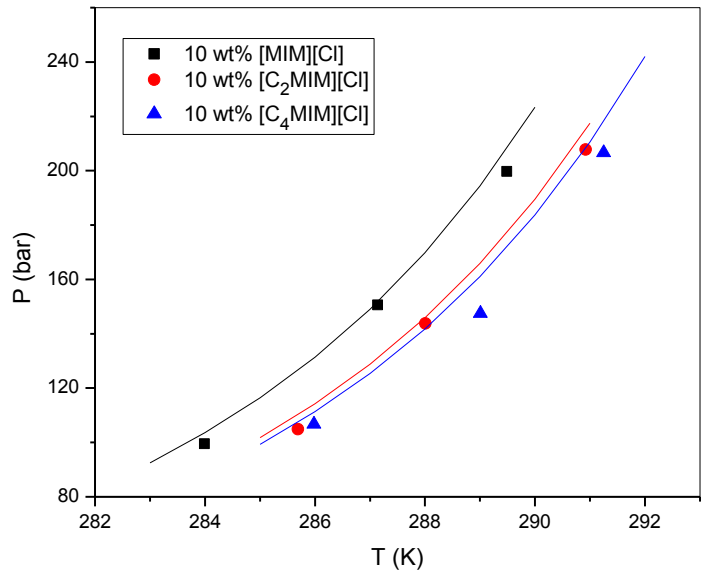

Figure 5.10. The effect of alkyl length in the cation of imidazolium IL on the effectiveness of imidazolium IL inhibitors, experimental (points)[2,41] and predicted (curves).

As seen in Figures 5.8, 5.9, 5.10, and 5.11, the model well captures the roles of pressure, anion type, alkyl length of the cation, and IL concentration on the hydrate inhibition performance of imidazolium ILs. Therefore, the parameters are indeed transferable, which demonstrates the advantage of this heterosegmented model of ionic liquids over homosegmented models.

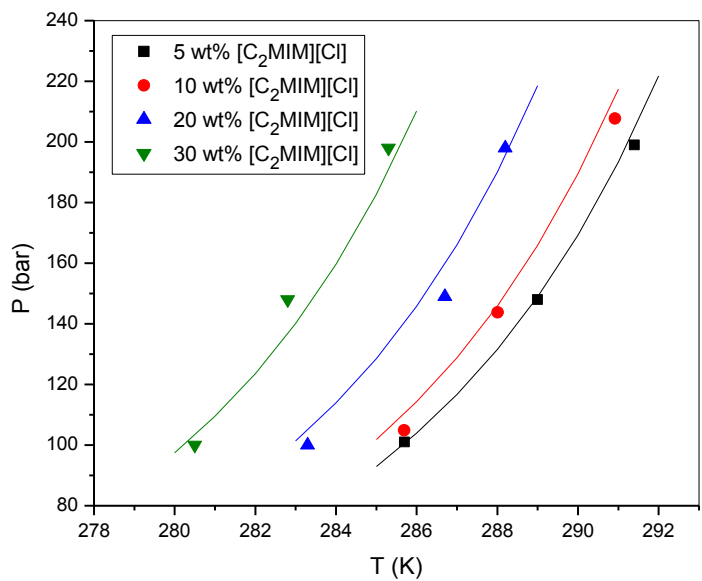

Figure 5.11. Methane hydrate dissociation conditions in the presence of [C₂MIM][Cl] at different concentrations, experimental (points)[2,3] and predicted (curves).

## 5.4 Conclusions

Heterosegmented statistical associating fluid theory (SAFT) is developed to model the thermodynamic properties of aqueous IL solutions containing [C₄MIM][Cl], [C₄MIM][BF₄], or [CₓMIM][Br] ($2 \leq x \leq 6$). The cation of IL is treated as a chain with a spherical segment representing the cation head and different groups of segments representing the alkyls. The positive charge is assumed to be located on the head segment of the cation chain. The anion of the ionic liquid is modeled as a spherical segment with a negative charge.

Since the systems investigated in this work contain a large amount of water, the model is developed to account for the long-range Coulomb interaction as well as the other interactions. We assume to a first approximation that RPM of MSA can be used to describe the electrostatic interaction between cation and anion in the presence of neutral segments. Following our previous works, the water-anion and water-cation hydrogen-bond interactions are not explicitly accounted for. The association considers only water-water and cation-anion interactions. One association site each is assigned to the cation head and anion, which can only cross associate.

A set of transferable parameters is obtained. The density, activity coefficients, and osmotic coefficients of aqueous IL solutions are in general well correlated and predicted by the proposed model. With this set of model parameters, we predict and study the methane hydrate dissociation conditions in the presence of [C$_4$MIM][BF$_4$], [C$_2$MIM][Br], and [C$_4$MIM][Cl], the density and activity coefficient data of which are used in the parameter estimation, and in the presence of [MIM][Cl] and [C$_2$MIM][Cl], the property data of which are not used in the parameter estimation. The model well captures the roles of pressure, anion type, alkyl length of the cation, and IL concentration on the hydrate inhibition performance of the imidazolium ILs. The transferable nature of the model parameters emphasizes the advantage of this heterosegmented model of ionic liquids over homosegmented models.

# Chapter 6. Study of Thermodynamic Properties of Symmetric and Asymmetric Electrolyte Systems in Mixture with Neutral Components: Monte Carlo Simulation Results and Integral Equation Predictions

## 6.1 Introduction

Understanding the properties of charged system is extremely important for the modeling of both chemical and biological solutions, such as those containing brine, ionic liquid, colloid, and protein. Theories have therefore been proposed to describe the thermodynamic properties of such systems.[1-6] Most of the theories use primitive model based on the McMillan-Mayer scheme, in which the solvent is represented by a continuous medium of uniform dielectric constant.

Based on this primitive model, the thermodynamic properties of charged systems, including internal energy, Helmholtz energy, osmotic coefficient, activity coefficient, specific heat, and radial distribution function have also been extensively investigated using Monte Carlo (MC) simulations.[7-17] In addition to the thermodynamic properties, the studies of phase behavior and critical parameters of charged systems using simulations are also available. For example, the vapor-liquid phase transition for ionic fluid was studied by Orkoulas and Panagiotopoulos using grand canonical and Gibbs ensemble MC simulations,[18] and the critical parameters of ionic fluid were reported by Panagiotopoulos.[19] In these works, ionic fluid was represented as charged hard spheres in the restricted primitive model, in which the hard spheres were assumed to be of the same size. Cheong and Panagiotopoulos later investigated the critical behavior of charge- and size-asymmetric primitive model electrolytes using grand canonical MC simulations.[20] Recently,

Lamperski et al. studied the fluid-solid phase transition of charged hard spheres of equal size using canonical MC simulation.[21] The fluid-solid phase transition, described by fluid- and solid-phase density limits, was estimated by investigating the change of compressibility factor with density.

There have also been a lot of interests in studying systems containing both charged and neutral species. By using primitive model, Outhwaite et al. conducted a series of MC studies on the structure and thermodynamic properties of charged and neutral hard-sphere mixtures,[22,23,24] in which the effects of neutral species on the structure, internal energy, and osmotic coefficient of charged system with ion size and charge asymmetry were investigated. The structure and thermodynamic properties calculated from the modified Poisson-Boltzmann (MPB) equation, hypernetted chain (HNC), and mean spherical approximation (MSA) were also compared with simulation results. Li and Wu[25] conducted a density functional theory (DFT) study on the thermodynamic properties of a highly asymmetric charged and neutral hard-sphere mixture. In Wu's work, DFT was proved to be accurate for such a mixture. In these previous works, the concentration of the neutral component was much higher than that of the charged component, which is common in most biological systems, such as DNA-binding protein.[26]

The knowledge of the structure and thermodynamic properties of charged and neutral hard-sphere mixtures, in which the concentrations of charged and neutral components are comparable, is also important, as we alluded in our recent work on ionic liquid modelling.[27] Despite their importance, however, systems containing charged and neutral components of comparable concentrations have never been studied.

Therefore, it is the purpose of this work to provide such knowledge by conducting canonical MC simulations, from which the structure, represented as radial distribution functions, and thermodynamic properties, including excess internal and Helmholtz energy, are obtained. For the record, to the best of our knowledge, Helmholtz energy, which is crucial for developing thermodynamic models, is reported for the first time for systems containing charged and neutral hard spheres using MC simulation.

Furthermore, in this work the Ornstein–Zernike (OZ) equation with HNC closure is also solved to obtain the structure and thermodynamic properties of systems containing charged and neutral components of comparable concentrations, which are then compared with the simulation results. Since our other goal is to make the results of MSA agree with those of MC simulation, i.e., to make the MSA representation of the excess internal and Helmholtz energy of charged hard-sphere fluid with or without neutral component accurate at high reduced density, an empirical modification of MSA formalism is then proposed.

## 6.2 Simulation and Theory

### 6.2.1 Monte Carlo Simulation

For a system containing charged and neutral hard spheres, the pair potential between particles $i$ and $j$ is given by,

$$u_{ij} = \begin{cases} \dfrac{z_i z_j e^2}{4\pi\varepsilon_0\varepsilon r} & r > \sigma_{ij} \\ \infty & r \leq \sigma_{ij} \end{cases} \tag{6.1}$$

where $r$ is the center-to-center distance between two particles, $z_i$ is the valence of the charged particle $i$, which is set to zero for neutral particles, $e$ is the elementary charge, $\varepsilon_0$ is the vacuum permittivity, $\varepsilon$ is the relative permittivity of the solvent, $\sigma_{ij}$ is defined as $\dfrac{\sigma_i + \sigma_j}{2}$, and $\sigma_i$ is the diameter of particle $i$. The relative permittivity $\varepsilon$ is set to 79.13, which mimics a water-like solution. There are two independent parameters of the studied system, i.e., the reduced temperature $T^* = \dfrac{4\pi\varepsilon\varepsilon_0\sigma k_B T}{e^2}$ and the packing fraction $\eta = \sum_i \dfrac{\pi}{6}\rho_i\sigma_i^3$ or the reduced density $\rho^* = \sum_i \rho_i\sigma_i^3$, where $\rho_i$ is the number density of particle $i$, $T$ is the temperature, and $k_B$ is the Boltzmann constant. Note that although the relative permittivity $\varepsilon$ is set to a certain value, using a different value of $T^*$ in the simulation in fact could also be equivalent to assigning a different value of $\varepsilon$.

MC simulation is performed in a cubic box using the Metropolis algorithm.[28] The Ewald summation technique[29] with periodic boundary condition is used to handle the long-range Coulomb interaction. Simulations to determine the excess internal energy, Helmholtz energy, and radial distribution function of fluids containing charged and neutral hard spheres for restricted and unrestricted primitive models with or without charge asymmetry are performed. The multistage Bennett's acceptance ratio method[30] is used to estimate the excess Helmholtz energy. An intermediate ensemble is created to bridge the energy gap between the system of interest, which is the charged and neutral hard-sphere mixture, and the reference system, which is the uncharged hard-sphere system. The potential of an intermediate ensemble ($u_i$) used in this work is defined in terms of the reference potential ($u_{hs}$) and the potential of the system of interest ($u$) as follows,

$$u_i = u_{hs} + \lambda \cdot (u - u_{hs}) \tag{6.2}$$

where $\lambda$ is the coupling parameter. Unless otherwise stated, one intermediate ensemble with $\lambda = 0.5$ is used.

In this work, we consider cases where the concentrations of charged and neutral hard spheres have the same order of magnitude, i.e., at $N_e/N_k = 1/3$, 1, and 3, which could be useful for the modelling of ionic liquids. Simulations with different total numbers of particles ($N$ = 384, 512, and 864) are first performed to investigate the effect of $N$ on the thermodynamic properties. In general, 5 million trial moves (translational moves) are performed to reach equilibration and 2.5 million trial moves (translational moves) are conducted to obtain the ensemble averages of the excess energy. The radial distribution functions are obtained after a simulation is completed by sorting the minimum image separations of pairs of particles.

### 6.2.2 OZ Equation with HNC Closure

For multi-component system, the OZ equation is given by,

$$\gamma_{\lambda\mu}(r) \equiv h_{\lambda\mu}(r) - c_{\lambda\mu}(r) = \sum_{\nu} \rho_{\nu} \int c_{\lambda\nu}\left(\left|r - r'\right|\right) \cdot h_{\nu\mu}\left(r'\right) \cdot dr' \tag{6.3}$$

where $\lambda$, $\mu$, and $\nu$ are the labels of components. The relation between the total correlation function $h(r)$ and the direct correlation function $c(r)$ is given by the HNC closure,

$$c_{\lambda\mu}(r) = \exp\left(-\beta \cdot u_{\lambda\mu}(r) + \gamma_{\lambda\mu}(r)\right) - \gamma_{\lambda\mu}(r) - 1 \tag{6.4}$$

where $\beta = 1/(k_B T)$. There are six distinct total correlation functions and direct correlation functions for a mixture of charged and neutral hard spheres.

In this work, the OZ equation with HNC closure is solved using Gillan's method.[31] A linear grid of 512 or 1024 points is used and the separation interval $\Delta r$ is carefully chosen to satisfy Stillinger-Lovett moment conditions.[32,33] With the knowledge of the radial distribution function $g_{ij}(r)$ calculated from the OZ equation with HNC closure, the dimensionless excess internal energy $U^*$ can be calculated as follows,[34]

$$U^* = 2\pi\rho\beta \sum_{i}\sum_{j} x_i x_j \int_{0}^{\infty} u_{ij}(r_{ij}) g_{ij}(r_{ij}) r_{ij}^2 dr_{ij} \tag{6.5}$$

where $\rho$ is the total number density, $x_i$ is the mole fraction of component $i$, and $r_{ij}$ is the separation distance between particles $i$ and $j$. The integration in Eq. 6.5 is evaluated numerically, where the upper integration limit is set to a separation distance at which each of $g_{ij}$ has reached a value of 1. The dimensionless excess Helmholtz energy $A^*$ is then obtained by thermodynamic integration,[35]

$$A^* = \int_{T^*}^{\infty} \frac{U^*}{T^*} dT^* \qquad (6.6)$$

where the integration is also evaluated numerically. Using a trapezoidal rule with 70 points and an upper integration limit of 50 can provide the desired number of significant figures.

### 6.2.3 Modification of MSA Formalism

Since the analytical solution of the MSA formalism is available, MSA is widely used to describe the electrostatic interaction in statistical associating fluid theory.[36,37] Although the structural information obtained from MSA is known to be unrealistic, the representation of excess energies using MSA at low density is relatively accurate due to fortuitous cancellation of error. However, it is well-known that MSA cannot provide accurate descriptions of charged systems without neutral component in the high density range or in the low reduced temperature range. As shown later in this work, the MSA formalism for charged systems with neutral component is also inaccurate in the intermediate to high density range ($\eta > 0.10$).

It is widely accepted that an accurate representation of the excess Helmholtz energy is crucial to the thermodynamic modelling of solution containing charged species. While HNC might be a good choice even for systems containing both charged and neutral components, it cannot be solved analytically, which makes it difficult to use for modeling purposes.

Therefore, in this work we attempt to modify the MSA formalism and obtain an analytical expression for the excess Helmholtz energy of charged systems that is accurate in the high density range. In this modification, an empirical parameter $K$ is introduced to the MSA formalism to correct the excess internal energy

$$U^* = -\frac{e^2}{4\pi\varepsilon\varepsilon_0 k_B T \sum_i \rho_i}\left\{\Gamma\sum_i \frac{\rho_i z_i^{\,2}}{1+\Gamma K\sigma_i} + \frac{\pi}{2\Delta}\Omega Pn^2\right\} \qquad (6.7)$$

where $\Gamma$ is the shielding parameter. The calculations of the parameters $Pn$, $\Omega$, $\Gamma$, and $\Delta$ are the same as the original MSA formalism,

$$Pn = \frac{1}{\Omega}\sum_i \frac{\rho_i \sigma_i z_i}{1+\Gamma\sigma_i} \qquad (6.8)$$

$$\Omega = 1 + \frac{\pi}{2\Delta}\sum_i \frac{\rho_i \sigma_i^{\,3}}{1+\Gamma\sigma_i} \qquad (6.9)$$

$$2\Gamma = \left\{\frac{e^2}{\varepsilon\varepsilon_0 k_B T}\cdot\sum_i \rho_i\left[\frac{z_i - (\pi/2\Delta)\sigma_i^{\,2}Pn}{1+\Gamma\sigma_i}\right]^2\right\}^{0.5} \qquad (6.10)$$

$$\Delta = 1 - \eta \qquad (6.11)$$

The dimensionless excess Helmholtz energy is then calculated as follows:

$$A^* = U^* + \frac{\Gamma^3}{3\pi\sum_i \rho_i} \qquad (6.12)$$

In the temperature range studied, the parameter $K$ is only density dependent and obtained by fitting the values of $U^*$ and $A^*$ for a certain system. The details of the parameter fitting are given in Section 6.3.

## 6.3 Results and Discussions

We first try to reproduce the literature data before generating any new MC data. For a 1:1 type electrolyte at $\eta$=0.1498 and $T^*$=0.531, the excess internal energy we obtain is −0.839±0.003, while the value estimated by Larsen[9] was −0.839±0.01. For a 1:1 type electrolyte at $\eta$=0.095 and $T^*$=0.589, the excess Helmholtz energy we obtain is −0.5280±0.0016, which is also in excellent agreement with the result obtained by Valleau and Card,[8] −0.5236±0.0035. These agreements thus serve well as the validation of our simulation method.

### 6.3.1 Structure of Charged and Neutral Hard-Sphere Mixtures

The structure of a charged and neutral hard-sphere mixture, which is presented here in terms of two-body distribution function ($g(r)$), is obtained from MC simulations and compared with the results from the OZ equation with HNC closure. Figure 6.1 shows the radial distribution functions of like ($g_{++}$ or $g_{--}$) and unlike ($g_{+-}$) charged hard spheres, and charged−neutral ($g_{ek}$) hard spheres for a restricted primitive system with $N_e/N_k$=1.0, $z_1$=1, and $z_2$=−1 at $\eta$=0.3 and $T^*$=0.5944. In Figure 6.1, the effect of neutral component on the structure of charged system is well represented by the HNC closure. The distribution function of neutral−neutral hard spheres ($g_{kk}$) is very similar to $g_{ek}$ and thus is not shown.

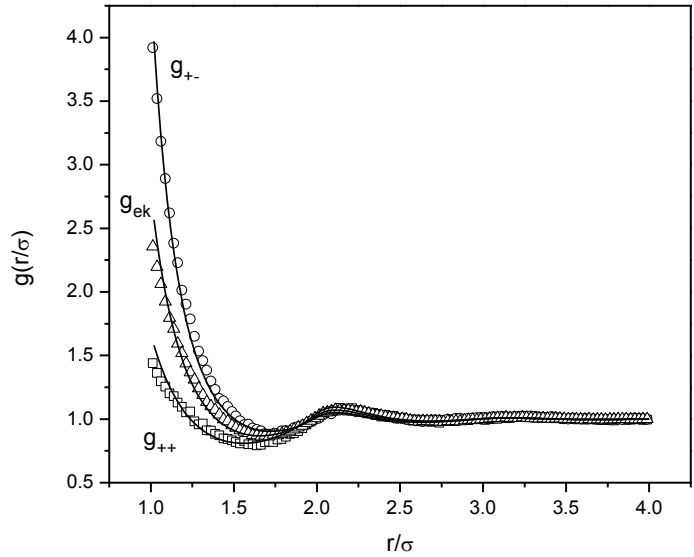

Figure 6.1. Like ($g_{++}$) and unlike ($g_{+-}$) charged hard-sphere, and charged–neutral hard-sphere ($g_{ek}$) distribution functions for an equal-sized mixture with $N_e/N_k$=1, $z_1$=1, and $z_2$=−1 at $\eta$=0.3 and $T^*$=0.5944; MC (symbols) and HNC (lines).

The distribution functions for mixtures of charged and neutral hard spheres of different size are also obtained by MC simulations. The positively and negatively charged hard spheres have the same size ($\sigma_e$), while the size of neutral hard sphere ($\sigma_k$) is 2/3 of $\sigma_e$. Figure 6.2 shows the like ($g_{++}$) and unlike ($g_{+-}$) charged hard-sphere distribution functions for this unequal-sized mixture with $N_e/N_k$=1, $z_1$=1, and $z_2$=−1 at $\eta$=0.3 and $T^*$=0.5944. Figure 6.3 shows the distribution functions of neutral–neutral ($g_{kk}$) and neutral–charged ($g_{ek}$) hard spheres. As shown in Figures 6.2 and 6.3, the size effect on the structure of this system is also well described using the HNC closure.

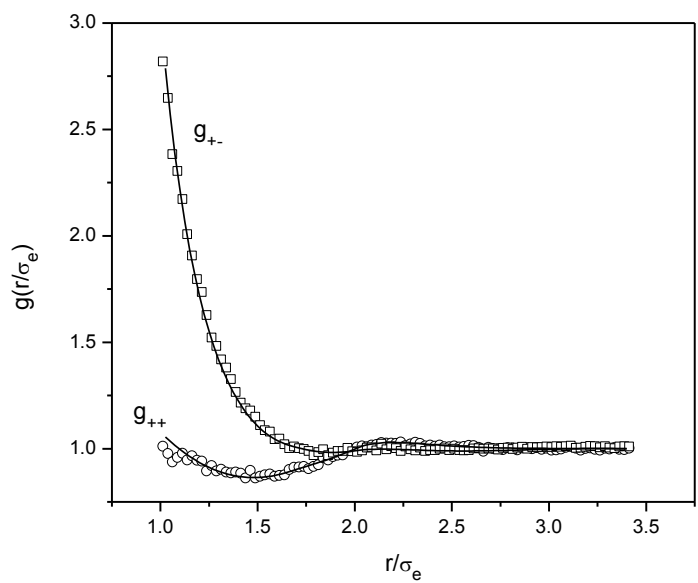

Figure 6.2. Like ($g_{++}$) and unlike ($g_{+-}$) charged hard-sphere distribution functions for an unequal-sized mixture with $N_e/N_k$= 1, $\sigma_e/\sigma_k$ =1:0.667, $z_1$=1, and $z_2$=−1 at $\eta$= 0.3 and $T^*$= 0.5944; MC (symbols) and HNC (lines).

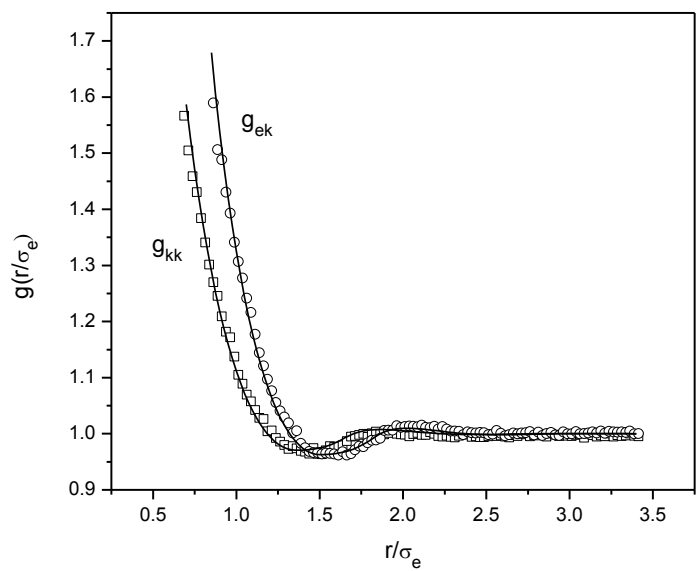

Figure 6.3. Neutral–neutral ($g_{kk}$) and neutral–charged ($g_{ek}$) hard-sphere distribution functions for an unequal-sized mixture with $N_e/N_k$= 1, $\sigma_e/\sigma_k$ =1:0.667, $z_1$=1, and $z_2$=−1 at $\eta$=0.3 and $T^*$= 0.5944; MC (symbols) and HNC (lines).

Figure 6.4 shows the distribution functions for equal-sized mixtures of charged and neutral hard spheres with $N_e/N_k$=1, $z_1$=2 and $z_2$=−1 at $\eta$=0.3 and $T^*$=0.5944. Due to the charge asymmetry, the distribution functions of like charged hard spheres ($g_{++}$ and $g_{--}$) are different. The distribution functions of positive-neutral, negative-neutral, and neutral-neutral hard spheres are very similar. As shown in Figure 6.4, HNC well represents the effect of charge asymmetry on the radial distribution functions of charged and neutral hard-sphere mixture.

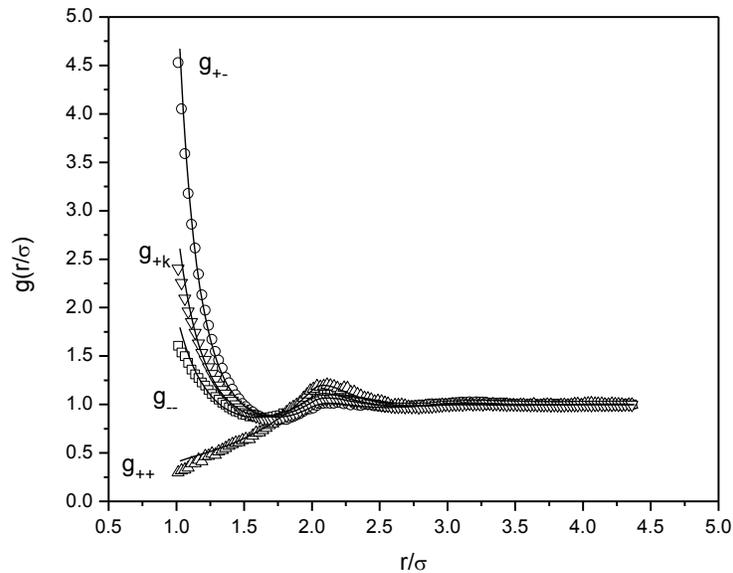

Figure 6.4. Charged-charged ($g_{++}$, $g_{--}$, $g_{+-}$) and charged-neutral ($g_{+k}$) hard-sphere distribution functions for an equal-sized mixture with $N_e/N_k$= 1, $\sigma_e/\sigma_k$ =1:1, $z_1$=2, and $z_2$ =−1 at $\eta$= 0.3 and $T^*$= 0.5944; MC (symbols) and HNC (lines).

Similar representations can also be found for different $N_e/N_k$ ratios. In general, for the structure information, HNC is in good agreement with MC simulations for all cases studied, i.e., equal- and unequal-sized, charge symmetric and asymmetric mixtures containing charged and neutral hard spheres at different concentration ratios ($N_e/N_k$). Moreover, HNC is not only able to represent the structures of systems at high concentration of the neutral component as concluded by Outhwaite and co-workers,[5] but also proved to be accurate for systems at high concentration of the charged component, as shown in this work.

## 6.3.2 Excess Internal and Helmholtz Energy

The values of $U^*$ and $A^*$ are obtained by simulations for mixtures containing charged and neutral hard spheres. The excess internal energy is calculated as the ensemble average during the production cycles, and the excess Helmholtz energy is obtained by the Bennett's Acceptance Ratio method.

It is found that the excess internal and Helmholtz energy are only weakly dependent on the total number of particles ($N$) used in simulation. For a mixture with a charged to neutral hard-sphere concentration ratio of 1/3, $z_1=1$, and $z_2 = -1$ at $\eta=0.4$ and $T^*=0.5944$, the values of $U^*$ obtained using $N=384$, 512, and 864 are $-0.1773\pm0.0013$, $-0.1768\pm0.0011$, and $-0.1754\pm0.0010$, respectively, and the values of $A^*$ are $-0.1418\pm0.0025$, $-0.1403\pm0.0022$, and $-0.1463\pm0.0021$, respectively. Unless otherwise stated, the number of particles used for the rest of the simulations is 512. The complete MC results are given in the supplementary materials accompanied this paper.

The thermodynamic properties calculated using the HNC approximation and MSA are compared with the MC data. Since the results of HNC are found to be in good agreement with MC results, the excess internal and Helmholtz energy of HNC for an equal-sized, type 1:1 charged hard-sphere system without neutral component at $T^*=0.5944$, which corresponds to a water-like solution with an ion size of 4.25 nm at 298.15 K, are then used to fit the parameter $K$ (in Eq. 6.7) in the modified MSA approximation, hereafter referred to as KMSA approximation. Instead of fitting to MC data, we choose to use HNC results because they are easier to obtain in terms of computational time.

The universal parameter $K$ obtained is as follows

$$K = 1 - 0.5786 \cdot \eta + 0.4825 \cdot \eta^2 \qquad (6.13)$$

and the fitting result is shown in Figure 6.5. The MSA results agree well with those of HNC only at low density ($\eta<0.10$). As shown in Figure 6.5, the parameter $K$ introduced significantly improves the MSA descriptions at higher density. Note that since the absolute value of the excess internal energy of MSA is always less than that of MC/HNC, the value of $K$ is always less than 1 in the fluid density range.

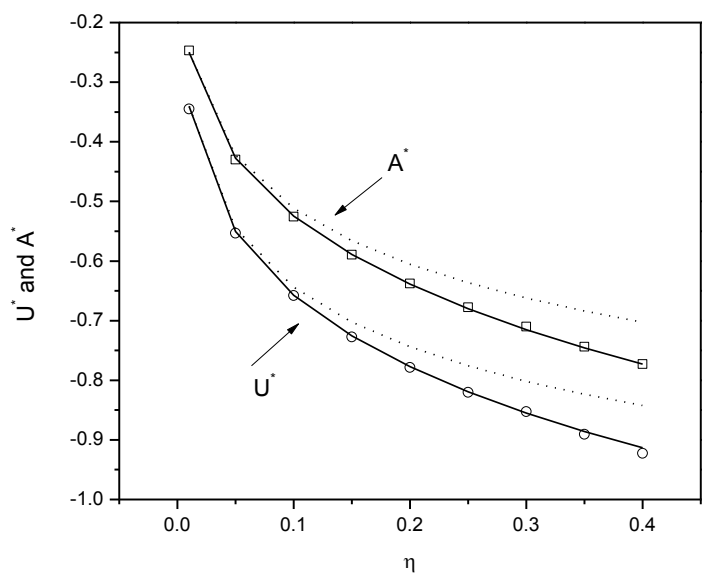

Figure 6.5. The values of $U^*$ and $A^*$ of HNC, MSA, and KMSA for 1:1 charged hard-sphere fluids of equal size at $T^*$=0.5944; HNC (symbols), KMSA (solid lines), and MSA (dotted lines).

The predictions of KMSA are then verified by comparing with those of MC, HNC, and MSA for charged and neutral hard-sphere fluid mixtures at different concentrations, ion sizes, and valences. The KMSA approximation is first tested to unrestricted primitive model, in which the $Pn$ term in Eq. 6.7 is not 0. Table 6.1 shows the values of $U^*$ and $A^*$ of KMSA, compared with those of MC, HNC, and MSA for systems of unequal-sized charged hard spheres without neutral component. It turns out that KMSA is also accurate for size asymmetric cases even at high fluid density.

Table 6.1. The values of $U^*$ and $A^*$ of MC, HNC, KMSA, and MSA for systems of 1:1 charged hard sphere of different size

| $\sigma_+/\sigma_-$ | $\eta$ | Excess internal energy ($U^*$) | | | | Excess Helmholtz energy ($A^*$) | | | |
|---|---|---|---|---|---|---|---|---|---|
| | | MC | HNC | KMSA | MSA | MC | HNC | KMSA | MSA |
| 1:1.5 | 0.2187 | -0.6013±0.0014 | -0.5991 | -0.6001 | -0.5746 | -0.4945±0.0017 | -0.4921 | -0.4900 | -0.4644 |
| 1:1.5 | 0.328 | -0.6685±0.0022 | -0.6778 | -0.6640 | -0.6229 | -0.5509±0.0015 | -0.5600 | -0.5514 | -0.5107 |
| 1:2 | 0.09 | -0.3675±0.0002 | -0.3676 | -0.3692 | -0.3640 | -0.2892±0.0018 | -0.2870 | -0.2877 | -0.2824 |
| 1:2 | 0.405 | -0.5653±0.0014 | -0.5700 | -0.5551 | -0.5194 | -0.4600±0.0017 | -0.4514 | -0.4610 | -0.4253 |

Figure 6.6 shows the values of $U^*$ and $A^*$ of MC, HNC, KMSA, and MSA for mixtures of equal-sized charged and neutral hard spheres with $N_e/N_k$=1, $z_1$=1, and $z_2$=−1 at $T^*$=0.5944. As shown in Figure 6.6, the thermodynamic properties calculated using HNC and KMSA are in good agreement with MC results in the studied range of reduced density, while those of MSA deviate from simulation results at high reduced density. Thus, we find that the empirical parameter $K$, which is fitted to the HNC results for charged systems without the presence of neutral component, can be used for mixtures containing both charged and neutral hard spheres. The details of MC results shown in Figure 6.6 can be found in Table 6.2.

Table 6.2. Excess internal ($U^*$) and Helmholtz ($A^*$) energy for mixtures with $N_e/N_k$ =1, $\sigma_e/\sigma_k$ =1, $z_1$=1, and $z_2$=−1 at $T^*$=0.5944

| η | Excess internal energy ($U^*$) | | | | Excess Helmholtz energy ($A^*$) | | | |
|---|---|---|---|---|---|---|---|---|
| | MC | HNC | KMSA | MSA | MC | HNC | KMSA | MSA |
| 0.01 | -0.1368±0.0004 | -0.1352 | -0.1338 | -0.1336 | -0.0995±0.0015 | -0.0977 | -0.0963 | -0.0962 |
| 0.1 | -0.2816±0.0009 | -0.2802 | -0.2776 | -0.2728 | -0.2210±0.0018 | -0.2162 | -0.2161 | -0.2110 |
| 0.2 | -0.3391±0.0010 | -0.3381 | -0.3343 | -0.3219 | -0.2676±0.0016 | -0.2682 | -0.2680 | -0.2557 |
| 0.3 | -0.3795±0.0014 | -0.3771 | -0.3731 | -0.3511 | -0.3061±0.0017 | -0.3039 | -0.3031 | -0.2830 |
| 0.4 | -0.4102±0.0015 | -0.4077 | -0.4001 | -0.3719 | -0.3361±0.0017 | -0.3320 | -0.3300 | -0.3027 |

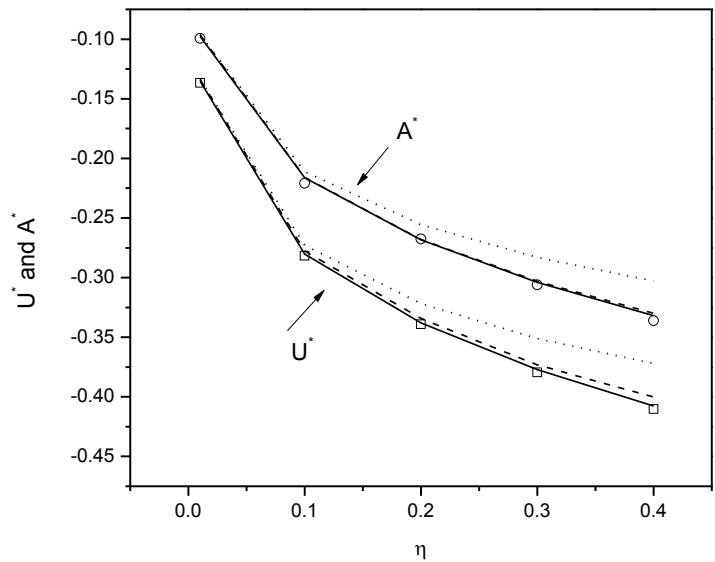

Figure 6.6. The values of $U^*$ and $A^*$ of MC, HNC, KMSA, and MSA for mixtures of equal-sized charged and neutral hard spheres with $N_e/N_k$=1, $z_1$=1, and $z_2$=−1 at $T^*$=0.5944; MC (symbols), HNC (dashed lines), KMSA (solid lines), and MSA (dotted lines).

Figure 6.7 shows the values of $A^*$ obtained from MC simulation and different approximations at different ratios of charged to neutral hard-sphere concentrations ($N_e/N_k$). The effect of adding neutral component on $A^*$ is well captured by HNC and KMSA in the studied density range. However, the deviation between the results of MSA and MC becomes larger as the number of charged hard spheres ($N_e$) increases.

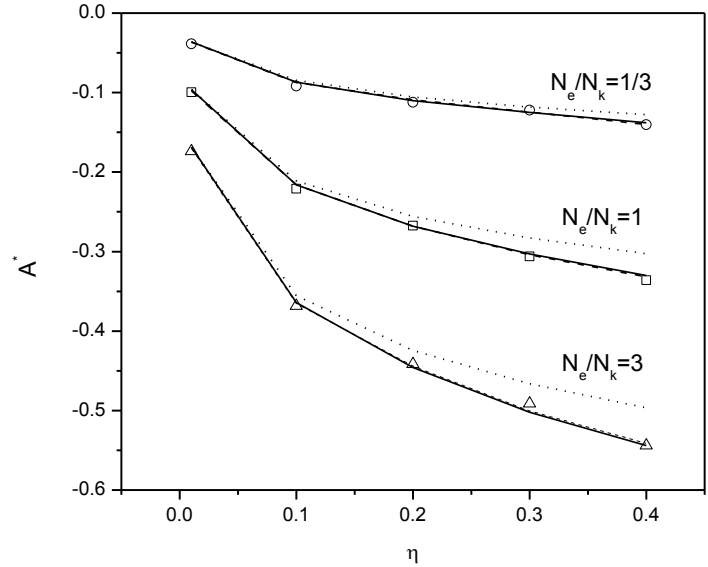

Figure 6.7. The values of $A^*$ of MC, HNC, KMSA, and MSA for mixtures of equal-sized charged and neutral hard spheres with $z_1$=1 and $z_2$=−1 at $T^*$=0.5944 and different $N_e/N_k$ ratios; MC (symbols), HNC (dashed lines), KMSA (solid lines), and MSA (dotted lines).

For the case when the charged and neutral hard spheres have different size ($\sigma_e/\sigma_k$=1:0.667, $N_e/N_k$=1, $T^*$=0.5944, $z_1$=1, and $z_2$=−1), the values of $U^*$ and $A^*$ are given in Figure 6.8, where the HNC approximation and KMSA well represent the MC data. This indicates that the density-dependent parameter $K$ is also applicable to size asymmetric cases.

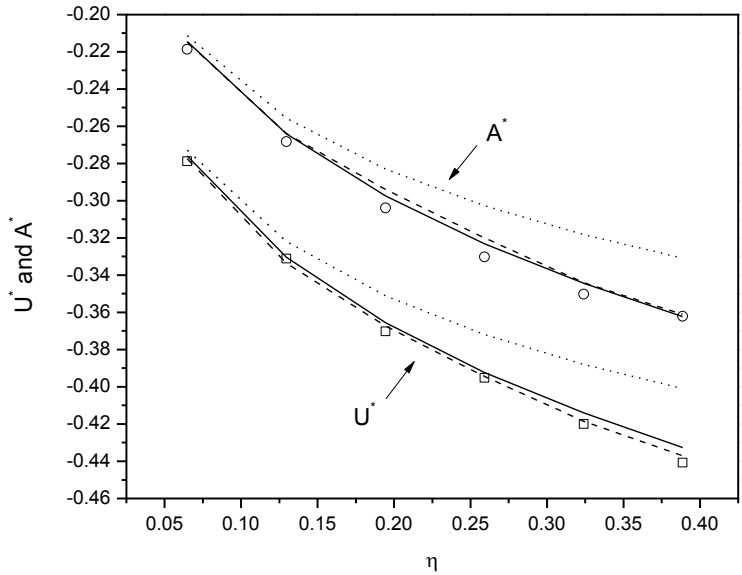

Figure 6.8. The values of $U^*$ and $A^*$ of MC, HNC, KMSA, and MSA for mixtures of unequal-sized charged and neutral hard spheres ($\sigma_e/\sigma_k$=1:0.667) with $N_e/N_k$ = 1.0, $z_1$=1 and $z_2$=−1 at $T^*$=0.5944; MC (symbols), HNC (dashed lines), KMSA (solid lines), and MSA (dotted lines).

Similar representations of excess energy using KMSA can also be found for charge asymmetric cases. Figure 6.9 gives the thermodynamic properties of mixtures of equal-sized charged and neutral hard spheres with $N_e/N_k$ = 1.0, $z_1$=2, and $z_2$=−1 at $T^*$=0.5944. The number of particles used in the simulation for this mixture is 384 due to the charge asymmetry. As shown in Figure 6.9, the KMSA approximation slightly underestimates the MC data of $U^*$ and $A^*$, but it still gives very satisfactory results.

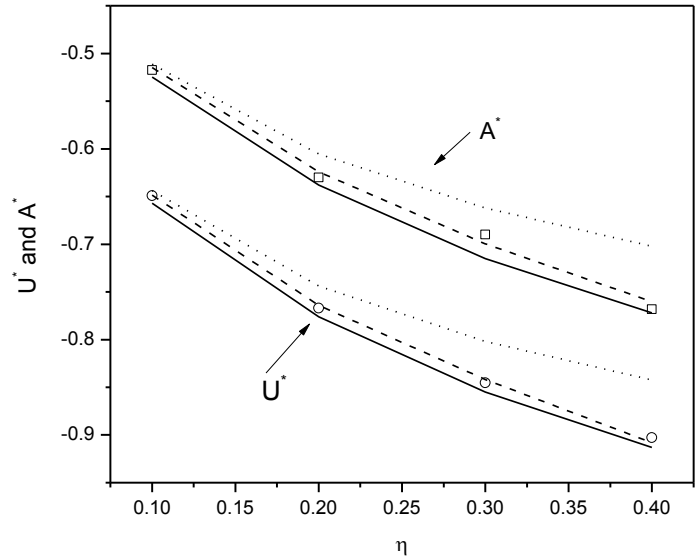

Figure 6.9. The values of $U^*$ and $A^*$ of MC, HNC, KMSA, and MSA for mixtures of equal-sized charged and neutral hard spheres with $N_e/N_k = 1.0$, $z_1=2$, and $z_2=-1$ at $T^*=0.5944$; MC (symbols), HNC (dashed lines), KMSA (solid lines), and MSA (dotted lines).

To test the accuracy of HNC, KMSA, and MSA in representing the excess energy for an electrolyte system having low permittivity, MC simulations at $T^*=0.2$ are conducted, which corresponds to an electrolyte system with an ion size of 6 nm and a relative permittivity of 18 at $T=298.15$ K. Three intermediate ensembles with $\lambda=0.25$, 0.5, and 0.75 are used to bridge the energy gap between the system of interest and the reference system. Figure 6.10 shows the values of $U^*$ and $A^*$ for mixtures of equal-sized charged and neutral hard spheres with $N_e/N_k=1$, $z_1=1$, and $z_2=-1$ at $T^*=0.2$. Note that the absolute values of $U^*$ and $A^*$ at $T^*=0.2$ are higher than those at $T^*=0.5944$ (see Figure 6.6) because of the stronger Coulomb interactions at low permittivity. Compared with the MC data, HNC and KMSA well represent the excess internal and Helmholtz energy of electrolyte system with small dielectric constant, while MSA significantly overestimates these excess properties. Although the empirical parameter $K$ (in Eq. 6.7) is obtained by fitting to the excess properties of a water-like solution, it can be applied to systems with low permittivity. Of course, the accurate representation of KMSA at low $T^*$ also means that KMSA can well capture the effect of temperature on the excess energy of charged systems in the presence of neutral component. The accurate excess energy representation of KMSA at $T^*$=0.6978 and 0.5445 are also reported in the supplementary materials accompanied this paper.

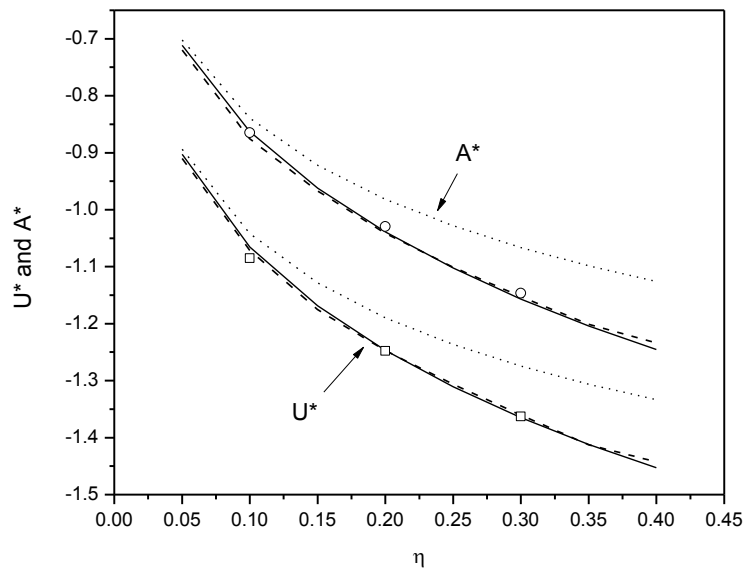

Figure 6.10. The values of $U^*$ and $A^*$ of MC, HNC, KMSA, and MSA for mixtures of equal-sized charged and neutral hard spheres with $N_e/N_k = 1.0$, $z_1$=1 and $z_2$=−1 at $T^*$=0.2; MC (symbols), HNC (dashed lines), KMSA (solid lines), and MSA (dotted lines).

## 6.4 Conclusions

The thermodynamic properties of symmetric and asymmetric electrolyte systems in mixture with neutral components are obtained by MC simulations at different reduced densities, temperatures, and concentration ratios of charged to neutral hard spheres. These MC results are also compared with the predictions obtained from integral equation using HNC and MSA closures.

For the radial distribution functions of charged and neutral hard-sphere mixtures, in general, the predictions from the OZ equation with HNC closure are in good agreement with MC results for all cases studied. The OZ equation with HNC closure is also proved to give satisfactory results for the excess internal and Helmholtz energy of all the systems studied in this work, while the MSA formalism is found to deviate from MC data in the intermediate to high density range. The representation of MSA also becomes worse as the concentration of charged particles increases or the reduced temperature decreases.

A simple empirical modification of MSA approximation, referred to as KMSA, is proposed. The modification is achieved by introducing a single parameter $K$, which is dependent only on density. Although this parameter $K$ is fitted only to the excess energy of HNC for an equal-sized, type 1:1 charged system without the presence of neutral component at one reduced temperature, KMSA is proved to capture the effects of neutral component, size and charge asymmetry, system temperature, and dielectric constant of the background solvent on the excess energy of electrolyte systems.

## Acknowledgement

Authors are grateful to Mr. Tao Li of the Department of Geology and Geophysics of the University of Wyoming for his useful advice on parallel computing.

# Supplementary Material

Table 6.3. Excess internal ($U^*$) and Helmholtz ($A^*$) energy for mixtures with $N_e/N_k$=1/3, $\sigma_e/\sigma_k$=1, $z_l$=1, and $z_2$=−1 at $T^*$=0.5944

| $\eta$ | Excess internal energy ($U^*$) | | | | Excess Helmholtz energy ($A^*$) | | | |
|---|---|---|---|---|---|---|---|---|
| | MC | HNC | KMSA | MSA | MC | HNC | KMSA | MSA |
| 0.01 | -0.0534±0.0003 | -0.0526 | -0.0515 | -0.0515 | -0.0386±0.0027 | -0.0367 | -0.0364 | -0.0364 |
| 0.1 | -0.1178±0.0004 | -0.1165 | -0.1146 | -0.1130 | -0.0918±0.0015 | -0.0873 | -0.0870 | -0.0845 |
| 0.2 | -0.1449±0.0008 | -0.1437 | -0.1410 | -0.1364 | -0.1122±0.0016 | -0.1090 | -0.1101 | -0.1056 |
| 0.3 | -0.1651±0.0012 | -0.1625 | -0.1600 | -0.1506 | -0.1221±0.0015 | -0.1247 | -0.1250 | -0.1184 |
| 0.4 | -0.1768±0.0011 | -0.1772 | -0.1711 | -0.1610 | -0.1403±0.0022 | -0.1400 | -0.1380 | -0.1279 |

Table 6.4. Excess internal ($U^*$) and Helmholtz ($A^*$) energy for mixtures with $N_e/N_k$=3, $\sigma_e/\sigma_k$=1, $z_l$=1, and $z_2$=−1 at $T^*$=0.5944

| $\eta$ | Excess internal energy ($U^*$) | | | | Excess Helmholtz energy ($A^*$) | | | |
|---|---|---|---|---|---|---|---|---|
| | MC | HNC | KMSA | MSA | MC | HNC | KMSA | MSA |
| 0.01 | -0.2362±0.0003 | -0.2349 | -0.2317 | -0.2314 | -0.1738±0.0017 | -0.1702 | -0.1687 | -0.1684 |
| 0.1 | -0.4645±0.0006 | -0.4634 | -0.4608 | -0.4520 | -0.3684±0.0016 | -0.3653 | -0.3641 | -0.3553 |
| 0.2 | -0.5537±0.0011 | -0.5523 | -0.5490 | -0.5267 | -0.4417±0.0015 | -0.4440 | -0.4460 | -0.4245 |
| 0.3 | -0.6120±0.0015 | -0.6110 | -0.6062 | -0.5705 | -0.4913±0.0015 | -0.5003 | -0.5020 | -0.4663 |
| 0.4 | -0.6603±0.0010 | -0.6519 | -0.6491 | -0.6014 | -0.5440±0.0015 | -0.5412 | -0.5442 | -0.4964 |

Table 6.5. Excess internal ($U^*$) and Helmholtz ($A^*$) energy for mixtures with $N_e/N_k$=1, $\sigma_e/\sigma_k$=1, $z_l$=1, and $z_2$=−1 at $T^*$=0.6978

| $\eta$ | Excess internal energy ($U^*$) | | | | Excess Helmholtz energy ($A^*$) | | | |
|---|---|---|---|---|---|---|---|---|
| | MC | HNC | KMSA | MSA | MC | HNC | KMSA | MSA |
| 0.01 | -0.1086±0.0002 | -0.1076 | -0.1075 | -0.1074 | -0.0809±0.0015 | -0.0773 | -0.0772 | -0.0769 |
| 0.1 | -0.2299±0.0005 | -0.2288 | -0.2267 | -0.2229 | -0.1800±0.0016 | -0.1755 | -0.1755 | -0.1717 |
| 0.2 | -0.2780±0.0007 | -0.2779 | -0.2743 | -0.2644 | -0.2132±0.0016 | -0.2156 | -0.2185 | -0.2088 |

| 0.3 | -0.3143±0.0011 | -0.3111 | -0.3053 | -0.2893 | -0.2486±0.0017 | -0.2480 | -0.2478 | -0.2318 |
| 0.4 | -0.3311±0.0013 | -0.3343 | -0.3288 | -0.3069 | -0.2663±0.0016 | -0.2685 | -0.2700 | -0.2485 |

Table 6.6. Excess internal ($U^*$) and Helmholtz ($A^*$) energy for mixtures with $N_e/N_k$ =1, $\sigma_e/\sigma_k$ =1, $z_1$=1, and $z_2$=−1 at $T^*$=0.5445

| η | Excess internal energy ($U^*$) | | | | Excess Helmholtz energy ($A^*$) | | | |
|---|---|---|---|---|---|---|---|---|
| | MC | HNC | KMSA | MSA | MC | HNC | KMSA | MSA |
| 0.01 | -0.1544±0.0002 | -0.1539 | -0.1507 | -0.1506 | -0.1038±0.0016 | -0.1098 | -0.1087 | -0.1086 |
| 0.1 | -0.3136±0.0005 | -0.3120 | -0.3100 | -0.3044 | -0.2486±0.0016 | -0.2420 | -0.2420 | -0.2366 |
| 0.2 | -0.3774±0.0007 | -0.3764 | -0.3724 | -0.3585 | -0.2923±0.0017 | -0.2966 | -0.2990 | -0.2855 |
| 0.3 | -0.4185±0.0014 | -0.4192 | -0.4130 | -0.3902 | -0.3366±0.0017 | -0.3374 | -0.3382 | -0.3154 |
| 0.4 | -0.4558±0.0012 | -0.4358 | -0.4436 | -0.4128 | -0.3766±0.0011 | -0.3675 | -0.3678 | -0.3371 |

Table 6.7. Excess internal ($U^*$) and Helmholtz ($A^*$) energy for mixtures with $N_e/N_k$ =1, $\sigma_e/\sigma_k$ =1:0.667, $z_1$=1, and $z_2$=−1 at $T^*$=0.5944

| η | Excess internal energy ($U^*$) | | | | Excess Helmholtz energy ($A^*$) | | | |
|---|---|---|---|---|---|---|---|---|
| | MC | HNC | KMSA | MSA | MC | HNC | KMSA | MSA |
| 0.0648 | -0.2788±0.0008 | -0.2774 | -0.2759 | -0.2728 | -0.2187±0.0018 | -0.2148 | -0.2145 | -0.2113 |
| 0.1296 | -0.3311±0.0007 | -0.3335 | -0.3304 | -0.3219 | -0.2682±0.0016 | -0.2639 | -0.2641 | -0.2557 |
| 0.1944 | -0.3703±0.0013 | -0.3674 | -0.3655 | -0.3511 | -0.3039±0.0017 | -0.2940 | -0.2974 | -0.2830 |
| 0.2592 | -0.3952±0.0015 | -0.3944 | -0.3923 | -0.3719 | -0.3303±0.0015 | -0.3200 | -0.3231 | -0.3027 |
| 0.3241 | -0.4200±0.0016 | -0.4185 | -0.4140 | -0.3879 | -0.3502±0.0015 | -0.3440 | -0.3443 | -0.3182 |
| 0.3888 | -0.4408±0.0023 | -0.4370 | -0.4326 | -0.4009 | -0.3621±0.0016 | -0.3610 | -0.3623 | -0.3309 |

Table 6.8. Excess internal ($U^*$) and Helmholtz ($A^*$) energy for mixtures with $N_e/N_k$ =1, $\sigma_e/\sigma_k$ =1.0, $z_1$=2, and $z_2$=−1 at $T^*$=0.5944

| η | Excess internal energy ($U^*$) | | | | Excess Helmholtz energy ($A^*$) | | | |
|---|---|---|---|---|---|---|---|---|
| | MC | HNC | KMSA | MSA | MC | HNC | KMSA | MSA |
| 0.1 | -0.6491±0.0008 | -0.6484 | -0.6570 | -0.6439 | -0.5175±0.0020 | -0.5149 | -0.5247 | -0.5114 |

| 0.2 | -0.76703±0.001 | -0.7641 | -0.7760 | -0.7438 | -0.6301±0.0017 | -0.6242 | -0.6280 | -0.6055 |
| 0.3 | -0.8453±0.0013 | -0.8418 | -0.8549 | -0.8018 | -0.6898±0.0015 | -0.6996 | -0.7100 | -0.6619 |
| 0.4 | -0.9027±0.0016 | -0.9080 | -0.9130 | -0.8424 | -0.7683±0.0017 | -0.7603 | -0.7720 | -0.7022 |

Table 6.9. Excess internal ($U^*$) and Helmholtz ($A^*$) energy for mixtures with $N_e/N_k$ =1, $\sigma_e/\sigma_k$ =1.0, $z_1$=1, and $z_2$=−1 at $T^*$=0.2

| η | Excess internal energy ($U^*$) | | | | Excess Helmholtz energy ($A^*$) | | | |
|---|---|---|---|---|---|---|---|---|
| | MC | HNC | KMSA | MSA | MC | HNC | KMSA | MSA |
| 0.1 | -1.0852±0.0014 | -1.0713 | -1.0653 | -1.0416 | -0.8650±0.0013 | -0.8757 | -0.8628 | -0.8392 |
| 0.2 | -1.2478±0.0017 | -1.2472 | -1.2468 | -1.1896 | -1.0295±0.0015 | -1.0411 | -1.0390 | -0.9818 |
| 0.3 | -1.3628±0.0039 | -1.3605 | -1.3650 | -1.2744 | -1.1467±0.0014 | -1.1413 | -1.1567 | -1.0662 |

# Chapter 7. Monte Carlo Simulation and Equation of State for Flexible Charged Hard-Sphere Chain Fluids: Polyampholyte and Polyelectrolyte Solutions

## 7.1 Introduction

Solutions containing charged chain molecules, such as polyelectrolyte and polyampholyte solutions, have a variety of applications in chemical and biological processes. They are used, for example, as dewatering agents, viscosity modifier, cosmetics additives, etc. Polyelectrolytes are polymers that form polyions (charged polymer molecules) in aqueous solutions, and polyampholyte are copolymers that have positively and negatively charged monomers interspersed within the same linear chain. The properties of polyelectrolytes and polyampholytes are dominated by the short-range excluded volume interactions of the polymer backbone and the long-range electrostatic interactions between charged monomers. The complex molecular structures and interactions present serious challenges to the modeling of polyelectrolyte and polyampholyte solutions.

Several models have been developed to investigate the thermodynamic properties of polyelectrolyte or polyampholyte solutions. In general, these models treat the polyions or polyampholyte as flexible or rod-like charged hard-sphere chains, while the solvent molecules are treated implicitly or explicitly. Das et al.[1] applied the modified Poisson-Boltzmann theory, which is based on a cell model, for polyelectrolyte solutions. Wang and Denton investigated the effect of electrostatic interactions on the structure of polyelectrolyte stars by density functional theory.[2] Integral equation based theories have also been developed to study the properties of polyelectrolyte. Solm and Chiew used the multi-density Ornstein-Zernike (OZ) equation within the mean spherical approximation (MSA) to study the osmotic pressure, internal energy, Helmholtz energy, and pair correlation functions of polyelectrolyte with counterions.[3,4] Kalyuzhnyi proposed an ideal chain polymer mean-spherical approximation theory for flexible hard-sphere chains with an arbitrary type of long-range site-site pair potential.[5] Kalyuzhnyi and Cummings then extended the theory to multicomponent mixture of charged hard-sphere chains with arbitrary distributions of monomer charge and size along the molecular backbone.[6] Shew and Yethiraj studied the osmotic pressure of charged hard-sphere chains with counterions using the polymer reference interaction site model (PRISM).[7] All of the charged hard-sphere chain models that are based on integral equations, except the PRISM theory by Shew and Yethiraj,[7] were inappropriately validated by the simulation data of charged chains with Lennard-Jones potential,[8] which makes the validation less convincing.

Besides the models based on integral equations mentioned above, the thermodynamic perturbation theory was also extended to polyelectrolyte and polyampholyte solutions. By using the thermodynamic perturbation theory of the first order, Jiang et al. studied the osmotic coefficients, activity coefficients, and phase equilibria of charged hard-sphere chains for polyelectrolyte[9,10] and polyampholyte solutions.[11] The model was expressed in terms of Helmholtz energy, which was divided into contributions from hard-sphere repulsion, electrostatic interaction, and chain connectivity. The hard-sphere contribution to Helmholtz energy was calculated using Boublik-Mansoori-Carnahan-Starling equation of state,[12] while the electrostatic interaction was accounted for using MSA in the primitive model. The contribution of chain connectivity to the Helmholtz energy was estimated by using the nearest-neighbor two-particle cavity correlation function of charged monomers derived from the combined hypernetted chain and mean spherical approximations (HNC/MSA) and by assuming a simple dimerization process from monomers. In this approach, the contribution to bond formation was independent of chain size, and the overall chain contribution was approximated by the contribution of one bond formation from two monomers multiplied by the total number of bonds in the chain. The approach can also be referred to as the monomer approach since the structure information needed in calculating the chain contribution is only that for monomer fluid. As we will show later that this monomer approach should not be used for the modeling of charged hard-sphere chains. The osmotic coefficient of polyelectrolyte predicted by this model[9,10] was also incorrectly compared to the simulation data of charged chain with Lennard-Jones potential, and the predicted vapor-liquid phase equilibria of polyampholyte[11] were not satisfactory when compared with simulation data.[13] Recently, Naeem and Sadowski proposed a pePC-SAFT model for polyelectrolyte solutions with counterions.[14,15] In the pePC-SAFT model, the formation of polyelectrolyte chains from charged monomers is explicitly taken into account by implementing the method of Jiang et al.[11] for flexible charged hard-sphere chain.

In addition to theories, several molecular simulations have been conducted to study the properties of polyelectrolyte and polyampholyte solutions. Bizjak et al.[16] used canonical Monte Carlo (MC) simulations to investigate the pair correlation functions, internal energies, and activity coefficients of fully flexible charged hard-sphere chains. Although the chain length studied was up to 128, the monomer reduced density considered in the simulations was only limited to $7.5 \cdot 10^{-3}$. Stevens and Kremer8 conducted molecular dynamics simulations to obtain the osmotic pressure for flexible linear polyelectrolytes with monomers interacting through Coulomb and Lennard-Jones potentials. It is this osmotic pressure data that is widely used to validate theories for charged hard-sphere chains regardless of the difference between the models used in theories and simulation. Liao et al.[17] studied the osmotic coefficient and counterion condensation of rod-like and flexible polyelectrolytes by molecular dynamics simulations. Similar to the work by Stevens and Kremer, the monomers of polyelectrolytes studied in their work also had a Lennard-Jones type potential. Chang and Yethiraj performed canonical Monte Carlo simulations to obtain the osmotic coefficients of polyelectrolyte solutions (polyions with counterions), and the excluded volume was accounted for through hard-sphere potential.[18] In their work, the systems studied ranged from dilute to concentrated solutions, and the chain length ranged from 16 to 64.

In general, most of molecular simulation studies focused on polyelectrolyte solutions with counterions, while the properties of polyampholyte solutions acquire much less attention and no suitable data can be found to validate theories for charged hard-sphere chain model of polyampholyte. Although Cheong and Panagiotopoulos[13] performed MC simulations for fully charged hard-sphere chain representing polyampholyte and investigated the effects of chain length and charge configurations on the critical properties, the critical property is usually not a good criterion to test molecular models, especially for models based on the thermodynamic perturbation theory. Hu et al.[19] obtained the osmotic coefficients of polyampholyte solutions by molecular dynamics simulations, but the charged chains simulated were also interacting with Coulomb and Lennard-Jones potentials.

In this work, we conduct canonical and isobaric-isothermal MC simulations for fully flexible charged hard-sphere chains. New MC data of the Helmholtz energy and osmotic coefficient, which is equivalent to compressibility factor, for solutions containing short polyampholytes are obtained; to the best of our knowledge, the Helmholtz energy and osmotic coefficients of polyampholyte solutions are obtained for the first time by MC simulations. New MC data of the osmotic coefficient for solutions containing short polyelectrolytes are also presented. This set of polyelectrolyte data complements the simulation data of Chang and Yethiraj.[18]

In addition, we develop an equation of state, which is based on the thermodynamic perturbation theory, for flexible charged hard-sphere chains. While the free energy contribution from the electrostatic interactions of monomers is based on the modified MSA (KMSA) obtained in our previous work,[20] the free energy contribution from the chain formation is carefully built with the help of MC simulation results. Instead of using the simple monomer approach (dimerization process) in calculating the chain contribution, as implemented in the model of Jiang et al.,[9,10,11] we explore the use of dimer and dimer-monomer approaches. At this stage, the equation of state can be applied to solutions of polyelectrolyte or polyampholyte as long as the charged monomers and counterions, if any, have the same size.

## 7.2 Theory

In solution, the polyion or polyampholyte molecule is treated as a flexible charged hard-sphere chain with a chain length of $n$ monomers, while the solvent is not explicitly accounted for and assumed to be a structureless continuous medium. The monomers of the charged chain and counterions, if any, are considered as spherical monovalent particles of the same size. For polyelectrolyte solutions, each polyion chain of length $n$ consists of $n$ negatively charged monomers ($N$), and there are $n$ positivity charged counterions ($P$) to ensure the electroneutrality. Each polyampholyte molecule consists of positively ($P$) and negatively ($N$) charged monomers, and thus for cases where there are no counterions considered, the chain length ($n$) of a polyampholyte molecule must be an even number. In this work, we do not consider a charged chain having neutral monomers.

The pair potential between particles $i$ and $j$ is given by,

$$u_{ij} = \begin{cases} \dfrac{z_i z_j e^2}{4\pi\varepsilon_0 \varepsilon r} & r > \sigma \\ \infty & r \le \sigma \end{cases} \tag{7.1}$$

where $z_i$ is the valence of particle $i$, $e$ is the elementary charge, $\varepsilon_0$ is the vacuum permittivity, $\varepsilon$ is the dielectric constant, which is set to 79.13 to mimic water-like solution, $\sigma$ is the diameter of particle, and $r$ is the center-to-center distance between two particles. In this work, particles include positively charged monomer, negatively charged monomer, and counterion. There are two independent parameters of the studied system, i.e., the reduced temperature $T^* = \dfrac{4\pi\varepsilon\varepsilon_0 \sigma k_B T}{e^2}$ and the reduced density $\rho^* = \rho\sigma^3$, where $\rho$ is the number density of particles, $T$ is the temperature, and $k_B$ is the Boltzmann constant. Note that although the relative permittivity $\varepsilon$ is set to a certain value, using a different value of $T^*$ in the simulation in fact could also be equivalent to assigning a different value of $\varepsilon$.

### 7.2.1 Equation of State

The equation of state for flexible charged hard-sphere chains is based on the thermodynamic perturbation theory, and expressed in terms of the excess Helmholtz energy,

$$a^{ex} = a^{hs} + a^{ion} + a^{chain} \tag{7.2}$$

where $a^{ex}$ is the dimensionless Helmholtz energy of charged hard-sphere chains in excess of that of ideal monomeric gas at the same temperature and monomer number density, $a^{hs}$ is the hard-sphere contribution, $a^{ion}$ is the contribution due to the electrostatic interactions of the particles, and $a^{chain}$ is the contribution due to the formation of charged hard-sphere chain.

### 7.2.1.1 Hard-Sphere and Ionic Terms

The hard sphere term $a^{hs}$ is given by Boublik-Mansoori-Carnahan-Starling equation of state,[12] which for an equal-sized system studied in this work reduces to

$$a^{hs} = \sum_k x_k n_k \cdot \frac{4\eta - 3\eta^2}{(1-\eta)^2} \tag{7.3}$$

where $x_k$ and $n_k$ are the mole fraction and the number of monomers of component $k$, respectively, and $\eta$ is the packing fraction ($= \frac{\pi}{6}\rho^* \sum_k x_k n_k$). A polyelectrolyte solution consists of two components, i.e., polyion and counterion, while a polyampholyte solution with no counterion has only a single component.

The contribution of the electrostatic interaction to the Helmholtz energy is calculated by using a modified MSA, referred to as KMSA, which was proposed in our previous work,[20] as follows

$$u^* = -\frac{e^2}{4\pi\varepsilon\varepsilon_0 k_B T \rho} \left\{ \Gamma \sum_i \frac{\rho_i z_i^2}{1 + \Gamma K \sigma} + \frac{\pi}{2\Delta} \Omega P_n^2 \right\} \tag{7.4}$$

$$a^{ion} = \sum_k x_k n_k \cdot \left( u^* + \frac{\Gamma^3}{3\pi \sum_i \rho_i} \right) \tag{7.5}$$

$$\Omega = 1 + \frac{\pi}{2\Delta} \sum_i \frac{\rho_i \sigma^3}{1 + \Gamma \sigma} \tag{7.6}$$

$$\Delta = 1 - \frac{\pi}{6} \sum_i \rho_i \sigma^3 \tag{7.7}$$

where $u^*$ is the excess internal energy per particle, $\rho_i$ is the number density of particle $i$, $\Gamma$ is the shielding parameter, and $K$ is an empirical parameter given by

$$K = 1 - 0.5786 \cdot \eta + 0.4825 \cdot \eta^2 \qquad (7.8)$$

Note that for the restricted primitive model, $P_n$ is equal to 0. This KMSA has been proved to accurately predict the excess energy of primitive charged hard-sphere systems, and the details can be found elsewhere.[20]

### 7.2.1.2 Chain Term: Dimer Approach

In this dimer approach, we consider the formation of charged hard-sphere chain from charged monomers as a two-step process, and use a charged dimer fluid as an intermediate state. Figure 7.1 shows the formation of flexible charged hard-sphere chains according to the two-step process. In the first step, we consider the charged monomers to form charged dimers, and the contribution to the Helmholtz energy for one bond formation is calculated from the Helmholtz energy of bringing two charged monomers together to form a charged dimer at the bond distance $\sigma$ in the bulk charged monomer fluid. In the second step, the charged dimers form the charged hard-sphere chain, and the Helmholtz energy change of one bond formation in the second step is the Helmholtz energy of bringing two charged dimers at arbitrarily large distance to the bond distance $\sigma$ in the bulk charged dimer fluid. It is obvious that this approach can only be used for chains containing even number of charged monomers.

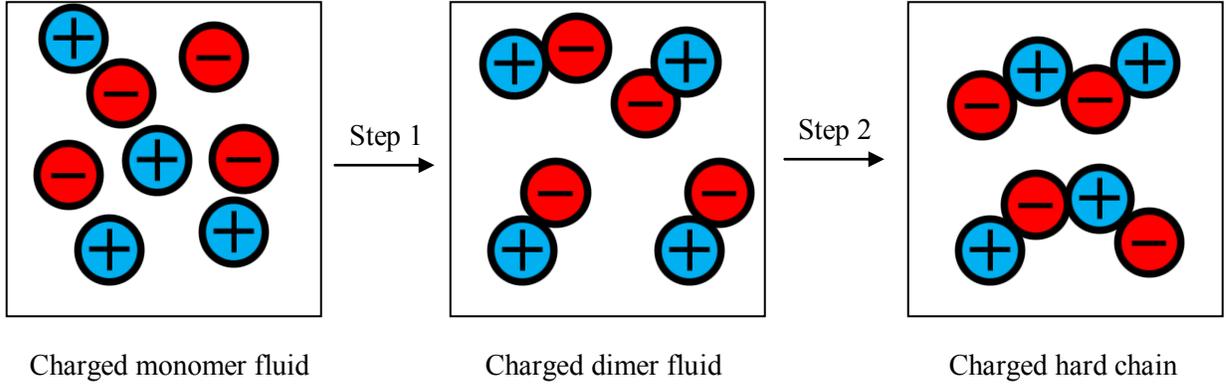

Figure 7.1. Two-step process to form charged hard-sphere chains from charged monomers.

The chain term based on this dimer approach is then expressed as,

$$a^{chain} = -x_{chain} \cdot \left( \sum_{j=1}^{\frac{n}{2}} \ln g_{2j-1,2j}(\sigma) + \sum_{j=1}^{\frac{n-2}{2}} \ln g_{(2j-1)2j,(2j+1)(2j+2)}(\sigma) \right) \tag{7.9}$$

where $x_{chain}$ is the mole fraction of chain molecules, $g_{2j-1,2j}(\sigma)$ is the pair correlation function between charged monomers $2j-1$ and $2j$ evaluated at contact, and $g_{(2j-1)2j,(2j+1)(2j+2)}(\sigma)$ is the site $2j$–site $(2j+1)$ pair correlation function between charged dimers $(2j-1)2j$ and $(2j+1)(2j+2)$ evaluated at contact. This approach is similar to that used by Sandler and coworkers to describe hard-sphere and square-well chain fluids.[21,22]

The pair correlation function at contact between two charged monomers $\alpha$ and $\beta$, i.e., $g_{\alpha,\beta}(\sigma)$ is calculated using the EXP method by Andersen and Chandler,[23,24]

$$g_{\alpha,\beta}(\sigma) = g_{hs}^{PY}(\sigma) \exp\left[ g_{\alpha,\beta}^{MSA}(\sigma) - g_{hs}^{PY}(\sigma) \right] \tag{7.10}$$

where $g_{hs}^{PY}(\sigma)$ is the pair correlation function at contact between uncharged hard spheres calculated using the Percus-Yevick (PY) approximation and $g_{\alpha,\beta}^{MSA}(\sigma)$ is the pair correlation function at contact between charged hard spheres calculated using MSA as follows,

$$g_{\alpha,\beta}^{MSA}(\sigma) = g_{hs}^{PY}(\sigma) - \frac{\Gamma^2 a_\alpha a_\beta}{4\pi^2 \sigma_{\alpha\beta} l_B} \qquad (7.11)$$

where $l_B$ is the Bjerrum length and $a_\alpha$ is calculated as,

$$a_\alpha = \frac{2\pi l_B (z_\alpha - \pi P_n \sigma_\alpha^2 / 2\Delta)}{\Gamma(1 + \Gamma\sigma_\alpha)} \qquad (7.12)$$

This EXP method has been proved to accurately approximate the contact value of the pair correlation function between charged particles,[9] and the details of the EXP method and $g^{MSA}(\sigma)$ can be found elsewhere.[9,23,24]

For the site-site pair correlation function at contact between charged dimers $g_{(2k-1)2k,(2k+1)(2k+2)}(\sigma)$, extensive Monte Carlo simulations are conducted to obtain 3 distinct types of site-site pair correlation function for the formation of block polyampholyte in a solution with no counterions, i.e., $g_{PN,PN}(\sigma)$, $g_{PP,NN}(\sigma)$, and $g_{PP,PP}(\sigma)$. Therefore, the proposed dimer approach for the chain term in this work cannot be applied to block polyampholyte with any charge configuration. The block polyampholyte must be one of the following types: *PNPN...PNPN* or $(P)_r(N)_r...(P)_r(N)_r$, where *r* is an even number. For a block polyampholyte with other charge configuration, such as *PPPNNN*, other types of site-site correlation function at contact, such as $g_{PN,NN}(\sigma)$, will be needed. The site-site pair correlation function between charged dimers for the formation of polyelectrolyte, i.e. $g_{NN,NN}(\sigma)$, is different from that of polyampholyte and also obtained from simulations; for polyampholyte, $g_{NN,NN}(\sigma)$ is equivalent to $g_{PP,PP}(\sigma)$.

### 7.2.1.3 Chain Term: Dimer-Monomer Approach

In this work, for this dimer-monomer approach, only polyampholyte is considered. The approach can be used for polyampholyte with any charge configuration and any number of monomers (even or odd), and any number of counterions. In the development of the change of the Helmholtz energy due to the formation of chain from charged monomers, the contribution due to the formation of the first bond is calculated from the Helmholtz energy of bringing two charged monomers together to form a charged dimer at the bond distance $\sigma$ in the bulk charged monomer fluid. The contribution due to the formation of the second and each of the successive bonds is then calculated from the Helmholtz energy of bringing one charged monomer from an arbitrarily large distance to form a charged trimer with the dimer already formed, canonically averaged over all configurations of the remaining monomer-dimer mixtures.

By using this dimer-monomer approach, the chain term is

$$a^{chain} = -x_{chain} \cdot \left[ \ln g_{12}(\sigma) + \sum_{j=2}^{n-1} \ln g_{(j-1)j,j+1}(\sigma) \right] \tag{7.13}$$

where $g_{1,2}(\sigma)$ is the pair correlation function between the first and second monomers of the chain evaluated at contact, which is calculated using Eq. 7.10, and $g_{(j-1)j,j+1}(\sigma)$ is the site $j$–monomer $(j+1)$ pair correlation function between dimer $(j-1)j$ and monomer $j+1$ evaluated at contact.

To account for the chain formation of a polyampholyte with any type of charge configuration, 4 distinct types of site-monomer pair correlation functions at contact are needed, i.e., $g_{PP,P}(\sigma)$, $g_{PN,P}(\sigma)$, $g_{PP,N}(\sigma)$, and $g_{PN,N}(\sigma)$. To obtain these pair correlation functions at contact, extensive Monte Carlo simulations for charged hard dimer-monomer systems are conducted.

### 7.2.2 Monte Carlo Simulation

Canonical Monte Carlo simulations are conducted to obtain the pair correlation functions of charged hard dimers and dimer-monomer systems, and the excess Helmholtz energy of charged 4-, 8-, and 16-mers of polyampholyte solutions without counterion. Isobaric-isothermal Monte Carlo simulations are also conducted to obtain the osmotic coefficients of 4-, 8-, and 16-mers of flexible charged hard-sphere chains with counterions for polyelectrolyte solutions and without counterion for polyampholyte solutions. To validate the dimer-monomer approach for polyampholyte solutions with odd number of monomers, isobaric-isothermal Monte Carlo simulations are also conducted to obtain the osmotic coefficient of 7-mers polyampholyte with counterions.

All Monte Carlo simulations are performed in cubic boxes. The long-range electrostatic interaction is handled using the Ewald summation[25] with periodic boundary condition. A two-step acceptance rule is used: if a particle overlaps with the other particle, the move is immediately rejected; otherwise, the electrostatic energy of the system is calculated for determining the acceptance probability. The number of monomers and counterions used in the simulations ranges from 512 to 2560 depending on the chain length and reduced density studied, and the Monte Carlo results are found to be weakly dependent on the size of simulation. For the determination of Helmholtz energy and osmotic coefficient, the reduced density range of our interest in this work is 0.1-0.5. For the canonical MC simulation, the chain molecules are regrown using the configurational-biased method[26] and also moved by translation, while the counterions are moved by translation. In general, 5 millions of configurational-biased moves and 5 millions of translational moves are performed to equilibrate the system, and another 5 million moves (half of them are configurational-biased moves) are performed to sample the ensemble averages. For the isobaric-isothermal simulations, in addition to the chain regrown and translational moves, volume change is also performed to change the system density.[26] In the isobaric-isothermal simulations, a total of 10000 volume changes are performed in the equilibration and sampling period, and the number of configurational-biased and translational moves are the same as that in the canonical MC simulations.

### 7.2.2.1 Pair Correlation Functions at Contact

The pair correlation functions are obtained by sorting the minimum separation distance between a pair of particles after the simulation is completed, and the contact values of the pair correlation functions are estimated by extrapolation.

To obtain the site-site pair correlation function between charged dimers, systems that reflect the composition of the bulk intermediate dimer fluid are used. As described previously, the intermediate dimer fluid will eventually form the charged hard-sphere chain of interest. For $g_{PN,PN}(\sigma)$, which is needed to describe the $PNPN...PNPN$-type block polyampholyte, a pure $PN$ dimer fluid is simulated. For $g_{PP,NN}(\sigma)$ and $g_{PP,PP}(\sigma)$, which are needed to describe the $(P)_r(N)_r...(P)_r(N)_r$-type block polyampholyte, a two-component fluid consists of $PP$ and $NN$ dimers is simulated, and for $g_{NN,NN}(\sigma)$ used for polyelectrolyte, a two-component fluid consists of $NN$ dimers and positively charged monomers is simulated. Since the electroneutrality must be satisfied, the composition of the two-component fluid simulated in this work is always fixed. Hence, these site-site correlation functions are only functions of the system reduced density and temperature.

As mentioned previously, for a block polyampholyte such as $PPPNNN$, other types of site-site correlation function at contact, i.e., $g_{PP,PN}(\sigma)$ and $g_{PN,NN}(\sigma)$, will be needed. To obtain such correlation functions, a fluid that contains three types of charged dimers (*PP, NN*, and *PN*) needs to be simulated, and the effect of system composition on the pair correlation function may need to be considered.

The site-monomer pair correlation functions between charged dimer and monomer are obtained by conducting simulation for symmetric systems. For example, to obtain the site-monomer pair correlation function between a positively charged dimer *PP* and a positively charged monomer *P*, i.e. $g_{PP,P}(\sigma)$, the simulation is conducted on a fluid mixture containing the same amount of negatively charged dimers *NN* and negatively charged monomers *N*, which ensure the electroneutrality and each component in the fluid mixture has the same concentration. Note that by using symmetric systems, the obtained values of $g_{PP,P}(\sigma)$, $g_{PN,P}(\sigma)$, $g_{PP,N}(\sigma)$, and $g_{PN,N}(\sigma)$ are equivalent to those of $g_{NN,N}(\sigma)$, $g_{NP,N}(\sigma)$, $g_{NN,P}(\sigma)$, and $g_{NP,P}(\sigma)$, respectively. Although the symmetric system does not necessarily reflect the real composition of the intermediate monomer-dimer fluid, we find that using the site-monomer pair correlation function obtained from such systems in Eq. 7.13 leads to an accurate description of the chain formation.

### 7.2.2.2 Helmholtz Energy

The dimensionless excess Helmholtz energy of charged-hard sphere chain is estimated by MC simulations as,

$$a^{ex} = a_1 + a_2 \qquad (7.14)$$

where $a_1$ is the dimensionless Helmholtz energy of the uncharged hard-sphere chains in excess of that of ideal monomeric gas at the same temperature and monomer density, and it can be estimated by the dimer approach of Chang and Sandler,[27]

$$a_1 = a_{hs} - \frac{n}{2} \ln g_{hs}(\sigma) - \frac{n-2}{2} \ln g_d(\sigma) \qquad (7.15)$$

where $a_{hs}$ is the hard-sphere contribution calculated by Eq. 7.3, $g_{hs}(\sigma)$ is the pair correlation function of hard-sphere fluid at contact calculated by the Carnahan Starling expression,[12] and $g_d(\sigma)$ is the site-site pair correlation function at contact between hard dimers calculated by an expression developed by Yethiraj and Hall,[28]

$$g_d(\sigma) = \frac{(1-0.5\eta)(0.534 + 0.414\eta)}{(1-\eta)^3} \qquad (7.16)$$

The equation of state for uncharged hard-sphere chain derived from Eqs. 7.15 and 7.16 was proved by Chang and Sandler[27] to be in good agreement with Monte Carlo simulation data, and hence Eq. 7.15 ($a_1$) provides an accurate estimation of the excess Helmholtz energy of uncharged hard-sphere chain.

In Eq. 7.14, $a_2$ is the dimensionless Helmholtz energy difference between the system of interest, which is the charged hard-sphere chain, and a system of reference, which is the uncharged hard-sphere chain. The value of $a_2$ is calculated using the multistage Bennett's acceptance ratio method[29]. Three intermediate stages are created to bridge the energy gap between the charged and uncharged systems. The details of the intermediate potential can be found in our earlier work.[20] After $a_1$ and $a_2$ are obtained, the dimensionless excess Helmholtz energy of charged hard-sphere chains can be calculated and compared with the prediction by Eq. 7.2.

### 7.2.2.3 Osmotic Coefficient

The osmotic coefficients (compressibility factors) of charged hard-sphere chains with or without counterions are defined as,

$$\phi = \frac{P}{\rho_m k_B T} \qquad (7.17)$$

where $P$ is the osmotic pressure and $\rho_m$ is the molecule number density, which is the sum of the number densities of all components in the solution. The osmotic coefficient data obtained from the isobaric-isothermal Monte Carlo simulation are then compared with the prediction of the proposed equation of state.

## 7.3 Results and Discussions

### 7.3.1 Pair Correlation Function at Contact

The reduced temperature and reduced density ranges of the Monte Carlo data are 0.2 to 1.0 and 0.01 to 0.5, respectively. The following correlation functions are proposed to correlate the site-site pair correlation functions at contact between charged dimers:

$$g_{PN,PN}(\sigma) = 1.0 + \sum_{\alpha=1}^{4} \sum_{\beta=1}^{4} D_{\alpha\beta} \cdot (\rho^*)^{\alpha-1} \cdot (T^*)^{1-\beta} \qquad (7.18)$$

$$g_{PP,NN}(\sigma) = g_{P,N}(\sigma)^{2.75} \cdot \sum_{\alpha=1}^{4} \sum_{\beta=1}^{4} D_{\alpha\beta} \cdot (\rho^*)^{\alpha-1} \cdot (T^*)^{1-\beta} \qquad (7.19)$$

$$g_{PP,PP}(\sigma) = g_{P,P}(\sigma)^{-1} \cdot \sum_{\alpha=1}^{4} \sum_{\beta=1}^{4} D_{\alpha\beta} \cdot (\rho^*)^{\alpha-1} \cdot (T^*)^{1-\beta} \qquad (7.20)$$

where $D_{\alpha\beta}$'s are the fitting parameters for a particular pair correlation function, $g_{PN}(\sigma)$ and $g_{NN}(\sigma)$ are the pair correlation functions between charged monomers at contact calculated by Eq. 7.10. The site-site pair correlation function $g_{NN,NN}(\sigma)$ for polyelectrolyte can be correlated by Eq. 7.18 but with different set of parameters $D_{\alpha\beta}$. The parameters $D_{\alpha\beta}$ obtained are listed in Tables 7.1–7.4 in the Appendix. We present the values of $g_{PN,PN}(\sigma)$ at $T^* = 0.2$ in Figure 7.2, $g_{PP,NN}(\sigma)$ at $T^* = 0.5944$ in Figure 7.3, and $g_{PP,PP}(\sigma)$ at $T^* = 1.0$ in Figure 7.4. Figure 7.4 also shows $g_{NN,NN}(\sigma)$ for polyelectrolyte at $T^* = 0.2$. As shown in Figures 7.2–7.4, the Monte Carlo data of the site-site pair correlation functions at contact between charged dimers can be well correlated by Eqs. 7.18 $-$ 7.20. It is worthy to mention that there could be different functions other than Eqs. 7.18 $-$ 7.20 that may be used to correlate the Monte Carlo data of site-site pair correlation functions at contact. Since the simulation data can be well correlated by Eqs. 7.18 $-$ 7.20, we do not try to explore other fitting functions.

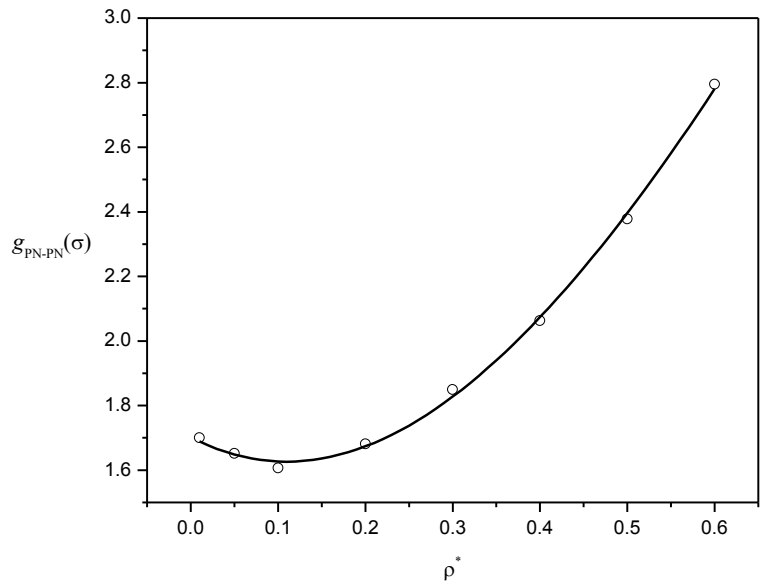

Figure 7.2. $g_{PN,PN}(\sigma)$ at $T^* = 0.2$; MC (symbols) and fitting curve - Eq. 7.18 (line).

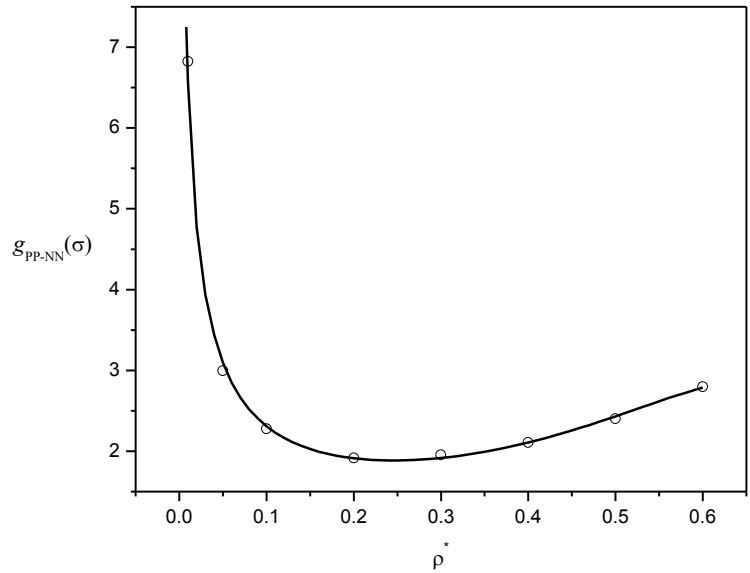

Figure 7.3. $g_{PP,NN}(\sigma)$ at $T^* = 0.5944$; MC (symbols) and fitting curve - Eq. 7.19 (line).

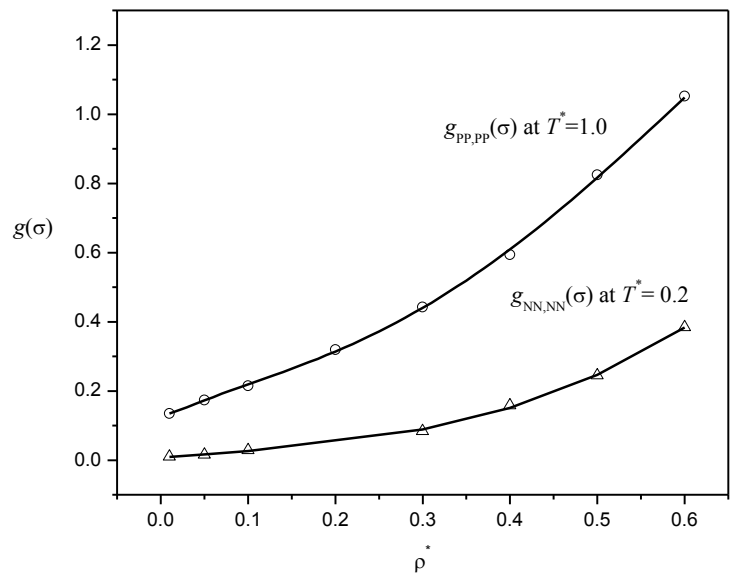

Figure 7.4. $g_{NN,NN}(\sigma)$ for polyelectrolyte at $T^* = 0.2$ and $g_{PP,PP}(\sigma)$ at $T^* = 1.0$; MC (symbols) and fitting curves - Eqs. 7.18 and 7.20 (line).

Similar correlations are also proposed to fit the Monte Carlo data of the site-monomer pair correlation functions at contact between charged dimer and monomer:

$$g_{PP,P}(\sigma) = 1.0 + \sum_{\alpha=1}^{4} \sum_{\beta=1}^{4} D_{\alpha\beta} \cdot (\rho^*)^{\alpha-1} \cdot (T^*)^{1-\beta} \tag{7.21}$$

$$g_{PP,N}(\sigma) = g_{PN}(\sigma) \cdot \sum_{\alpha=1}^{4} \sum_{\beta=1}^{4} D_{\alpha\beta} \cdot (\rho^*)^{\alpha-1} \cdot (T^*)^{1-\beta} \tag{7.22}$$

where $D_{\alpha\beta}$'s are the fitting parameters specific for a particular pair correlation function and $g_{PN}(\sigma)$ is the pair correlation function at contact between positively and negatively charged monomers, calculated by Eq. 7.10. For the other site-monomer pair correlation functions, i.e., $g_{PN,P}(\sigma)$ and $g_{PN,N}(\sigma)$, they can be correlated using Eq. 7.21 with different sets of parameters $D_{\alpha\beta}$. Good correlation of MC data can also be found by using Eqs. 7.21 and 7.22. The parameters $D_{\alpha\beta}$ and the MC data are given in the Appendix.

### 7.3.2 Excess Helmholtz Energy

From simulations, it is found that $a_2$ is only weakly dependent on the system size. For a PNPN...PN-type polyampholyte with $n = 8$ at $\rho^* = 0.1$ and $T^* = 0.5944$, the values of $a_2$ obtained using 512 and 1024 monomers are $-7.9632\pm0.0027$ and $-7.9720\pm0.0022$, respectively.

Figure 7.5 shows the dimensionless excess Helmholtz energy of the PNPN...PN-type polyampholyte solutions with $n = 4$, 8, and 16 at $T^* = 0.5944$. As shown in Figure 7.5, the excess Helmholtz energy increases with reduced density. At low to intermediate density, the excess Helmholtz energy is negative, while it becomes positive at high density, which indicates that the properties of the charged hard-sphere chain is dominated by the electrostatic interactions at low density and the hard-sphere potential at high density. For the PNPN...PN-type polyampholyte with $n = 16$, the excess Helmholtz energy predicted by the dimer approach and the dimer-monomer approach are similar and both are in good agreement with the MC results.

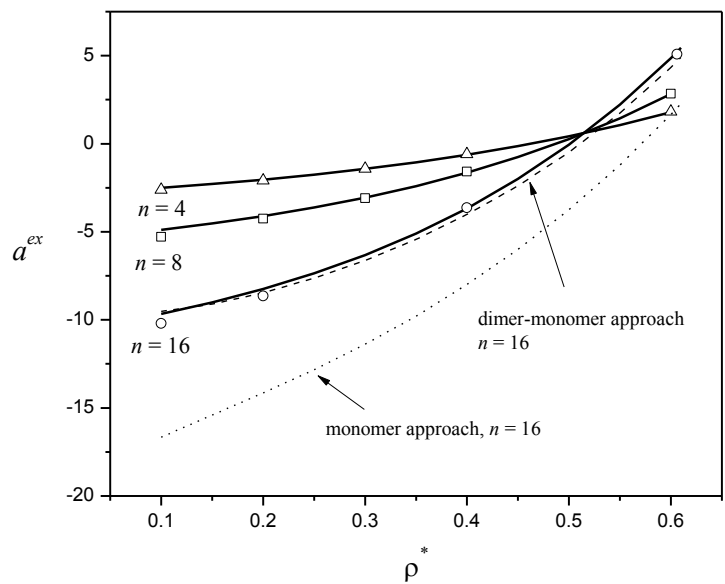

Figure 7.5. Dimensionless excess Helmholtz energy of the *PNPN...PN*-type polyampholyte solutions at $T^*=0.5944$; MC (symbols), predicted – dimer approach (solid lines), predicted – dimer-monomer approach for $n = 16$ (dashed line), and predicted – monomer approach for $n = 16$ (dotted line).

By using the monomer approach, in which the chain contribution is represented by $-(n-1) \cdot \ln[g_{PN}(\sigma)]$, the excess Helmholtz energy for the 16-mer polyampholyte is also calculated and compared with the results obtained using the dimer and dimer-monomer approaches. As shown in Figure 7.5, the excess Helmholtz energy of the 16-mer polyampholyte is significantly underestimated if such an approach is used, which indicates more detailed structure information is important for the description of charged chain formation.

The dimensionless excess Helmholtz energy of the *PNPN...PN*-type polyampholyte solutions at $T^* = 0.2$ are also obtained from simulations, and as shown in Figure 7.6, the model predictions using the dimer approach are in good agreement with the MC results. The prediction using the dimer-monomer approach (not shown) is similar to that using the dimer approach. The absolute values of the excess Helmholtz energy at $T^*$=0.2 are higher than those at $T^* = 0.5944$ (see Figure 7.5) due to stronger Coulomb interactions at low temperature.

Figure 7.7 shows the dimensionless excess Helmholtz energy of solutions containing diblock polyampholytes with $n$ = 4, 8, and 16 at $T^* = 0.5944$. The prediction of the excess Helmholtz energy using the dimer approach, which is similar to that of the dimer-monomer approach (not shown), is satisfactory except for $n$ = 8 at $\rho^* = 0.1$. It is worthy to notice that at $\rho^* = 0.1$, the average end-to-end distance of the diblock polyampholyte with $n$ = 8 obtained from simulation is 2.88$\sigma$. An end-to-end distance that is much smaller than the chain length indicates a strongly nonlinear conformation due to the intramolecular charge interactions. It is widely accepted that an approach that is built on the Wertheim's first order perturbation theory is less accurate for chains with such a conformation, and the use of the perturbation theory of higher order would be needed. Similarly, the excess Helmholtz energy for the diblock polyampholyte of length $n$ = 16 at low density is expected to be inaccurate, although no simulation at this low density is performed.

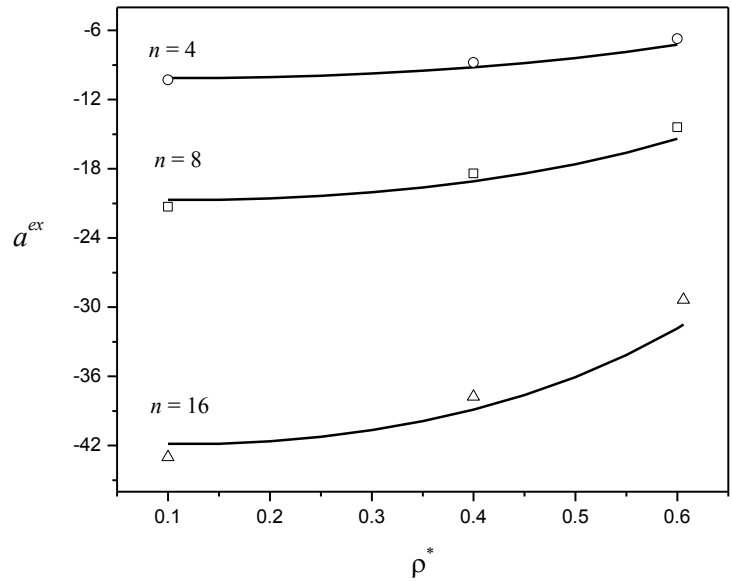

Figure 7.6. Dimensionless excess Helmholtz energy of the *PNPN...PN*-type polyampholyte solutions at $T^*$=0.2; MC (symbols), predicted – dimer approach (lines).

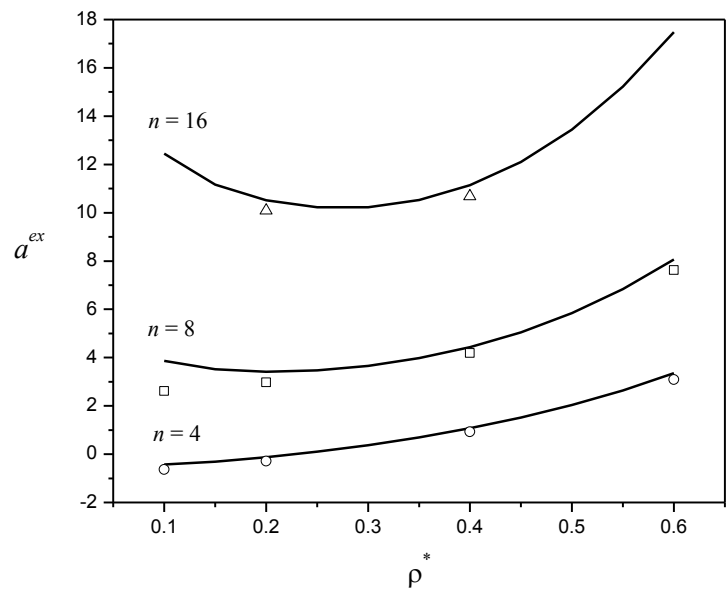

Figure 7.7. Dimensionless excess Helmholtz energy of solutions containing diblock polyampholytes $[(P)_r(N)_r]$ at $T^*$=0.5944; MC (symbols), predicted (lines).

### 7.3.3 Osmotic Coefficient

The osmotic coefficients of polyampholyte solutions are calculated using the proposed equation of state and compared with the results obtained from the isobaric-isothermal Monte Carlo simulations. Figure 7.8 shows the osmotic coefficients of the *PNPN...PN*-type polyampholyte solutions at $T^* = 0.5944$. As expected, the osmotic coefficient increases with density and chain length. The predictions of osmotic coefficient using the dimer and dimer-monomer approaches are in agreement with the simulation results. The osmotic coefficient is overestimated if the monomer approach is used, although a correct trend can be observed. However, the model proposed by Jiang et al.[11] significantly underestimates the osmotic coefficient and cannot even follow the trend of the simulation data. In that model, the charged hard-sphere chain formation is accounted for by using the cavity correlation function $y_{PN}(\sigma)$, which is estimated from the combined HNC/MSA. The underestimation of the osmotic coefficient is due to the inaccurate HNC/MSA, which leads to the incorrect cavity correlation function $y_{PN}(\sigma)$. Similar representation of the osmotic coefficient can be found for the *PPNN...PPNN*-type polyampholyte solutions, as shown in Figure 7.9.

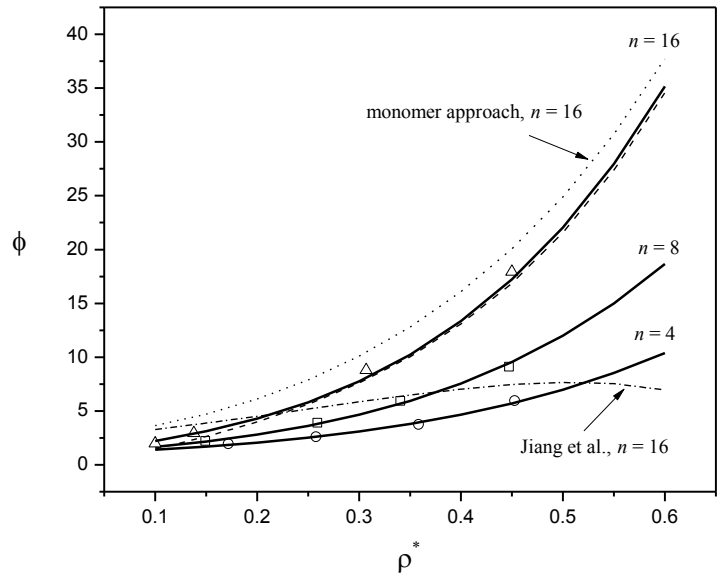

Figure 7.8. Osmotic coefficients of the *PNPN...PN*-type polyampholyte solutions at $T^* = 0.5944$; MC (symbols), predicted – dimer approach (solid lines), predicted – dimer-monomer approach for $n = 16$ (dashed line), predicted – model by Jiang et al.[9,10,11] for $n = 16$ (dashed-dotted line), predicted – monomer approach for $n = 16$ (dotted line).

Figure 7.10 shows the osmotic coefficients of solutions containing a diblock polyampholyte with $n = 8$ at $T^* = 0.5944$. The osmotic coefficient predicted by the proposed model with the dimer approach is not satisfactory in the low density range ($\rho^* < 0.2$). As analyzed in the previous section, the deviation between the model prediction and the simulation results is attributed to the strongly nonlinear conformation of the chain molecule at low density, for which the first order thermodynamic perturbation theory fails. The prediction using the dimer-monomer approach (not shown) is similar to that of the dimer approach.

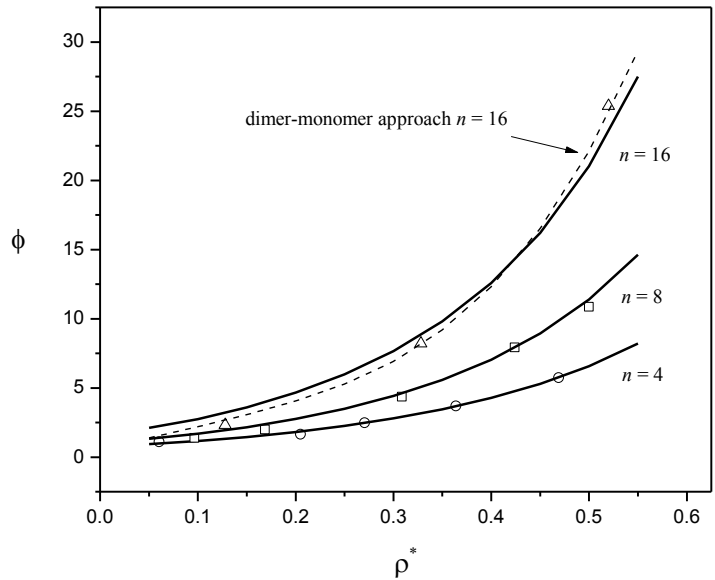

Figure 7.9. Osmotic coefficients of the *PPNN...PPNN*-type polyampholyte solutions at $T^* = 0.5944$; MC (symbols), predicted – dimer approach (solid lines), predicted – dimer-monomer approach for a charged chain with $n = 16$ (dashed line).

The proposed dimer approach is limited to block polyampholytes with an even number of monomers and a charge configuration of *PNPN...PNPN* or $(P)_r(N)_r...(P)_r(N)_r$ ($r$ is an even number), while the dimer-monomer approach can be applied to polyampholytes with any type of charge configuration and an odd number of monomers. Figure 7.11 shows the osmotic coefficients of solutions containing a mixture of *PNPPNPN* polyampholyte and negatively charged counterions. The prediction using the dimer-monomer approach is in good agreement with the simulation data. We find that polyampholytes with random charge configuration and arbitrary chain length can be accurately described by the dimer-monomer approach, provided the conformation of the chain molecule is not strongly nonlinear.

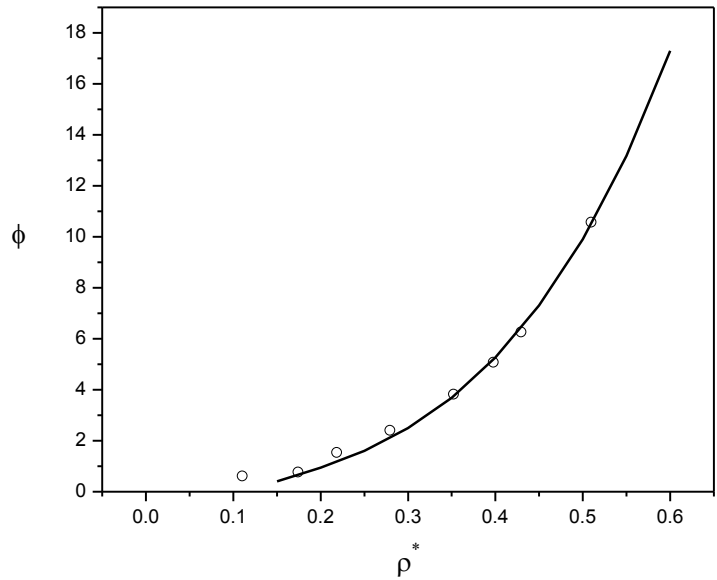

Figure 7.10. Osmotic coefficients of solutions containing a diblock polyampholyte with $n = 8$ $[(P)_4(N)_4]$ at $T^* = 0.5944$; MC (symbols), predicted – dimer approach (line).

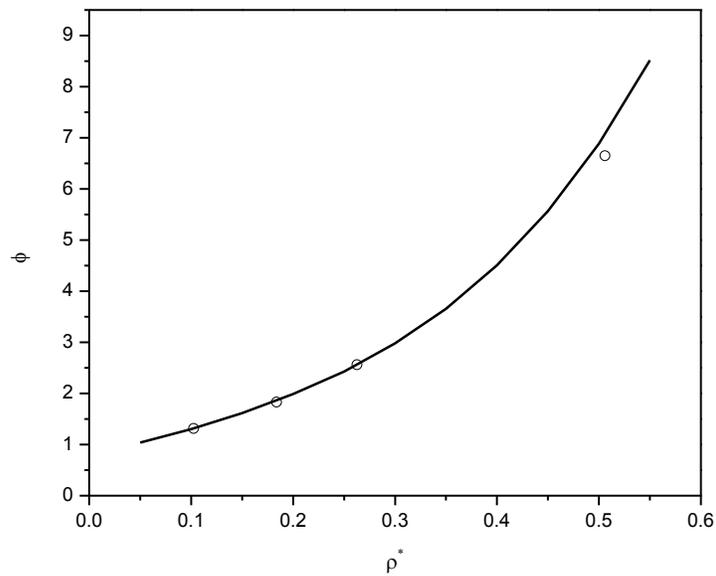

Figure 7.11. Osmotic coefficients of solutions containing a mixture of *PNPPNPN* polyampholyte and counterions at $T^* = 0.5944$; MC (symbols), predicted – dimer-monomer approach (line).

The osmotic coefficients of polyelectrolyte solutions are also predicted by the proposed equation of state and compared with the simulation results. Figure 7.12 shows the osmotic coefficients of the polyelectrolyte solutions at $T^* = 0.5944$. The osmotic coefficient of polyelectrolyte solutions in the high reduced density range is significantly smaller than that of polyampholyte, and a recognizable crossover of curves of the osmotic coefficients of polyelectrolyte solutions for different chain lengths can be observed from both simulations and model predictions (using either the dimer or dimer-monomer approach). The crossover indicates that the hard-sphere interaction (excluded volume) contribution to the osmotic coefficient is stronger for longer chains. The predictions by using the dimer and dimer-monomer approaches are indistinguishable and in good agreement with simulation results, while the model of Jiang et al.[9,10,11] that uses the combined HNC/MSA to calculate the cavity correlation function overestimates the osmotic coefficients and cannot correctly predict the crossover behavior.

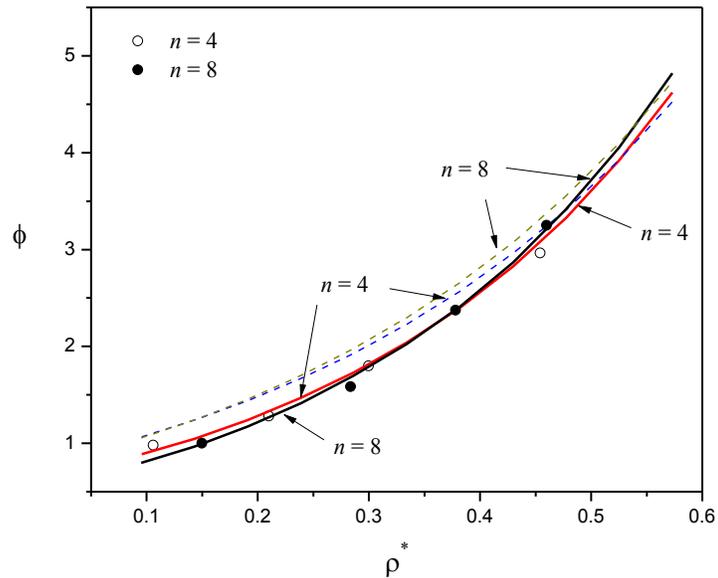

Figure 7.12. Osmotic coefficients of polyelectrolyte solutions at $T^* = 0.5944$; MC (symbols), predicted – dimer approach (solid lines), and predicted – model by Jiang et al.[9,10,11] (dashed lines).

Figure 7.13 shows the osmotic coefficients of polyelectrolyte solutions at $T^* = 0.4$. Compared to the simulation data, the osmotic coefficient is well predicted by the dimer approach and the dimer-monomer approach (not shown), indicating that the effect of temperature can be well captured by the proposed equation of state.

Figure 7.14 shows the osmotic coefficients of polyelectrolyte solutions with $n = 16$ at $T^* = 1.0$. The open symbols are the canonical MC data taken from Chang and Yethiraj[18] and the filled symbol is the osmotic coefficient obtained in this work from the isobaric-isothermal MC simulation. A good agreement between our simulation result and the MC data of Chang and Yethiraj is found, which serves well as the validation of our simulation. It is also worthy to mention that the osmotic coefficients of polyelectrolyte solutions with chain lengths higher than 16 can hardly be distinguished by the proposed equation of state and the model by Jiang et al.[10]

because the mole fraction of the chain molecules ($= \frac{1}{n+1}$) is so small that the property of such a system is dominated by the counterions.

We also examine how the proposed equation of state performs in the dilute solution region. As shown in Figure 7.14, the existence of a minimum value of the osmotic coefficient indicates the osmotic coefficient is affected by the competing repulsive hard sphere potential and attractive Coulomb potential. The proposed equation of state overestimates the osmotic coefficients in the dilute solution region, which is believed to be attributed to the counterion condensation on polyion. As stated by Chang and Yethiraj,[18] from the dilute to the semi dilute region of polyelectrolyte solution, the counterion condensation is the primary contribution to the non-ideality of the osmotic coefficient. Since the effect of counterion condensation is not explicitly taken into account in the proposed equation of state, the electrostatic interaction between counterions and polyion is underestimated, which leads to the overestimation of the osmotic coefficient. Nevertheless, the prediction of the proposed equation of state is more accurate than that of the models by Jiang et al.[9,10,11] and by Shew and Yethiraj[7,18] when $\rho^* > 0.015$.

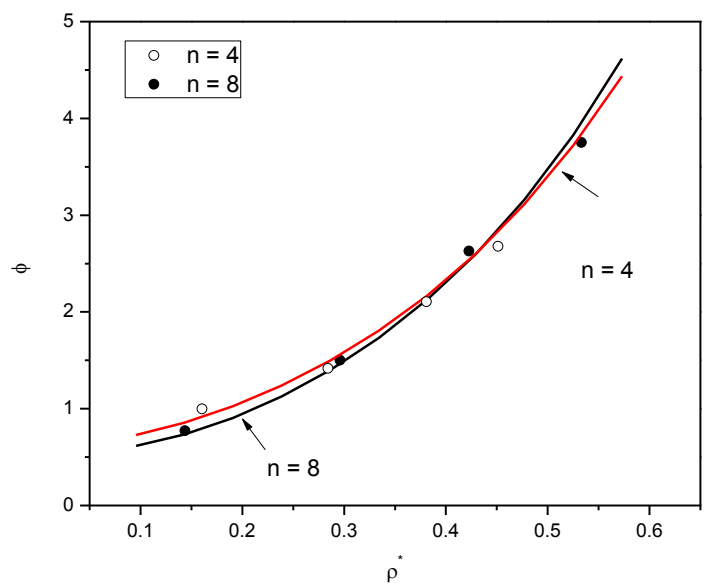

Figure 7.13. Osmotic coefficients of polyelectrolyte solutions at $T^* = 0.4$; MC (symbols) and predicted – dimer approach (solid lines).

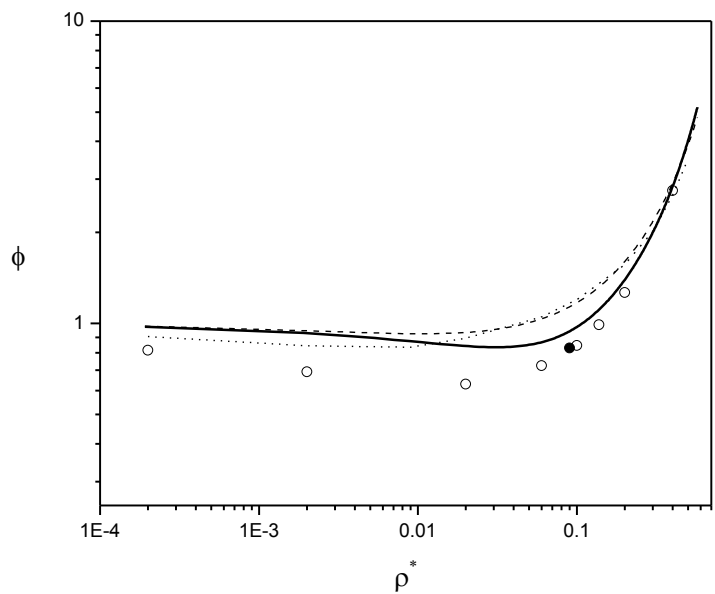

Figure 7.14. Osmotic coefficients of polyelectrolyte solutions with $n = 16$ at $T^* = 1.0$; MC data (open symbols – Chang and Yethiraj,[18] filled symbol – this work), predicted – dimer approach (solid line), predicted – model by Jiang et al.[9,10,11] (dashed line), and predicted – model by Shew and Yethiraj[7,18] (dotted line).

## 7.4 Conclusions

The thermodynamic modeling of polyelectrolyte and polyampholyte solutions has been considered. The polyion or polyampholyte molecule in solution is modeled as a flexible charged hard-sphere chain, while the solvent is not explicitly accounted for and assumed to be a structureless continuous medium. The monomers of the charged chain and counterions are considered as spherical monovalent particles of the same size. By performing canonical and isobaric-isothermal MC simulations, the excess Helmholtz energy and osmotic coefficient of short polyampholyte solutions are obtained for the first time and the osmotic coefficient for solutions containing short polyelectrolytes are presented.

A new equation of state based on the thermodynamic perturbation theory is also proposed for flexible charged hard-sphere chains representing polyelectrolyte or polyampholyte molecules in solution. The equation of state is expressed in terms of the excess Helmholtz energy, which can be divided into three parts: hard-sphere, electrostatic, and chain contributions. For the modeling of charged hard-sphere chains, the use of solely the structure information of monomer fluid for calculating the chain contribution is found to be insufficient and thus more detailed structure information must be considered. Two approaches, i.e., the dimer and dimer-monomer approaches, are explored to obtain the contribution of the chain formation to the Helmholtz energy. The pair correlation functions at contact needed in these two approaches are derived from the MC data obtained in this work. By comparing with the simulation results, the proposed equation of state with either the dimer or dimer-monomer approach accurately predicts the excess Helmholtz energy and osmotic coefficients of polyampholyte and polyelectrolyte solutions except at very low density, where the chain conformation is strongly nonlinear or the counterion condensation may occur. The equation of state also well captures the effect of temperature on the thermodynamic properties of these solutions. Furthermore, the crossover of curves of the osmotic coefficients of polyelectrolyte solutions as a function of density for different chain lengths, which indicates that the excluded volume contribution to the osmotic coefficient is larger for longer chains, can be accurately predicted. The advantage of the equation of state with the dimer-monomer approach is that it can be used to describe the thermodynamic properties of solutions containing polyampholyte with any charge configuration and any chain length.

# Appendix

Four types of pair correlation functions at contact between charged dimers, i.e., $g_{PN,PN}(\sigma)$, $g_{PP,NN}(\sigma)$, and $g_{PP,PP}(\sigma)$ for polyampholyte, and $g_{NN,NN}(\sigma)$ for polyelectrolyte, are obtained from Monte Carlo simulation. The Monte Carlo data are correlated using Eqs. 7.18 − 7.20 with the parameters $D_{\alpha\beta}$ given in Tables 7.1 to 7.4

Table 7.1. $D_{\alpha\beta}$ in Eq. 7.18 for $g_{PN,PN}(\sigma)$

| $D_{\alpha\beta}$ | $\alpha=1$ | $\alpha=2$ | $\alpha=3$ | $\alpha=4$ |
|---|---|---|---|---|
| $\beta=1$ | 0.6883 | 0.4124 | -0.1016 | 0.0149 |
| $\beta=2$ | 3.4818 | -3.1982 | 1.1660 | -0.1445 |
| $\beta=3$ | -8.6617 | 12.2556 | -4.2755 | 0.4895 |
| $\beta=4$ | 12.4352 | -13.6917 | 4.9323 | -0.5588 |

Table 7.2. $D_{\alpha\beta}$ in Eq. 7.19 for $g_{PP,NN}(\sigma)$

| $D_{\alpha\beta}$ | $\alpha=1$ | $\alpha=2$ | $\alpha=3$ | $\alpha=4$ |
|---|---|---|---|---|
| $\beta=1$ | 0.3146 | 0.0225 | -0.0463 | 0.0062 |
| $\beta=2$ | 1.0264 | -2.1093 | 0.8254 | -0.0884 |
| $\beta=3$ | -5.5235 | 7.9355 | -3.022262 | 0.3286 |
| $\beta=4$ | 5.3962 | -7.5051 | 2.8940 | -0.3197 |

Table 7.3. $D_{\alpha\beta}$ in Eq. 7.20 for $g_{PP,PP}(\sigma)$

| $D_{\alpha\beta}$ | $\alpha=1$ | $\alpha=2$ | $\alpha=3$ | $\alpha=4$ |
|---|---|---|---|---|
| $\beta=1$ | 0.2093 | -0.2227 | 0.0763 | -0.0079 |
| $\beta=2$ | 1.5795 | -0.6866 | 0.1113 | -0.0064 |
| $\beta=3$ | -1.1636 | 0.0687 | -0.1177 | 0.0257 |
| $\beta=4$ | 11.4853 | -5.1262 | 1.4458 | -0.1578 |

Table 7.4. $D_{\alpha\beta}$ in Eq. 7.21 for $g_{NN,NN}(\sigma)$

| $D_{\alpha\beta}$ | $\alpha=1$ | $\alpha=2$ | $\alpha=3$ | $\alpha=4$ |
|---|---|---|---|---|
| $\beta=1$ | -0.7041 | -0.3087 | 0.1104 | -0.0120 |
| $\beta=2$ | 0.5945 | 1.4495 | -0.9430 | 0.1273 |
| $\beta=3$ | 3.0292 | -7.5841 | 3.9973 | -0.5217 |
| $\beta=4$ | -1.6616 | 8.3002 | -4.3840 | 0.5702 |

The pair correlation functions at contact between charged dimer and charged monomer are also obtained from MC simulations. Four types of pair correlation function at contact are correlated by Eqs. 7.21 and 7.22 with the parameters $D_{\alpha\beta}$ given in Tables 7.5 to 7.8. We present the values of $g_{PP,N}(\sigma)$ at $T^* = 0.2$ in Figure 7.18, $g_{PN,P}(\sigma)$ and $g_{PN,N}(\sigma)$ at $T^* = 0.5944$ in Figure 7.19, and $g_{PP,N}(\sigma)$ and $g_{PP,P}(\sigma)$ at $T^* = 0.8$ in Figure 7.20. As shown in Figures 7.18–7.20, the Monte Carlo data of pair correlation functions at contact between charged dimers and monomers can be well correlated by Eqs. 7.21 and 7.22.

Table 7.5. $D_{\alpha\beta}$ in Eq. 7.21 for $g_{PN,P}(\sigma)$

| $D_{\alpha\beta}$ | $\alpha=1$ | $\alpha=2$ | $\alpha=3$ | $\alpha=4$ |
|---|---|---|---|---|
| $\beta=1$ | -0.9005 | 1.2914 | -0.4373 | 0.0643 |
| $\beta=2$ | 3.4469 | -3.5783 | 1.8076 | -0.3338 |
| $\beta=3$ | -10.9648 | 19.4541 | -9.5357 | 1.4772 |
| $\beta=4$ | 13.5447 | -18.8455 | 9.6078 | -1.4697 |

Table 7.6. $D_{\alpha\beta}$ in Eq. 7.21 for $g_{PN,N}(\sigma)$

| $D_{\alpha\beta}$ | $\alpha=1$ | $\alpha=2$ | $\alpha=3$ | $\alpha=4$ |
|---|---|---|---|---|
| $\beta=1$ | -0.6140 | 0.2992 | -0.2127 | 0.0289 |
| $\beta=2$ | 2.5209 | -3.3636 | 2.0046 | -0.2827 |
| $\beta=3$ | -1.9778 | 9.9352 | -6.8747 | 1.0015 |
| $\beta=4$ | 4.5708 | -11.6973 | 7.7416 | -1.1126 |

Table 7.7. $D_{\alpha\beta}$ in Eq. 7.21 for $g_{PP,P}(\sigma)$

| $D_{\alpha\beta}$ | $\alpha=1$ | $\alpha=2$ | $\alpha=3$ | $\alpha=4$ |
|---|---|---|---|---|
| $\beta=1$ | -0.8206 | 0.1244 | -0.0929 | 0.0128 |
| $\beta=2$ | 8.9765 | -10.0129 | 3.8566 | -0.4393 |
| $\beta=3$ | -30.1340 | 39.5035 | -15.6287 | 1.7889 |
| $\beta=4$ | 34.7831 | -42.8788 | 16.9490 | -1.9460 |

Table 7.8. $D_{\alpha\beta}$ in Eq. 7.22 for $g_{PP,N}(\sigma)$

| $D_{\alpha\beta}$ | $\alpha=1$ | $\alpha=2$ | $\alpha=3$ | $\alpha=4$ |
|---|---|---|---|---|
| $\beta=1$ | 0.2741 | 1.1577 | -0.5012 | 0.06112 |
| $\beta=2$ | 9.2118 | -16.5997 | 7.17068 | -0.87232 |
| $\beta=3$ | -32.9784 | 56.7025 | -24.4489 | 2.9537 |
| $\beta=4$ | 32.0622 | -53.8868 | 23.1086 | -2.7769 |

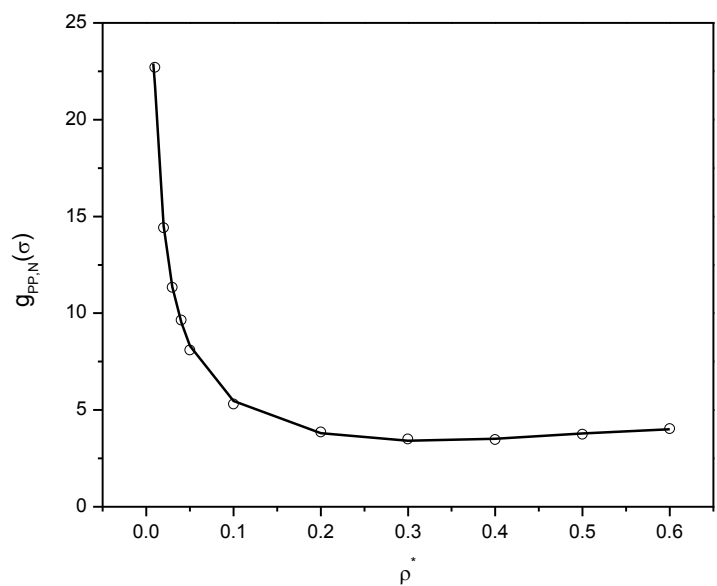

Figure 7.15. Pair correlation function at contact for a charged dimer (*PP*) and a negatively charged monomer (*N*) at $T^* = 0.2$; MC (symbols) and fitting curve - Eq. 7.21 (line).

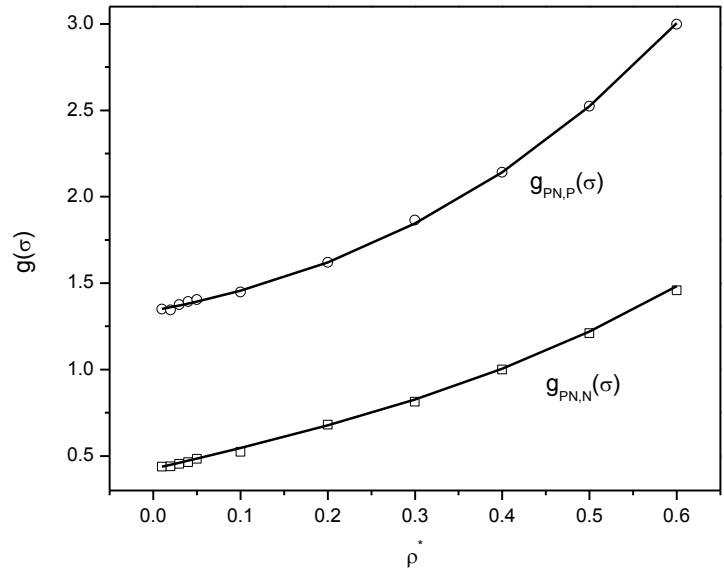

Figure 7.16. Pair correlation functions at contact for a charged dimer (*PN*) and a charged monomer (*P* or *N*) at $T^* = 0.5944$; MC (symbols) and fitting curve - Eq. 7.21 (lines).

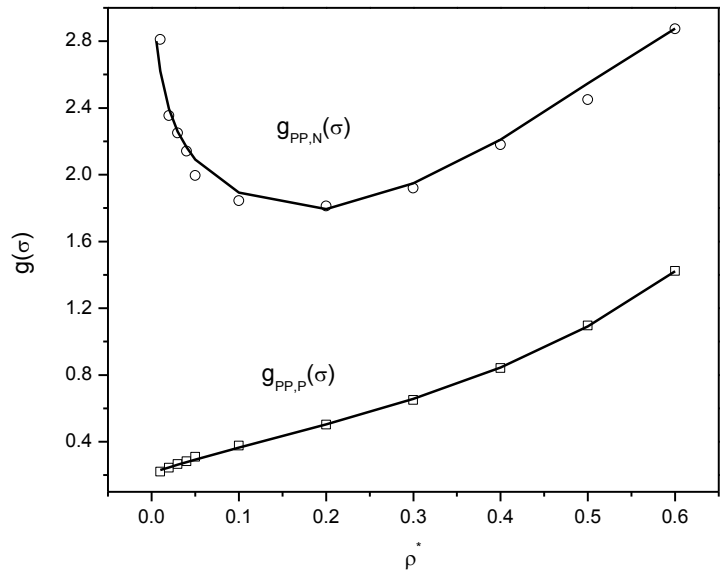

Figure 7.17. Pair correlation functions at contact for a charged dimer (*PP*) and a charged monomer (*P* or *N*) at $T^* = 0.8$; MC (symbols) and fitting curve – Eqs. 7.21 and 7.22 (lines).

# Chapter 8. A Modified Heterosegmented SAFT Equation of State for Calculating the Thermodynamic Properties of Ionic Liquids in Aqueous Solutions and Their Methane Hydrate Inhibition Performance

## 8.1 Introduction

Ionic liquid (IL) has been identified as a dual function gas hydrate inhibitor, and the screening of efficient and environmentally friendly IL-based hydrate inhibitor is of great significance. However, due to the huge number of ILs that can be potentially used as hydrate inhibitors, experimental study is not adequate and also expensive. Hence, it is desirable to develop a reliable model that can represent the thermodynamic properties of IL, especially its aqueous solutions. Various approaches have been developed to model the thermodynamic properties of ILs and fluid mixtures containing ILs. Statistical associating fluid theory (SAFT), as a statistical mechanical based equation of state, is promising for the modeling of complex fluids, such as ILs. A brief review of SAFT models used for the modeling of ILs can be found in our previous work.[1]

In our previous work,[1] the thermodynamic properties of aqueous imidazolium ILs had been investigated by the heterosegmented SAFT equation of state proposed by Ji and Adidharma.[2] In general, the liquid densities, activity coefficients, and osmotic coefficients of aqueous imidazolium IL solutions were correlated or predicted well with a set of transferrable parameters. In addition, the methane hydrate dissociation conditions in the presence of imidazolium ILs were successfully predicted by the heterosegmented SAFT coupled with van der Waals and Platteeuw theory.[3] The effects of anion types, alkyl length of the cation, and concentration of ILs on the methane hydrate phase equilibrium conditions were well captured. However, for some aqueous IL solutions in the high concentration range, the representation of the thermodynamic properties were not satisfactory, which indicates that some important features of the molecular interactions and structure of ILs cannot be accurately described by the heterosegmented SAFT equation of state proposed by Ji and Adidharma.[1,2] Although neutral alkyl branches exist in the cation of ILs, the heterosegmented SAFT[1] implemented the mean spherical approximation (MSA) to handle the Helmholtz energy contribution of the electrostatic interactions regardless that the MSA is derived without considering the effect of neutral component on the electrostatic interactions. In addition, the heterosegmented SAFT[1] modeled the cation head and anion of ILs as spherical charged segments, although some of the cation heads and anions could be better represented as charged chains. Therefore, a heterosegmented SAFT model is believed to be more realistic if it accounts for the effects of neutral component on the electrostatic interactions and models cation heads and anions as charged chains.

In Chapter 6, Monte Carlo simulations are conducted on charged hard sphere fluids in mixture with neutral hard spheres to understand the role of neutral species on the free energy and structure of electrolyte system.[4] An empirical modification, referred to as KMSA, is made to the analytical solution of MSA. The KMSA is proved to accurately predict the excess Helmholtz energy of charged system in the presence of neutral components, and it can be applied with confidence to systems with size and charge asymmetry at different temperatures. In Chapter 7, Monte Carlo simulations are conducted on charged hard-sphere chain fluids to understand the role of chain connectivity on the thermodynamic properties of charged chain molecules.[5] An equation of state for flexible charged hard-sphere chain molecules is proposed, and the excess

Helmholtz energy and osmotic coefficient of charged hard-sphere chain fluid can be well predicted.

In this chapter, a modified heterosegmented SAFT is proposed by incorporating the ionic term of KMSA obtained in Chapter 6 and the chain term of the charged hard-sphere chain equation of state obtained in Chapter 7. Then, the modified SAFT equation of state is used to correlate and predict the liquid density, activity coefficient, and osmotic coefficient of aqueous imidazolium and ammonium IL solutions. The cations studied in this chapter include $[C_xMIM]^+$ ($2 \leq x \leq 6$), $[(C_3H_7)_4N]^+$, and $[(C_4H_9)_4N]^+$, while anions studied include $Cl^-$, $Br^-$, $BF_4^-$, $MSO_4^-$, and $ESO_4^-$. With the modified heterosegmented SAFT, the modeling of aqueous ILs solutions now is extended to a wider range of cations and anions, and some of these IL species cannot be modeled by the original heterosegmented SAFT.[1] The modified heterosegmented SAFT equation of state is then coupled with van der Waals and Platteeuw theory[3] to predict the methane hydrate dissociation condition in the presence of ammonium aqueous IL solutions.

## 8.2 Thermodynamic Model

The cation of IL is composed of a cation head (aromatic ring for imidazolium IL and nitrogen atom for ammonium IL) and several alkyls. In this modified heterosegmented SAFT model, the cation head is treated as a charged chain instead of a charged spherical segment. For the cation of imidazolium ILs, there are 2 different groups of neutral segments representing two alkyl branches, while for the cation of ammonium ILs studied in this work, there are several identical groups of neutral segments representing the alkyls since the alkyl branches are symmetric. The anions $BF_4^-$, $MSO_4^-$, and $ESO_4^-$ are modeled as charged chains, while $Cl^-$ and $Br^-$ are modeled as spherical segments. The electrical charges of the cation head and anion are assumed to be evenly distributed along the charged chain.

## 8.2.1 Modified Heterosegmented SAFT

The modified heterosegmented SAFT equation of state is expressed in terms of residual Helmholtz energy,

$$\widetilde{a}^{res} \cong \widetilde{a}^{hs} + \widetilde{a}^{disp} + \widetilde{a}^{chain} + \widetilde{a}^{assoc} + \widetilde{a}^{ion} \qquad (8.1)$$

where the superscripts on the right hand side refer to terms accounting for the hard sphere, dispersion, chain, association, and ion interactions, respectively. The details of calculations of hard-sphere and dispersion terms can be found in our previous works.[6] Only water-water and cation-anion interactions are considered in the association term in Eq. 8.1. One association site each is assigned to the cation head and anion chain, which indicates that only the cross association is considered. The water parameters and the details of the association term can be found from our previous work.[6,7]

As stated in Chapter 5,[1] the electrostatic interaction should be considered due to the presence of a large amount of water in the solution, which makes the IL dissociate into solvated cation and anion. For the ion term accounting for the electrostatic interactions, similar to that used in Chapter 5,[1] the primitive model is used, in which the solvent is represented by a continuous medium of uniform dielectric constant. In Chapter 5,[1] the ion term is represented by MSA in the restricted primitive model (RPM), while KMSA[4] is used here as an improvement to the original heterosegmented SAFT; KMSA accurately predicts the excess Helmholtz energy for electrolyte systems in mixture with neutral species. The ion term is given by KMSA as follows,

$$a^{ion} = U^* + \frac{\Gamma^3}{3\pi \sum\limits_{i} \rho_i} \tag{8.2}$$

$$U^* = -\frac{e^2}{4\pi\varepsilon\varepsilon_0 k_B T \sum\limits_{i} \rho_i} \left\{ \Gamma \sum\limits_{i} \frac{\rho_i z_i^2}{1 + \Gamma K \sigma_i} + \frac{\pi}{2\Delta} \Omega P n^2 \right\} \tag{8.3}$$

where $U^*$ is the excess internal energy, $\rho_i$ is the molar density of component $i$, $\Gamma$ is the shielding parameter, $K$ is the empirical parameter introduced in Chapter 6 to account for the effect of neutral component on the electrostatic interactions, and $\sigma_i$ is the diameter of ion which is the same as the segment diameter. The hydrated diameter used in MSA-RPM is not needed. The details of the KMSA approach can be found in Chapter 6.[4]

The chain term is expressed as,

$$\tilde{a}^{chain} = -\sum\limits_{i} X_i (m_i - 1) \left[ \ln \overline{g}_i(\sigma_{\alpha\beta}) - \ln \overline{g}_{0,i}(\sigma_{\alpha\beta}) \right] \tag{8.4}$$

$$\ln \overline{g}_i(\sigma_{\alpha\beta}) = \sum\limits_{\beta \geq \alpha} B_{\alpha\beta,i} \ln g_{\alpha\beta}(\sigma_{\alpha\beta}) \tag{8.5}$$

where $X_i$ is the mole fraction of component $i$, $m_i$ is the segment number of component $i$, and $g_{\alpha\beta}(\sigma_{\alpha\beta})$ is the radial distribution function between segments $\alpha$ and $\beta$ evaluated at the contact distance, $\overline{g}_0$ is $\overline{g}$ evaluated at zero density, and $B_{\alpha\beta,i}$ is the bond fraction of type $\alpha\beta$ in molecule of component $i$. For the heterosegmented cation chain, the bond fraction $B_{\alpha\beta,i}$ is calculated based on the effective bond number ($n_B$) of each type of bond. For anion, the bond fraction is 1.0 since it is modeled as a homosegmented charged chain.

The cation heads and some of the anions of ILs studied in this work are represented by charged hard-sphere chains composed of several charged hard-sphere segments, while the alkyl branches are treated as neutral square-well chains. Note that the electrical charge of the cation head and anion is assumed to be evenly distributed on the charged chains. It is found that the number of segments in the cation head and anion chains for the ILs studied is close to 1 (the number of segments will be given in Section 8.3, hence the dimer or dimer-monomer approach proposed in Chapter 7[5] for a relatively long (number of segments in the chain is more than 4) charged hard-sphere chain are not needed here, and the monomer structure information is believed to be sufficient to calculate the chain term. The Helmholtz energy change caused by the formation of cation head and anion chains is calculated using the cavity correlation function at contact $y_{\alpha\beta}(\sigma_{\alpha\beta})$ of charged hard-sphere monomer fluid, which can be estimated after the pair correlation function at contact is obtained from the EXP method by Andersen and Chandler,[8,9]

$$g_{\alpha\beta}\left(\sigma_{\alpha\beta}\right) = g_{hs}^{PY}\left(\sigma_{\alpha\beta}\right)\exp\left[g_{\alpha\beta}^{MSA}\left(\sigma_{\alpha\beta}\right) - g_{hs}^{PY}\left(\sigma_{\alpha\beta}\right)\right] \tag{8.6}$$

$$y_{\alpha\beta}\left(\sigma_{\alpha\beta}\right) = g_{\alpha\beta}\left(\sigma_{\alpha\beta}\right)\cdot\exp\left[\frac{-u_{\alpha\beta}\left(\sigma_{\alpha\beta}\right)}{k_b T}\right] \tag{8.7}$$

where $g_{\alpha\beta}\left(\sigma_{\alpha\beta}\right)$ is the pair correlation function of charged hard spheres at contact and $u_{\alpha\beta}\left(\sigma_{\alpha\beta}\right)$ is the Coulomb potential between charged hard spheres. The details of the EXP method are given in Chapter 7.[5] It is worthy to mention that the charged hard-sphere chain model is only used to account for the formation of the cation head and anion chains. The square-well potential is assigned to the segments of cation head and anion for the calculations needed in the dispersion term, while the intramolecular dispersive interactions inside the cation head and anion chains are not considered. Since inconsistency still exists in the modeling of cation head and anion, it will be our future work to develop a charged square-well chain model for cation heads and anions.

As mentioned above, the cation of IL is modeled as a chain molecule with a charged chain representing the cation head, and different groups of neutral segments representing the alkyl branches. Similarly, the anion, except $Cl^-$ and $Br^-$, is modeled as a charged chain. Each of these groups of segments has four parameters: segment number ($m$), segment volume ($v^{oo}$), segment energy ($u/k$), and reduced width of the square well potential ($\lambda$). For the segments of cation head and alkyl, the bond number ($n_B$) is needed, which is used to calculate the bond fraction of the cation chain. The bond fraction of anion chain is 1.0 and the bond number of anion segments is not needed. For imidazolium ILs, it is arbitrarily assumed that 2/5 of the number of bonds of the cation head connect to the alkyls, while for ammonium ILs, 1/2 of the number of bonds of the cation head is arbitrarily assumed to connect to the alkyls.

The hydrated diameter of cation head or anion is not needed since KMSA is used instead of MSA-RPM. Besides these segment parameters, there are also two additional cross association parameters, i.e., association energy ($\varepsilon/k$) and association bonding volume parameter ($\kappa$), which are the properties of the cation head-anion pair. The bonding volume of cross association ($\kappa$) is set to 0.05 to reduce the number of parameters needed in the model. In this work, the binary interaction parameter ($k_{ij}$) is used to correct the short range alkyl-water interaction, while the binary interaction parameter between cation head and water, which has to be used in the original heterosegmented SAFT,[1] is not needed here.

## 8.2.2 Models for Chemical Potential Differences

The modified heterosegmented SAFT equation of state is coupled with van der Waals and Platteeuw theory[3] to predict the methane hydrate dissociation condition in the presence of imidazolium and ammonium aqueous IL solutions. In the hydrate phase equilibrium modeling, the chemical potential of water in different phases should be equal, and the detailed description of the chemical potential model for the hydrate phase can be found in our previous work.[6]

## 8.3 Results and Discussions

### 8.3.1 Parameter Estimation

In the modified heterosegmented SAFT, the parameters are still assigned to each group of segments, not molecules. The model parameters are believed to be transferrable and can be applied to other ILs having ions or alkyls that have been included in this work.

Some of the model parameters had been obtained. The segment energy ($u/k$) and reduced width of the square-well potential ($\lambda$) of alkyl groups are inherited from the original heterosegmented SAFT.[1] The following parameters are still needed: (1) segment parameters of cation heads and anions, (2) segment number ($m$), segment volume ($v^{oo}$) and bond number ($n_B$) of alkyls (3) the cross association energy between cation heads and anions ($\varepsilon/k$), and (4) the binary interaction parameters ($k_{ij}$'s) between alkyl groups and water. The liquid density ($\rho$), mean activity coefficient ($\gamma_{\pm}$), and osmotic coefficient ($\varphi$) data of aqueous IL solutions are used to fit these parameters. In addition to the experimental data used in Chapter 5,[1] the osmotic coefficients of [$C_2$MIM][$ESO_4$] and [$C_4$MIM][$MSO_4$] solutions,[10] and activity coefficients of [($C_3H_7$)$_4$N][Cl], [($C_4H_9$)$_4$N][Cl], [($C_3H_7$)$_4$N][Br], and [($C_4H_9$)$_4$N][Br] solutions[11] are included here to obtain the model parameters. Similar to our previous work, whenever possible, mean activity coefficient data is used instead of osmotic coefficient data ($\phi$) in the parameter estimation since the EOS parameters are more sensitive to activity coefficient data.[12]

The segment number ($m$), segment volume ($v^{oo}$), and bond number ($n_B$) of alkyls are found to be linear functions of alkyls' carbon number. The slope and intercept of these linear functions are given in Table 8.1, and all the other model parameters are listed in Tables 8.2, 8.3, and 8.4. Note that the number of segments of cation head and anion is close to 1, hence the dimer or dimer-monomer approach proposed in Chapter 7[5] for long flexible charge hard-sphere chain is not used here.

Table 8.1. Segment parameters for alkyl group ($=a \cdot n + b$); $n$ is the alkyl's carbon number.

|  | $a$ | $b$ |
| --- | --- | --- |
| $m$ | 0.179 | 0.104 |
| $v^{oo}$ (cc/mol) | 1.783 | 20.481 |
| $n_B$ | 0.0908 | -0.054 |

Table 8.2. Parameters of cation heads and anions.

|  | $m$ | $v^{oo}$ (cc/mol) | $u/k$ (K) | $\lambda$ | $n_B$ |
| --- | --- | --- | --- | --- | --- |
| $IMI^+$ | 1.016 | 33.09 | 534.06 | 1.520 | 0.03 |
| $N^+$ | 1.026 | 10.77 | 2083.08 | 1.800 | 0.07 |
| $Cl^-$ | 1.0 | 17.02 | 497.38 | 1.530 | - |
| $Br^-$ | 1.0 | 11.95 | 256.71 | 1.719 | - |
| $BF_4^-$ | 1.089 | 17.86 | 182.44 | 1.455 | - |
| $MSO_4^-$ | 1.0 | 11.00 | 237.34 | 1.789 | - |
| $ESO_4^-$ | 1.102 | 16.70 | 173.32 | 1.391 | - |

Table 8.3. Cross association parameters between cation heads and anions

| | $IMI^+$ -$Cl^-$ | $IMI^+$ -$Br^-$ | $IMI^+$ -$BF_4^-$ | $IMI^+$ -$MSO_4^-$ | $IMI^+$ -$ESO_4^-$ | $N^+$ -$Cl^-$ | $N^+$ -$Br^-$ |
|---|---|---|---|---|---|---|---|
| $\varepsilon/k$ (K) | 1404.44 | 1549.69 | 1360.25 | 1539.69 | 1390.99 | 1314.82 | 1641.72 |

Table 8.4. Binary interaction parameters ($k_{ij}$'s)

| | methyl-$H_2O$ | ethyl-$H_2O$ | propyl-$H_2O$ | butyl-$H_2O$ | pentyl-$H_2O$ | hexyl-$H_2O$ |
|---|---|---|---|---|---|---|
| $k_{ij}$ | 0 | −0.075 | 0.02 | 0 | −0.025 | 0.02 |

### 8.3.2 Liquid Density and Activity Coefficient

Due to the presence of a large amount of water, the accuracy of the density calculation of an aqueous solution is mainly attributed to that of pure water. The modification of the heterosegmented SAFT equation of state does not have an effect on the model representation of the density of aqueous IL solution. Similar to the results in our previous work,[1] a good representation of aqueous solution density can also be achieved by the modified heterosegmented SAFT. Therefore, the representation of the density of aqueous ILs solutions is not shown here.

The activity coefficients of [C$_4$MIM][Cl] and [C$_4$MIM][BF$_4$] solutions calculated by the modified heterosegmented SAFT are well represented, as shown in Figure 8.1. Figure 8.2 shows the activity coefficients of [C$_2$MIM][Br] and [C$_4$MIM][Br] solutions at 298.15 K and 318.15 K. The dashed line in Figure 8.2 is the activity coefficient of [C$_2$MIM][Br] solution calculated by the original heterosegmented SAFT in our previous work,[1] which is only qualitatively correct. With the modifications discussed in Section 8.2, a good correlation of activity coefficient of [C$_2$MIM][Br] solution can be obtained.

The modified heterosegmented SAFT is extended to the modeling of ammonium IL solutions, which is difficult to handle by the original heterosegmented SAFT.[1] Figure 8.3 shows the activity coefficient of ammonium IL solutions, including [(C$_3$H$_7$)$_4$N][Cl], [(C$_4$H$_9$)$_4$N][Cl], [(C$_3$H$_7$)$_4$N][Br], and [(C$_4$H$_9$)$_4$N][Br]. For the activity coefficients of [(C$_3$H$_7$)$_4$N][Br] and [(C$_4$H$_9$)$_4$N][Br] solutions, the calculation by the modified heterosegmented SAFT is quite satisfactory. For [(C$_3$H$_7$)$_4$N][Cl] and [(C$_4$H$_9$)$_4$N][Cl] solutions, although there is some discrepancy between the model calculation and experimental data, the data trend is correctly followed. In addition, the crossover of curves of the activity coefficients of [(C$_3$H$_7$)$_4$N][Cl] and [(C$_4$H$_9$)$_4$N][Cl] solutions can be captured by the modified heterosegmented SAFT. It is worthy to mention that the cations of ammonium ILs and imidazolium ILs share the same alkyl parameters, which indicates that the model parameters are indeed transferrable.

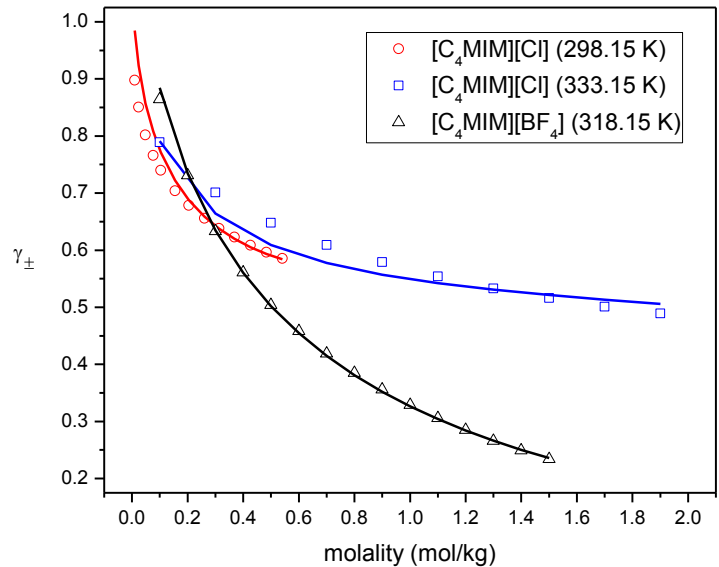

Figure 8.1. Mean activity coefficient of [C₄MIM][Cl] solution at 298.15 K and 333.15 K, and [C₄MIM][BF₄] solution at 318.15 K, experimental (points)[13,14], calculated (curves).

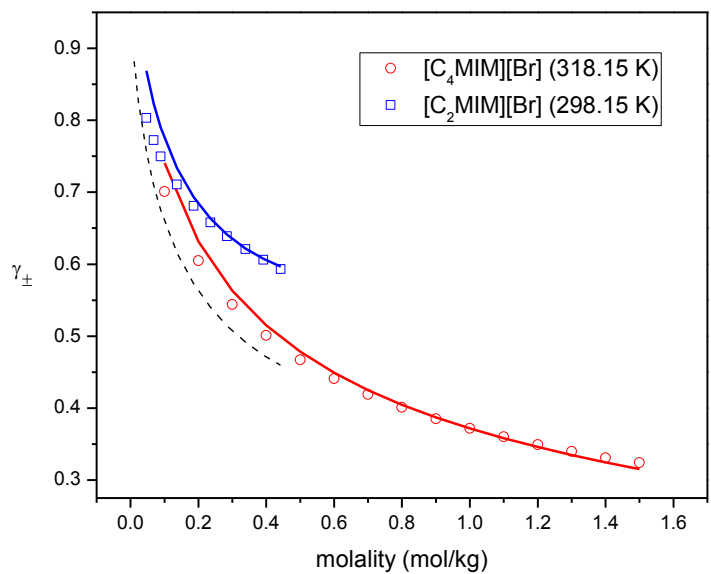

Figure 8.2. Mean activity coefficients of [C₂MIM][Br] solution at 298.15 K, and [C₄MIM][Br] solution at 318.15 K, experimental (points)[14,15], calculated (curves).

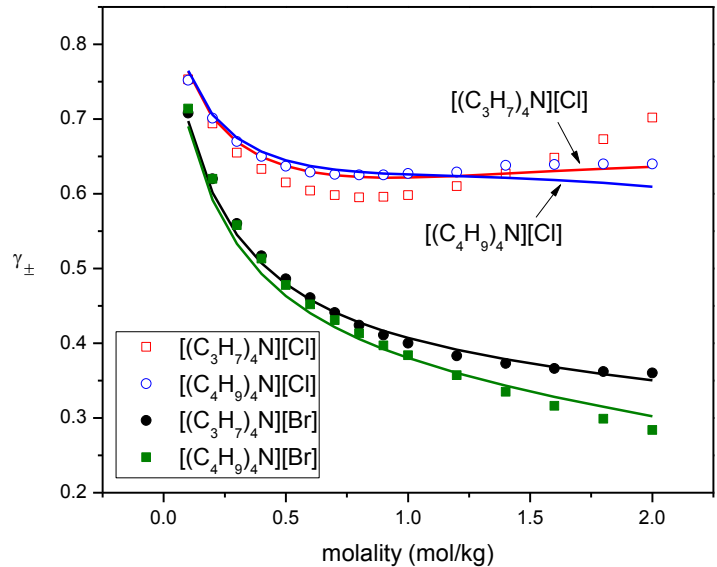

Figure 8.3. Mean activity coefficients of [(C₃H₇)₄N][Cl], [(C₄H₉)₄N][Cl], [(C₃H₇)₄N][Br] and [(C₄H₉)₄N][Br] solutions at 298.15 K, experimental (points),[11] calculated (curves).

### 8.3.3 Osmotic Coefficient

Figure 8.4 shows the osmotic coefficients of [$C_x$MIM][Br] ($3 \le x \le 6$) solutions. The osmotic coefficient of [$C_4$MIM][Br] solution at 298.15 K is predicted by using the parameters fitted to the corresponding activity coefficient data, while other osmotic coefficients shown in Figure 8.4 are correlated by the modified heterosegmented SAFT. For [$C_x$MIM][Br] ($3 \le x \le 6$) solutions, the osmotic coefficient decreases as the alkyl length increases, which is well captured by the model. Figure 8.5 shows the temperature effect on the osmotic coefficient of [$C_6$MIM][Br] solution. The osmotic coefficient of [$C_6$MIM][Br] solution at 298.15 K is used in the parameter fitting, while the osmotic coefficients at temperatures higher than 298.15 K are well predicted, except at high concentrations.

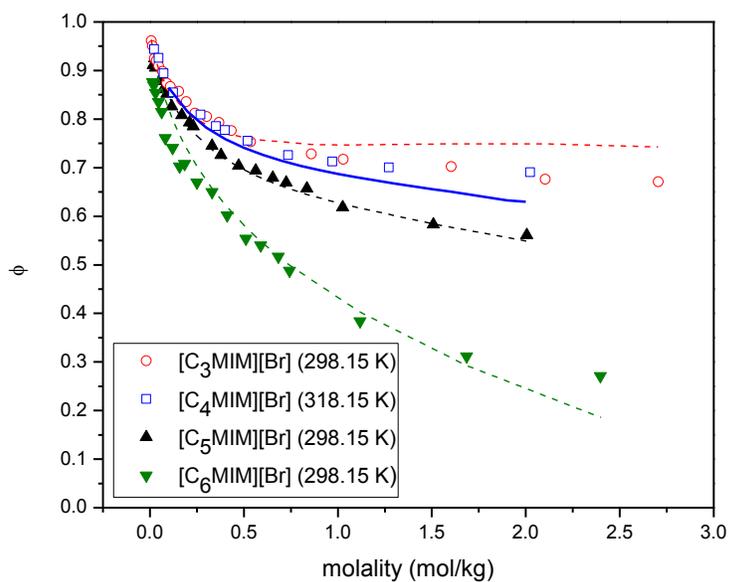

Figure 8.4. Osmotic coefficients of [C$_x$MIM][Br] (3 ≤ x ≤ 6) solutions at 298.15 K and 318.15 K, experimental (points),[10,16] predicted (solid line) and calculated (dashed lines).

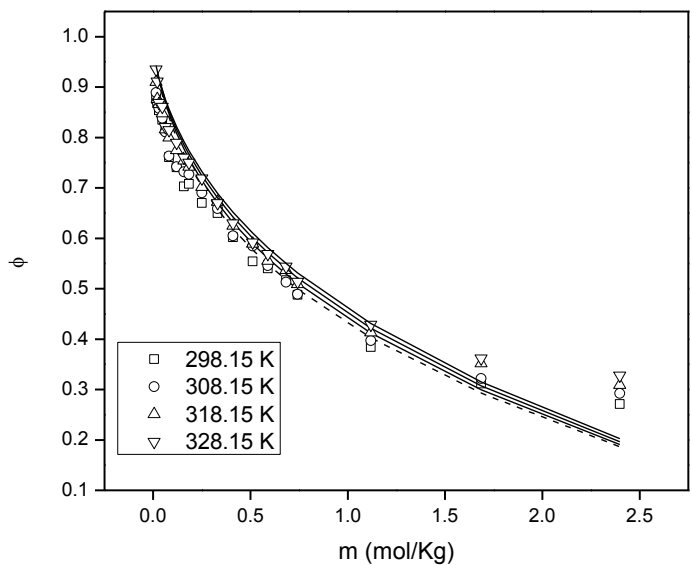

Figure 8.5. Osmotic coefficients of [C$_6$MIM][Br] solution at 298.15 K, 308.15 K, 318.15 K, and 328.15 K, experimental (points),[16] predicted (solid lines) and calculated (dashed line).

With the modification discussed in Section 8.2, the heterosegmented SAFT can be easily extended to the modeling of imidazolium ILs containing organic anions. Figure 8.6 shows the osmotic coefficients of [C$_2$MIM][ESO$_4$] and [C$_4$MIM][MSO$_4$] solutions at 298.15 K. As shown in Figure 8.6, a good agreement between the model calculation and the experimental data can be found.

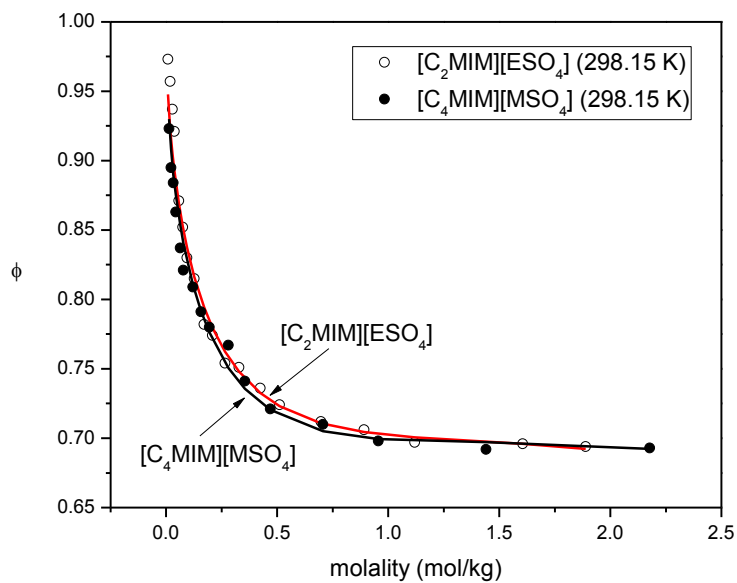

Figure 8.6. Osmotic coefficients of [C$_2$MIM][ESO$_4$] and [C$_4$MIM][MSO$_4$] solution at 298.15 K, experimental (points),[10] calculated (lines).

### 8.3.4 Methane Hydrate Dissociation Conditions in the Presence of Ionic Liquid Inhibitors

The methane hydrate dissociation conditions in the presence of ILs have been investigated in Chapter 5 by the original heterosegmented SAFT coupled with var der Waals and Platteeuw theory.[1] The roles of pressure, anion type, alkyl length of the cation, and IL concentration on the hydrate inhibition performance of imidazolium ILs have been investigated. However, only the inhibition effect of imidazolium ILs is studied due to the limited capacity of the original heterosegmented SAFT in representing the thermodynamic properties of aqueous IL solutions. In this chapter, with the modified SAFT equation of state, the inhibition effect of ammonium IL on methane hydrate is investigated and compared to that of imidazolium ILs.

Figure 8.7 shows the methane hydrate dissociation conditions in the presence of 10 wt% $[(CH_3)_4N][Cl]$, $[(CH_3)_4N][Br]$ and $[(C_2H_5)_4N][Cl]$ solutions predicted by the modified heterosegmented SAFT and the van der Waals and Platteeuw theory.[3] The properties of these ILs are not included in the parameter estimation, and their parameters are taken from Tables 8.1 to 8.4. The methane hydrate dissociation pressure in the presence of 10 wt% $[(CH_3)_4N][Cl]$ solution predicted by the model is in good agreement with the experimental data,[17] which indicates the model parameters are indeed transferrable. As shown in Figure 8.7, compared with $[(CH_3)_4N][Cl]$ solution, the $[(CH_3)_4N][Br]$ and $[(C_2H_5)_4N][Cl]$ solutions are similarly less effective in inhibiting methane hydrate. Therefore, it is believed that an ammonium IL with chloride anion is more effective than that with bromide anion, and an ammonium IL with shorter alkyl branch has a stronger inhibition effect on the methane hydrate. These findings for ammonium ILs are in agreement with the conclusions on the role of the anion type and alkyl length of imidazolium ILs.

Figure 8.8 gives a comparison of the inhibition effects of several imidazolium and ammonium ILs on methane hydrate. The inhibition effect of the 10 wt% $[(CH_3)_2NH_2][Cl]$ solution on methane hydrate dissociation pressure is predicted with the transferrable parameters in Tables 8.1 to 8.4. For both imidazolium and ammonium ILs, the inhibition effect increases as the length of alkyl group decreases. The [MIM][Cl] is found to be the most effective imidazolium based IL studied in this dissertation, while a more significant inhibition effect can be achieved by $[(CH_3)_2NH_2][Cl]$. The $[(CH_3)_2NH_2][Cl]$ solution is found to be the most effective

IL based hydrate inhibitor in this dissertation and further experimental investigation on $[(CH_3)_2NH_2][Cl]$ solution is desirable to validate the result predicted by the model.

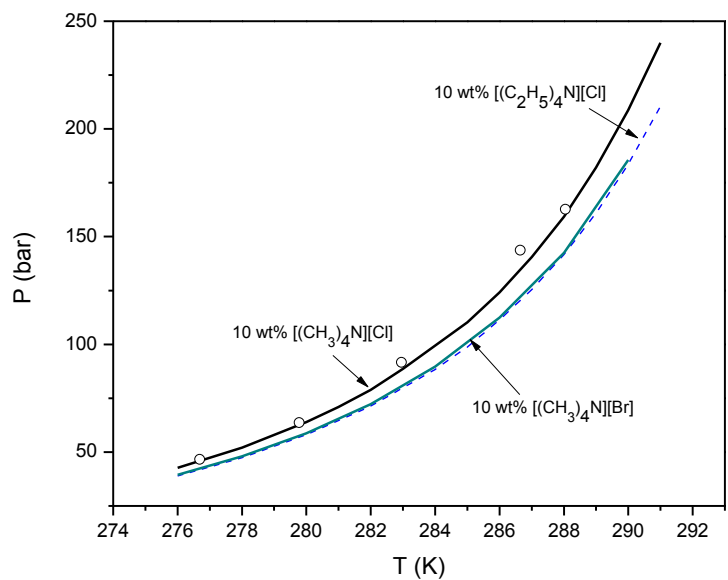

Figure 8.7. The effects of anion type and alkyl length on the effectiveness of ammonium IL inhibitors, experimental (points),[17] predicted (lines).

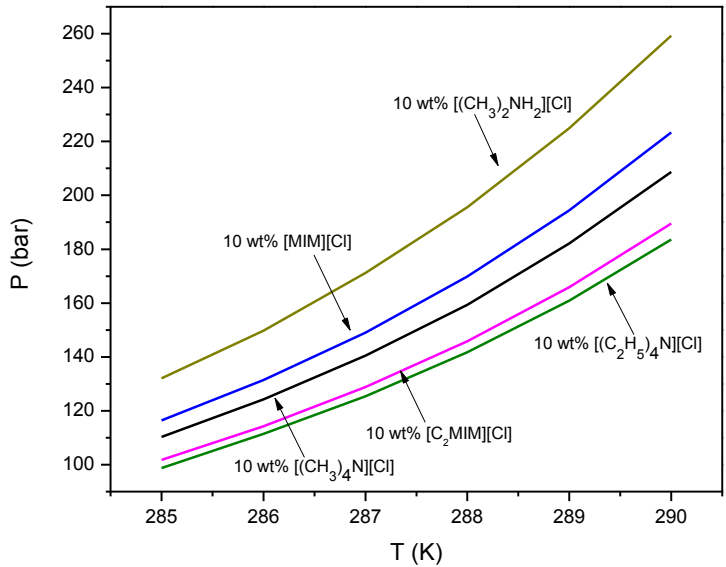

Figure 8.8. Comparison of inhibition effects of several ammonium and imidazolium ILs on methane hydrate.

## 8.4 Conclusions

Heterosegmented statistical associating fluid theory (SAFT) is developed to model the thermodynamic properties of aqueous ILs solutions. Two modifications have been made to the heterosegmented SAFT equation of state: (1) MSA is replaced by KMSA, which can capture the effect of the neutral segments of the alkyl branches on the electrostatic interactions of ions, and (2) the cation heads and some anions are modeled as charged chains instead of charged spherical segments, where the electrical charge of cation and anion is assumed to be evenly distributed on the charged chain. With these modifications, the activity coefficient and osmotic coefficient of aqueous imidazolium ILs can be better represented. In addition, the aqueous solutions of ammonium ILs and imidazolium ILs with organic anions can be successfully modeled by the modified heterosegmented SAFT. The representation of the thermodynamic properties of some ILs, especially in the high density range, could still be improved further, for example by removing the inconsistency within the model, i.e., the charged chain is still assumed to be a charged hard-sphere chain, while the other segments are interacting with square-well potential.

The inhibition effects of ammonium ILs on methane hydrate phase equilibrium are also studied by the heterosegmented SAFT and van der Waals and Platteeuw theory. The effects of anion type and alkyl length of the cation on the hydrate inhibition performance of ammonium ILs are discussed. Similar to imidazolium ILs, the ammonium ILs with chloride anion is more effective than that with bromide anion, and as the length of alkyls in the cation decreases, the inhibition effect of ammonium IL increases. $[(CH_3)_2NH_2][Cl]$ solution is found to be the most effective hydrate inhibitor studied in this dissertation. The modified heterosegmented SAFT, when coupled with the van der Waals and Platteeuw theory, is believed to be a useful tool for the screening of more effective IL-based hydrate inhibitor.

# Chapter 9. Conclusions and Future Work

## 9.1 Conclusions

By using statistical associating fluid theory (SAFT), the thermodynamic properties of traditional and innovative hydrate inhibitors, including electrolytes, alcohols, mixture of electrolytes and alcohols, and ionic liquids (ILs), are investigated. With the van der Waals and Platteeuw theory to describe the solid hydrate phase, SAFT equation of state can be used to model the alkane hydrate phase equilibrium conditions in the absence of inhibitors and in the presence of inhibitors mentioned above. Emphasis is placed on the modeling of aqueous IL solutions and their effects on methane hydrate dissociation conditions. To develop a more realistic SAFT model for the aqueous solutions of ILs, Monte Carlo simulations are conducted on fluids mixtures containing charged and neutral particles, and on flexible charged hard-sphere chain fluids.

In Chapter 2, a recent version of SAFT, referred to as SAFT2, is coupled with van der Waals and Platteeuw theory to model the phase equilibrium conditions for pure, binary, and ternary alkane hydrates in the absence of inhibitors. The liquid-vapor-hydrate (LVH) and ice-vapor-hydrate (IVH) boundaries of structure I hydrate dissociation are successfully predicted by the model. Since the van der Waals and Platteeuw theory neglects the interactions between gas molecules, a correction factor is introduced for some structure II hydrate formed by alkane mixture, and this correction factor is independent of the alkane mixture composition. In Chapter 3, ion-based SAFT2 is proved to be accurate in representing the liquid densities and activity coefficients of NaCl, KCl, and CaCl$_2$ solutions. The inhibition effect of these electrolytes on alkane hydrates dissociation conditions are also well predicted, and the model can be easily extended to mixed electrolyte solutions without introducing any additional empirical fitting parameter. In Chapter 4, ion-based SAFT2 is used to study the alkane hydrate phase equilibrium conditions in the presence of aqueous ethanol, ethylene glycol (MEG), and glycerol solutions, as well as mixed MEG and electrolyte solutions, including NaCl, KCl and CaCl$_2$. With the alcohol parameters fitted to liquid density and vapor pressure data of pure alcohol, the inhibition effect of these alcohols are accurately predicted. Moreover, it is found that the ion parameters obtained from the properties of aqueous electrolyte solutions can be successfully used to predict the alkane hydrate dissociation pressures in the presence of mixed MEG and electrolytes. This indicates that the ion parameters of the ion-based SAFT model are not solvent specific and can be used in different solvent provided the dielectric constant of that solvent is known.

The thermodynamic properties of aqueous imidazolium ILs solutions are investigated in Chapter 5 using the heterosegmented version of ion-based SAFT2. In that Chapter, the cation of these imidazolium ILs is modeled as a chain molecule with one spherical segment representing the cation head and several groups of segments representing the neutral alkyl branches. The anion of ILs are modeled as one spherical segment. The cross association between cation head and anion is considered to account for the formation of ion pairs, while the water-cation and water-anion hydrogen-bond interaction are not explicitly accounted for. The restricted primitive model (RPM) of the mean spherical approximation (MSA) is used to handle the electrostatic interactions, although the MSA was originally derived without considering the effects of neutral components on the electrostatic interactions (in the case of ILs, neutral components are alkyl branches). A set of transferrable parameters are obtained for imidazolium ILs, and these transferrable parameters are assigned to segments instead of IL molecules, which reduce the number of parameters needed for a group of IL solutions. In general, the liquid densities, activity coefficients, and osmotic coefficients of aqueous imidazolium IL solutions are well correlated or predicted by heterosegmented SAFT2. In addition, the methane hydrate dissociation pressures in the presence of imidazolium ILs are studied by heterosegmented SAFT2 coupled with the van der Waals and Platteeuw theory. Good agreement between model predictions and experimental data can be found for the methane hydrate dissociation conditions in the presence of imidazolium ILs. The effects of concentration, type of anion or cation, and alkyl length of ILs on the inhibition effects are successfully captured. However, for some aqueous ILs solutions in high concentration range, there is still obvious discrepancy between the calculated results of heterosegmented SAFT2 and experimental data, which indicates that a more realistic model is needed.

To improve the heterosegmented SAFT model for aqueous IL solutions, the effect of neutral alkyl groups on the electrostatic interactions should be considered, and the cation heads and anions of some ILs may need to be modeled as charged chains instead of charged single segments. In Chapter 6, the effect of neutral alkyl branches on the electrostatic interactions are studied by Monte Carlo simulations and Ornstein–Zernike (OZ) equation. The radial distribution function, excess internal energy, and excess Helmholtz energy of symmetric and asymmetric electrolyte system in mixture with neutral components are obtained by Monte Carlo simulation, and compared with the prediction from solving the OZ equation with the hypernetted chain closure (HNC) and MSA. The predictions of radial distribution function and excess energies from the OZ equation with the HNC closure are in excellent agreement with simulation results, while the excess energies calculated from MSA deviate from simulation results in the intermediate to high density range. An empirical modification to MSA, referred to as KMSA, is proposed by introducing a density-dependent parameter $K$. The parameter $K$, which is obtained by fitting the excess energies of HNC for an equal-sized, type 1:1 charged system without neutral component, significantly improves the accuracy of MSA and can be used to capture the effects of neutral component, size and charge asymmetry, system temperature, and dielectric constant of the background solvent on the excess energy of electrolyte systems. In Chapter 7, a thermodynamic perturbation theory based equation of state is proposed for flexible charged hard-sphere chains representing polyelectrolytes and polyampholytes, and Monte Carlo simulations are conducted to obtain the excess Helmholtz energies and osmotic coefficients of charged hard-sphere chain fluids. For the modeling of the charged hard-sphere chains, dimer and dimer-monomer approaches are proposed since the use of solely monomer structure information is found to be insufficient. The excess Helmholtz energies and osmotic coefficients predicted by the proposed equation of state using dimer or dimer-monomer approach are found to be in excellent agreement with Monte Carlo results, and the dimer-monomer approach is applicable to the modeling of polyampholyte molecule with any charge configuration and chain length.

Based on the findings in Chapters 6 and 7, an improved heterosegmented SAFT equation of state is proposed for aqueous IL solutions in Chapter 8. In this modified SAFT equation of state, MSA is replaced by KMSA to capture the effects of neutral components, and the cation heads and anions of some ILs are modeled as charged hard-sphere chains instead of spherical segments. The electrical charge of cation head and anion is assumed to be evenly distributed on the chain. With the modified SAFT equation of state, the liquid densities, activity coefficients, and osmotic coefficients of aqueous imidazolium and ammonium ILs are calculated, and a good representation of experiment data can be found. The inhibition effects of ammonium ILs on methane hydrate are also predicted by the modified SAFT equation of state coupled with van der Waals and Platteeuw theory. For the ILs studied in this dissertation, it is found that ILs with shorter alkyl branches have stronger inhibition effects on gas hydrate. The proposed SAFT equation of state, when coupled with van der Waals and Platteeuw theory, is reliable to model the inhibition effects of charged systems on gas hydrate.

## 9.2 Future Work

In this dissertation, the thermodynamic properties of charged systems, especially ionic liquid (IL), are studied by statistical associating fluid theory (SAFT) and Monte Carlo simulations. When coupled with the van der Waals and Platteeuw theory, the SAFT equation of state is a reliable and predictive tool to model the hydrate phase equilibrium in the presence of charged inhibitors. Although significant improvements have been made for the heterosegmented version of SAFT, the modeling of aqueous IL solutions is still a challenge and further refinements of the SAFT equation of state are needed. In the current model of heterosegmented SAFT, the cation heads and anions of some ILs are treated as charged hard-sphere chain. Although the charged hard-sphere chain model is believed to be more realistic than the charged segment model, which is used in Chapter 5, it neglects the dispersive interactions of cation head and anion. Therefore, a charged square-well chain model is needed, and an equation of state for charged square-well chain fluids will be developed in the future. Monte Carlo simulations will also be conducted for the flexible charged square-well chain fluids.

After the equation of state for charged square-well chain fluid is developed and incorporated in the heterosegmented SAFT, a wider variety of ILs will be studied with the new SAFT model and a screening for more effective IL based hydrate inhibitors will be conducted.